\documentclass{article}
\PassOptionsToPackage{numbers}{natbib}
\usepackage[main,final]{neurips_2026}
\usepackage[utf8]{inputenc}
\usepackage[T1]{fontenc}
\usepackage{newunicodechar}
\newunicodechar{ł}{\l}
\usepackage{microtype}
\usepackage{amsmath,amssymb}
\usepackage{graphicx}
\usepackage{wrapfig}
\usepackage[table]{xcolor}
\usepackage{booktabs,multirow,array,tabularx,makecell}
\usepackage{adjustbox}
\usepackage{placeins}
\usepackage{float}
\usepackage{enumitem}
\usepackage{algorithm}
\usepackage{algpseudocode}
\usepackage{xurl}
\usepackage[hidelinks]{hyperref}
\hypersetup{pdftitle={AttnDiff: Attention-based Differential Fingerprinting for Large Language Models}}
\makeatletter
\if@anonymous
  \hypersetup{pdfauthor={Anonymous Author(s)}}
\else
  \hypersetup{pdfauthor={Haobo Zhang, Zhenhua Xu, Junxian Li, Shangfeng Sheng, Dezhang Kong, Meng Han}}
\fi
\makeatother
\algrenewcommand\algorithmicrequire{\textbf{Require:}}
\algrenewcommand\algorithmicensure{\textbf{Return:}}

\definecolor{BestCell}{RGB}{248,206,204}
\definecolor{SecondCell}{RGB}{213,232,212}
\definecolor{ThirdCell}{RGB}{236,243,253}
\definecolor{VictimCell}{RGB}{255,242,204}
\newcommand{\bestcell}[1]{\cellcolor{BestCell}#1}
\newcommand{\secondcell}[1]{\cellcolor{SecondCell}#1}
\newcommand{\thirdcell}[1]{\cellcolor{ThirdCell}#1}
\newcommand{\legendblock}[1]{\begingroup\setlength{\fboxsep}{0pt}\colorbox{#1}{\phantom{\rule{0.6em}{0.6em}}}\endgroup}

\newcommand{\attndiff}{\textsc{AttnDiff}}

\title{\attndiff: Attention-based Differential Fingerprinting for Large Language Models}
\author{%
  Haobo Zhang\textsuperscript{2,3}\thanks{Equal contribution.}\quad
  Zhenhua Xu\textsuperscript{1,3}\footnotemark[1]\quad
  Junxian Li\textsuperscript{4} \\
  \textbf{Shangfeng Sheng}\textsuperscript{5}\quad
  \textbf{Dezhang Kong}\textsuperscript{1,3}\thanks{Corresponding authors.}\quad
  \textbf{Meng Han}\textsuperscript{1,3,6}\footnotemark[2] \\
  \textsuperscript{1}Zhejiang University\quad
  \textsuperscript{2}Renmin University of China \\
  \textsuperscript{3}Binjiang Institute of Zhejiang University \\
  \textsuperscript{4}Shanghai Jiao Tong University \\
  \textsuperscript{5}University of Science and Technology of China\quad
  \textsuperscript{6}Gentel.io \\
  \texttt{haobo0119@gmail.com}\quad
  \texttt{\{xuzhenhua0326, mhan\}@zju.edu.cn}
}

\begin{document}
\nocite{akiba2024evomodelmerge,azerbayev2023llemma,bai2023qwen,bai2022constitutional,cai2025utf,chiang2023vicuna,cui2023chinese-llama,davari2023model,deep2024della,fernandez2023functional,frantar2023gptq,ganguli2022redteaming,grattafiori2024llama3herdmodels,gubri2024trap,guo2025invariant,hotelling1936relations,ilharco2022editing,jiang2023mistral7b,jiang2024mixtralexperts,jin2024proflingo,kornblith2019similarity,kwiatkowski2019natural,lhoest2021datasets,li2023plmmark,li2024doublei,lin2021truthfulqa,deepseek2024v3,luan2025robust,luo2023wizardmath,ma2023llmpruner,meta2023llama,nasery2025robust,nasery2025scalable,ouyang2022instructgpt,pasquini2025llmmap,qwen2025qwen25technicalreport,rafailov2023direct,raghu2017svcca,roziere2023codellama,russinovich2024hey,tang2025resistanceneuralnetworkwatermarking,gemmateam2024gemma,tong2025seedprints,touvron2023llama2,Wan2024KnowledgeFusion,wan2024fusechat,wang2026srafstealthyrobustadversarial,wu2025llm,xu2024instructional,xu2025evertracer,xuInStyRobustMultilevel2025,xu2025fingerprintvectorenablingscalable,xuUnlockingEffectivenessLoRAFP2025,xu2025ctcc,Xu2025CopyrightSurvey,xu2026forgetmarkstealthyfingerprintembedding,xu2026dnfduallayernestedfingerprinting,yadav2023ties,yang2024model,yang2022metafinger,YangWu2024Fingerprint,yu2023language,yuePREEHarmlessAdaptive2025,zeng2024huref,zhang2024reef,zhang2025meraser,zhang2024emmark}

\maketitle
\begin{abstract}
Protecting the intellectual property of open-weight large language models (LLMs) requires verifying whether a suspect model is derived from a victim model despite common laundering operations such as fine-tuning (including PPO/DPO), pruning/compression, and model merging. We propose \attndiff, a data-efficient \textbf{white-box} framework that extracts fingerprints from models via intrinsic information-routing behavior. \attndiff{} probes minimally edited prompt pairs that induce controlled semantic conflicts, captures \emph{differential attention} patterns, summarizes them with compact spectral descriptors, and compares models using CKA. Across Llama-2/3 and Qwen2.5 (3B--14B) and additional open-source families, it yields high similarity for related derivatives while separating unrelated model families (e.g., $>0.98$ vs.\ $<0.22$ with $M=60$ probes). With 5--60 multi-domain probes, it supports practical provenance verification and accountability; our open-source implementation is available at \url{https://github.com/zhb0119/AttnDiff}.
\end{abstract}

\section{Introduction}
Large Language Models (LLMs), including Llama~\citep{touvron2023llama2}, Qwen~\citep{bai2023qwen}, and DeepSeek~\citep{deepseek2024v3}, require substantial data and computation to develop, making their weights valuable intellectual property (IP). Open-weight releases face risks of unauthorized redistribution and model laundering~\citep{Xu2025CopyrightSurvey,zhang2024reef}. Model provenance verification therefore seeks to determine whether a suspect derives from a given reference model despite subsequent modification~\citep{Xu2025CopyrightSurvey}.

\begin{figure}[t]
  \centering
  \includegraphics[width=\linewidth]{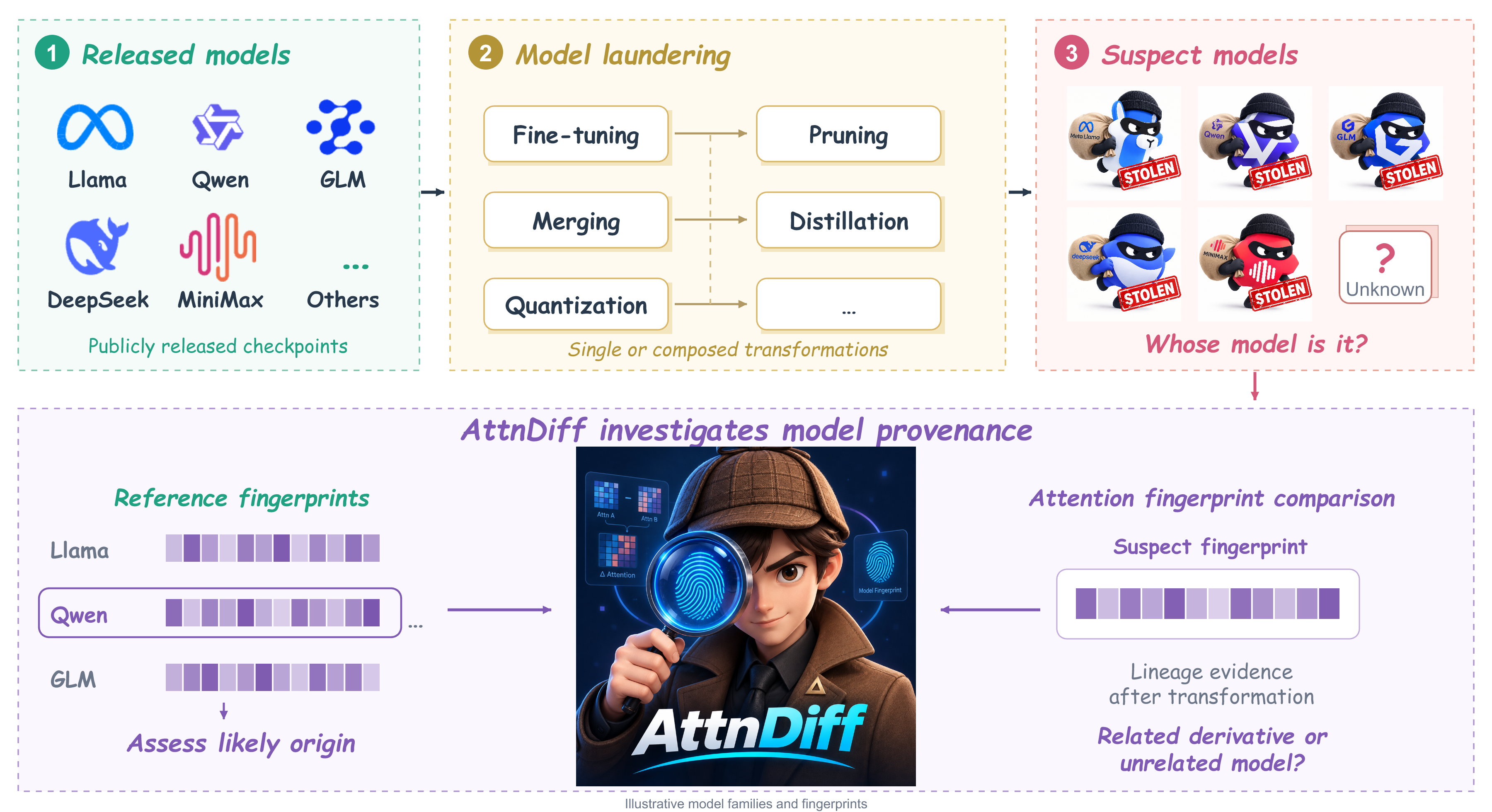}
  \caption{Model provenance verification with \attndiff. \textbf{Top:} publicly released checkpoints may undergo individual or composed transformations, resulting in suspect models whose origins are uncertain. \textbf{Bottom:} \attndiff{} independently extracts differential attention fingerprints from reference and suspect models using the same probe pairs, then compares them to assess evidence of shared lineage. Multiple references illustrate repeated pairwise comparisons. Model families, fingerprint patterns, and the highlighted match are illustrative.}
  \label{fig:attndiff_teaser}
\end{figure}

Figure~\ref{fig:attndiff_teaser} illustrates this problem: released checkpoints may undergo fine-tuning, pruning, merging, distillation, or quantization before appearing as suspects with uncertain origins. Verification requires fingerprints that retain evidence of lineage after these transformations.

Ownership verification methods include invasive and non-invasive approaches~\citep{Xu2025CopyrightSurvey}. Invasive fingerprints embed signals through training-time modifications, including backdoor watermarks~\citep{xu2025ctcc,xu2024instructional,li2023plmmark,nasery2025scalable}. They alter model weights, creating trade-offs between fingerprint capacity and utility and potentially introducing vulnerabilities~\citep{wang2026sightmindunveilinglatent}, spurious triggers, or degraded generalization~\citep{xu2025evertracer,xu2025ctcc,xu2024instructional}. They also cannot retrospectively protect a checkpoint stolen before signal embedding.

Non-invasive, or \emph{intrinsic}, fingerprints extract signatures without modifying parameters. Four families have distinct robustness gaps: \textbf{parameter-based} fingerprints~\citep{zeng2024huref} are fragile under structural edits (Sec.~\ref{sec:exp_pruning}); \textbf{representation-based} fingerprints~\citep{YangWu2024Fingerprint,zhang2024reef} can drift under architecture changes, distillation (Appendix~\ref{sec:appendix_distill}), and preference optimization (Appendix~\ref{sec:appendix_po}); \textbf{adversarial-example-based} fingerprints~\citep{jin2024proflingo} rely on potentially conspicuous triggers that input perturbations can neutralize~\citep{xu2025ctcc,xu2025evertracer}; and \textbf{semantic} fingerprints~\citep{pasquini2025llmmap} depend on generated text and decoding, making them sensitive to paraphrasing and rewriting (Appendix~\ref{sec:appendix_llmmap_attacks}).

We propose \attndiff, a data-efficient white-box framework based on \emph{differential attention dynamics} under controlled semantic conflicts. Its central hypothesis is that attention re-routing encodes lineage-specific structure that persists under model modification. As illustrated in the lower panel of Figure~\ref{fig:attndiff_teaser}, independently extracted reference and suspect fingerprints provide evidence for assessing shared lineage; comparisons can be repeated across candidate references. Concretely, single-word pivots create minimally edited prompt pairs, such as ``parallel lines never intersect'' versus ``always intersect''. \attndiff{} extracts their attention differences across layers and heads, compresses these into spectral descriptors, and compares models using centered linear CKA~\citep{kornblith2019similarity}. This comparison captures inter-probe structure while allowing different feature dimensions. Our experiments assess whether these fingerprints remain similar for related derivatives while separating unrelated models under common transformations.

\section{Related work}

\subsection{Invasive fingerprints}
Invasive fingerprints embed ownership signals by modifying models during (pre-)training, typically via weight watermarks or backdoor triggers. Weight watermark-based schemes encode verifiable patterns in parameter space~\citep{zhang2024emmark,guo2025invariant,fernandez2023functional,luan2025robust,tang2025resistanceneuralnetworkwatermarking} but face capacity--utility trade-offs and may lose robustness under pruning, quantization, or strong fine-tuning~\citep{Xu2025CopyrightSurvey}. Backdoor-based fingerprints use trigger datasets for black-box attribution~\citep{xu2024instructional,cai2025utf,xu2026dnfduallayernestedfingerprinting,xu2025ctcc,xuInStyRobustMultilevel2025,xu2026forgetmarkstealthyfingerprintembedding,li2024doublei,russinovich2024hey}, yet require training-time control and can be removed by targeted erasure~\citep{zhang2025meraser}. Recent work explores scalable fingerprint transfer~\citep{xuUnlockingEffectivenessLoRAFP2025,xu2025fingerprintvectorenablingscalable}, harmless editing~\citep{yuePREEHarmlessAdaptive2025}, and adversarial robustness~\citep{wang2026srafstealthyrobustadversarial,mao2026starestepwisetemporalalignment}.
\subsection{Intrinsic fingerprints}

 Non-invasive (intrinsic) fingerprints do not inject external signals and instead mine model-inherent properties for provenance verification~\citep{Xu2025CopyrightSurvey}. Representative families include parameter-space statistics in weight space~\citep{zeng2024huref}, representation-based descriptors over hidden states/logits~\citep{zhang2024reef,YangWu2024Fingerprint,tong2025seedprints,zhang2026memmarkstateevolutionattributionwatermarking}, semantic output traces~\citep{pasquini2025llmmap,wu2025llm}, and adversarial-example-based triggers~\citep{jin2024proflingo,gubri2024trap}. However, these signals can be brittle under realistic transformations, e.g., pruning/architecture edits, distillation, or preference optimization, and/or sensitive to decoding and input perturbations, limiting robustness across fine-tuning, pruning/merging, and distillation.

 Among intrinsic approaches, \attndiff{} departs from parameter-, representation-, and output-level signatures by fingerprinting models through their \emph{differential attention} responses to controlled semantic conflicts, with similarity measured via CKA.

\section{Threat model}

We study a post-hoc model forensics setting where a \textit{defender} (the legitimate model owner) verifies the provenance of a released or seized \textit{suspect} model, while an \textit{attacker} seeks to evade attribution after illicitly obtaining the model weights. We focus on realistic \textbf{white-box} model laundering operations that modify a stolen model's parameters and/or architecture while maintaining utility.

\textbf{Attacker capabilities.} With full white-box access, the attacker can apply arbitrary transformations, including supervised fine-tuning (SFT), alignment and PO (e.g., reinforcement learning from human feedback (RLHF) with Proximal Policy Optimization (PPO) or Direct Preference Optimization (DPO)), pruning, quantization, distillation to a different architecture, and model merging or weight interpolation. This setting is consistent with prior work on model fingerprinting~\citep{yang2022metafinger,jin2024proflingo}, where attackers are assumed to have access to the original model and can fine-tune it using validation data to evade attribution.

We also evaluate an adaptive attacker who knows the extraction procedure and obtains some defender probes, then trains the suspect to suppress differential attention (Sec.~\ref{sec:exp_adaptive}).

\textbf{Defender capabilities.} The defender has access to the original reference model and is granted white-box access to the suspect model during verification, consistent with legal/auditing workflows. This allows the defender to inspect internal states and compute fingerprints for both models under the same probing protocol. Verification requires inference-time attention access, but no gradients or model updates.

\section{Design of \attndiff}
\subsection{Motivation}
In post-hoc LLM forensics, a suspect is rarely an ``exact-copy'': stolen checkpoints are routinely laundered through fine-tuning/alignment, pruning, merging, or distillation. A practical fingerprint should therefore remain stable under such transformations, where weight-space statistics can break under structural edits and hidden-state/logit signatures can drift after strong adaptation.

We encountered an empirical turning point while stress-testing small, highly controlled probes. We constructed minimally edited origin/corrupted prompt pairs that preserve surface form while flipping semantics via a single-word lexical pivot. Under these ``controlled semantic conflicts'', attention did not drift arbitrarily; models re-allocated attention in a structured manner. More importantly, this re-allocation appeared family-consistent: Llama-2-7B~\citep{touvron2023llama2} derivatives exhibited closely matched profiles, whereas an unrelated model family such as Qwen2.5-7B~\citep{qwen2025qwen25technicalreport} followed a systematically different regime. Together, these observations suggested that a model's intrinsic information-routing strategy---how it redistributes self-attention when resolving a semantic contradiction---may encode lineage-specific structure that survives common ``laundering'' operations. A detailed pilot study quantifying these re-allocation patterns via interpretable structural statistics (concentration, entropy, locality) is presented in Appendix~\ref{sec:supplementary_materials}.

These observations motivate \attndiff: by probing a model with minimally edited prompt pairs that induce controlled semantic conflicts, we can elicit a characteristic re-routing response that is stable within a model lineage yet distinct across unrelated families. The next subsection describes how we operationalize this idea into a practical fingerprinting pipeline.

\subsection{\attndiff\ workflow}
\label{sec:probe_construction}
\textbf{Notation.} We denote a model with $L$ layers and $H$ attention heads per layer. For a prompt $p$ tokenized into $N$ tokens, $A^{(i)}_{l,h}\in\mathbb{R}^{N\times N}$ represents the post-softmax causal self-attention probability matrix at layer $l$ and head $h$ for the $i$-th probe pair. We use $\tilde{p}$ and $\tilde{A}$ to denote the corrupted prompt and its corresponding attention map. The differential attention matrix is $\Delta A=\tilde{A}-A$. The fingerprint matrix $F\in\mathbb{R}^{M\times D}$ stacks spectral descriptors from $M$ probe pairs, where $D=LHK$ and $K$ is the spectral rank.

\begin{figure}[t]
  \centering
  \includegraphics[width=\linewidth]{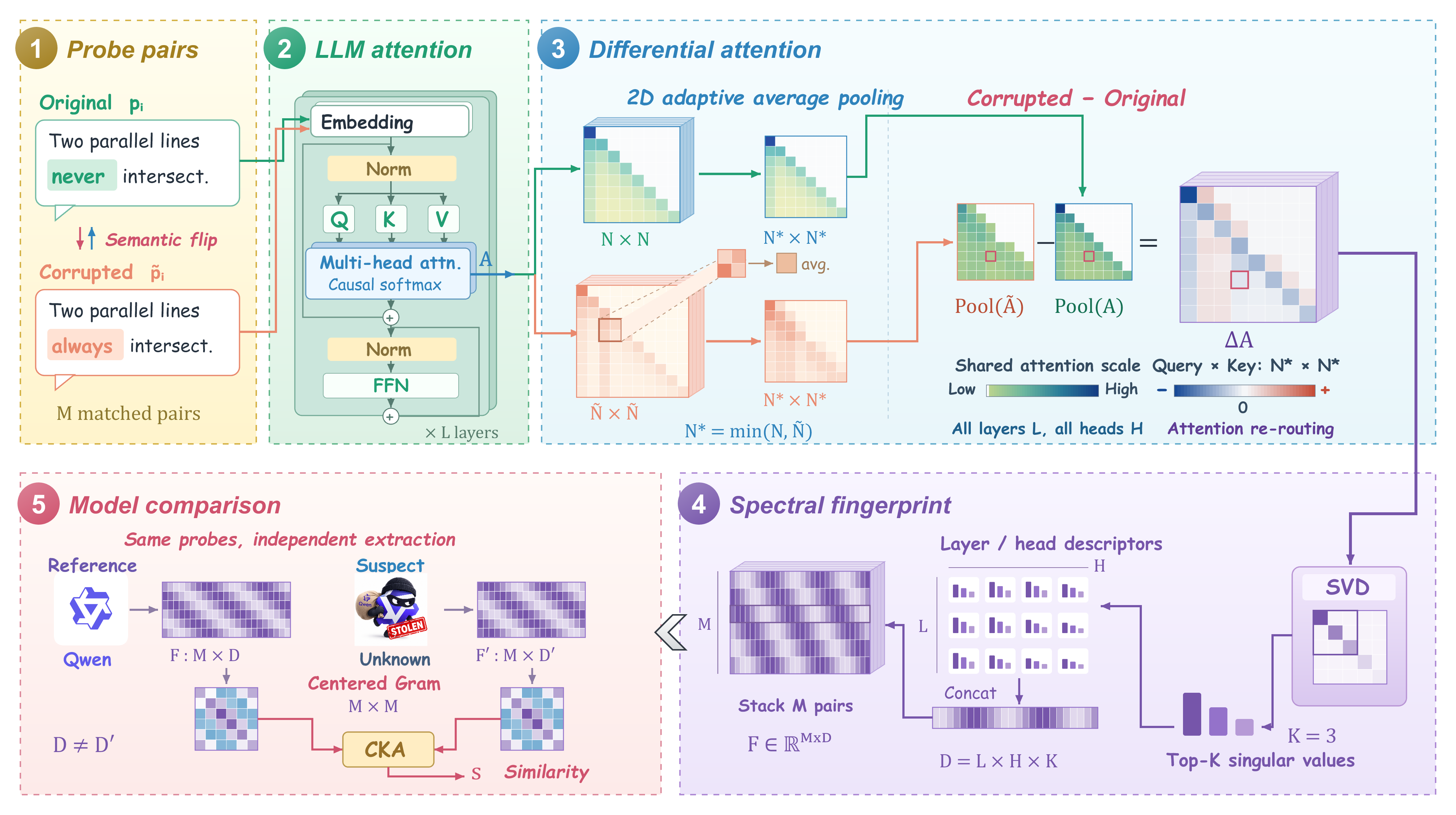}
  \caption{\attndiff\ workflow, following steps \textbf{1--3} across the top, then \textbf{4} at bottom right and \textbf{5} at bottom left. A single-word pivot creates a semantic conflict. Causal attention maps are aligned to $N^\ast=\min(N,\tilde{N})$ before computing $\Delta A=\mathrm{Pool}(\tilde{A})-\mathrm{Pool}(A)$. SVD yields the top-$K$ singular values per layer/head ($K=3$), which are concatenated and stacked into $F\in\mathbb{R}^{M\times D}$, $D=LHK$. Reference and suspect fingerprints use the same ordered probes and are extracted independently. Their centered $M\times M$ Gram matrices support CKA comparison even when $D\neq D'$. Qwen is an illustrative reference model; heatmaps and spectra are schematic.}
  \label{fig:attndiff_workflow}
\end{figure}

\attndiff\ fingerprints LLMs by capturing their \textbf{differential attention dynamics} under semantic conflict. Figure~\ref{fig:attndiff_workflow} presents five steps: (1) \textbf{Probe pairs:} construct $M$ origin/corrupted pairs via single-word pivot substitutions; (2) \textbf{LLM attention:} extract post-softmax causal attention maps for each prompt; (3) \textbf{Differential attention:} align each pair of maps by adaptive average pooling when needed, then subtract origin from corrupted attention; (4) \textbf{Spectral fingerprint:} retain the top-$K$ singular values per layer/head ($K=3$ by default), concatenate the descriptors, and stack all pairs into $F\in\mathbb{R}^{M\times D}$; (5) \textbf{Model comparison:} independently extract $F$ and $F'$ from the reference (victim) and suspect models using the same ordered probe pairs, then compare their centered Gram matrices using CKA. The score $s=\mathrm{CKA}(F,F')$ measures similarity in differential routing behavior. Implementation details are provided in Appendix~\ref{sec:attndiff_details}.

\subsubsection{Probe dataset}
We construct a probe set of $M$ prompts spanning six domains: \textit{Code}, \textit{Math}, \textit{Economics}, \textit{Medicine}, \textit{Daily QA}, and \textit{Safe Alignment}. Since post-training typically specializes a model toward a particular domain, we diversify probe domains to reduce domain-specific drift in internal routing behaviors and to provide complementary anchors for fingerprint comparison. In addition, \textit{Safe Alignment} is included to cover policy-gated behaviors (e.g., refusal or cautious responses) that can follow a distinct internal routing regime compared to capability-oriented queries, providing a complementary anchor under realistic instruction/alignment settings.

We then construct $M$ origin/corrupted probe pairs for fingerprint extraction, as detailed below.

\subsubsection{Differential fingerprint}
\label{sec:differential_fingerprint}
For the $i$-th probe pair $(p_i,\tilde{p}_i)$, we generate a corrupted prompt $\tilde{p}_i$ from the origin prompt $p_i$ via a \emph{single-word (lexical) pivot} substitution, yielding a \emph{semantic conflict}. Tokenization is performed using the target model's tokenizer on \emph{only the prompt string itself} (excluding chat templates or system prefixes). Typically $\tilde{N}=N$, with rare mismatches due to tokenizer segmentation.

For each layer and head, we extract the post-softmax causal self-attention probability matrices $A^{(i)}_{l,h}\in\mathbb{R}^{N\times N}$ and $\tilde{A}^{(i)}_{l,h}\in\mathbb{R}^{\tilde{N}\times\tilde{N}}$, where strictly upper-triangular entries are masked to zero. We refer to the stacked tensors as the \textbf{origin attention map} $\mathcal{A}^{(i)}\in\mathbb{R}^{L\times H\times N\times N}$ and the \textbf{corrupted attention map} $\tilde{\mathcal{A}}^{(i)}\in\mathbb{R}^{L\times H\times\tilde{N}\times\tilde{N}}$.

\begin{wrapfigure}{r}{0.45\textwidth}
  \centering
  \includegraphics[width=\linewidth]{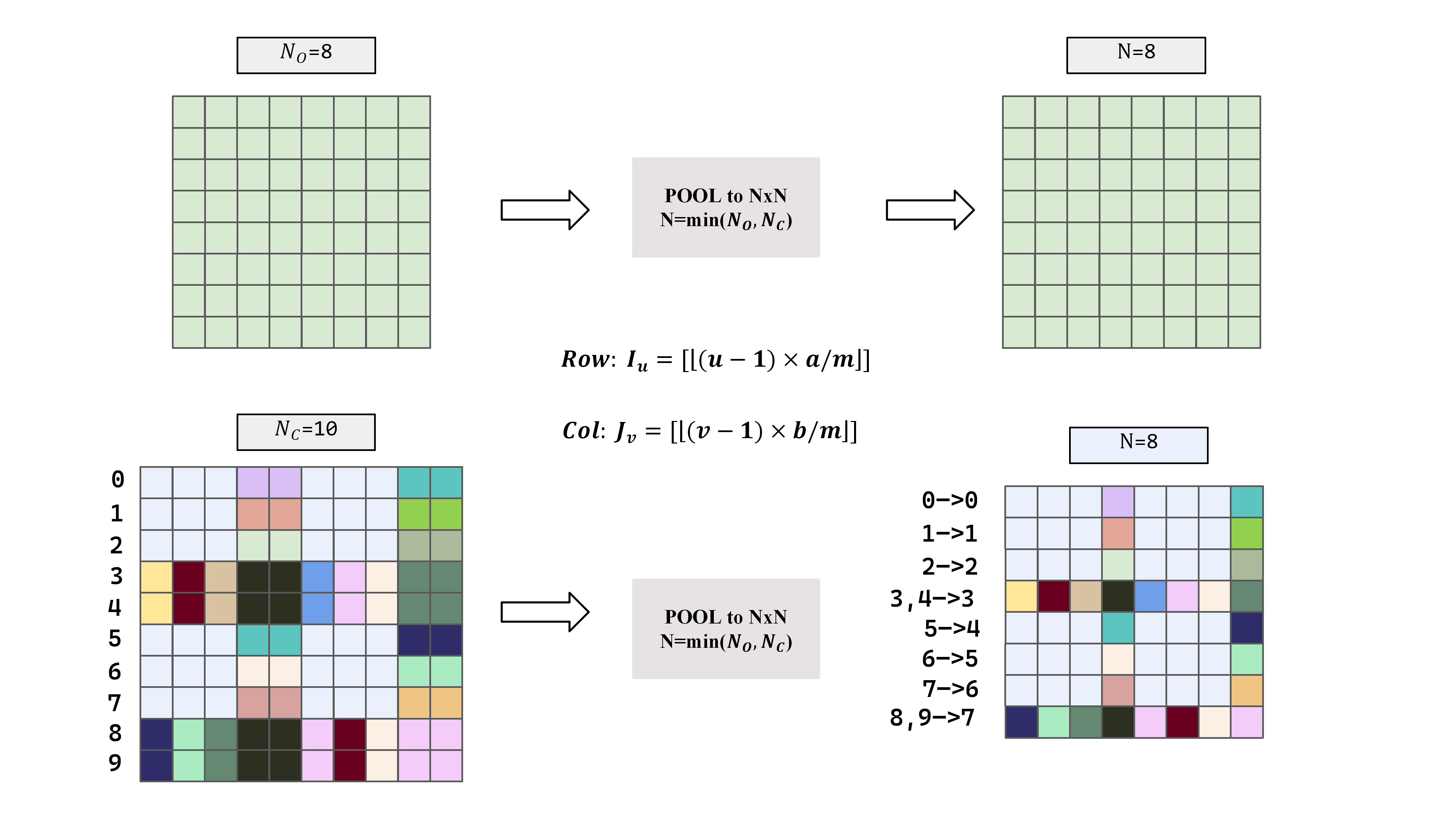}
  \caption{Attention map alignment via 2D adaptive average pooling to resolution $N^{\ast}=\min(N,\tilde{N})$. This preserves coarse-grained routing structure without boundary artifacts when tokenization induces length mismatch.}
  \label{fig:pool_visualization}
\end{wrapfigure}

To handle the rare case $\tilde{N}\neq N$, we align attention maps to a common resolution $N^{\ast}=\min(N,\tilde{N})$ via \textbf{2D adaptive average pooling}, as illustrated in Figure~\ref{fig:pool_visualization}. This operator partitions the input matrix into bins and averages within each bin to produce the target resolution, preserving coarse-grained attention distribution without introducing boundary artifacts. We apply this operator to pool each $A^{(i)}_{l,h}$ and $\tilde{A}^{(i)}_{l,h}$ to $\bar{A}^{(i)}_{l,h},\bar{\tilde{A}}^{(i)}_{l,h}\in\mathbb{R}^{N^{\ast}\times N^{\ast}}$.

We then compute the \textbf{differential attention matrix} $\Delta A^{(i)}_{l,h}=\bar{\tilde{A}}^{(i)}_{l,h}-\bar{A}^{(i)}_{l,h}$ (i.e., $\mathrm{Pool}(\tilde{A}^{(i)}_{l,h})-\mathrm{Pool}(A^{(i)}_{l,h})$), which highlights attention re-routing induced by semantic conflict under aligned token topology. Step~3 in Figure~\ref{fig:attndiff_workflow} uses a shared scale for origin/corrupted attention and a zero-centered diverging scale for their signed difference. To obtain a compact, length-invariant signature, we apply truncated SVD to $\Delta A^{(i)}_{l,h}$ and retain the top-$K$ singular values as a spectral descriptor $\mathbf{s}^{(i)}_{l,h}\in\mathbb{R}^{K}$. Step~4 labels this operation ``Top-$K$ singular values'' and illustrates the default $K=3$. Concatenating across all layers and heads yields a one-dimensional instance-level fingerprint vector $\mathbf{x}^{(i)}\in\mathbb{R}^{LHK}$, and stacking over $M$ probes forms the fingerprint matrix $F\in\mathbb{R}^{M\times LHK}$.

\subsubsection{Similarity measurement by CKA}
To compare two models, we independently compute fingerprints $F$ and $F'$ on the same ordered probe pairs and measure similarity using \textbf{CKA}, as shown in step~5 of Figure~\ref{fig:attndiff_workflow}. Specifically, we form linear Gram matrices $K=FF^\top$ and $K'=F'{F'}^\top\in\mathbb{R}^{M\times M}$ that encode inter-probe relational geometry. Let $H=I-\tfrac{1}{M}\mathbf{1}\mathbf{1}^\top$ be the centering matrix and define centered Gram matrices $\bar{K}=HKH$ and $\bar{K}'=HK'H$. We compute CKA via the Hilbert-Schmidt Independence Criterion (HSIC), where $\mathrm{HSIC}(K,K')\propto\mathrm{tr}(\bar{K}\,\bar{K}')$: $\mathrm{CKA}(F,F')=\frac{\mathrm{HSIC}(K,K')}{\sqrt{\mathrm{HSIC}(K,K)\mathrm{HSIC}(K',K')}}$. We use centered Gram matrices in HSIC to remove mean effects and focus the comparison on similarity structure across probes.

Critically, practical transformations such as structured pruning, distillation, or architecture edits may change model structure (e.g., removing layers or attention heads), yielding fingerprints with different feature dimensions (e.g., $F\in\mathbb{R}^{M\times D}$ and $F'\in\mathbb{R}^{M\times D'}$ with $D\neq D'$ due to different $L$ or $H$). This makes direct feature matching ill-posed. In contrast, CKA operates on probe-indexed Gram matrices in $\mathbb{R}^{M\times M}$ and requires matching probe identities and ordering, enabling direct comparison without explicit layer/head alignment. Consequently, \attndiff\ remains applicable across models with heterogeneous depths and head counts, while retaining robustness to re-parameterizations and other real-world modifications.

\section{Experiment}
\label{sec:experiment}
\begin{wrapfigure}{R}{0.48\textwidth}
  \centering
  \includegraphics[width=0.95\linewidth]{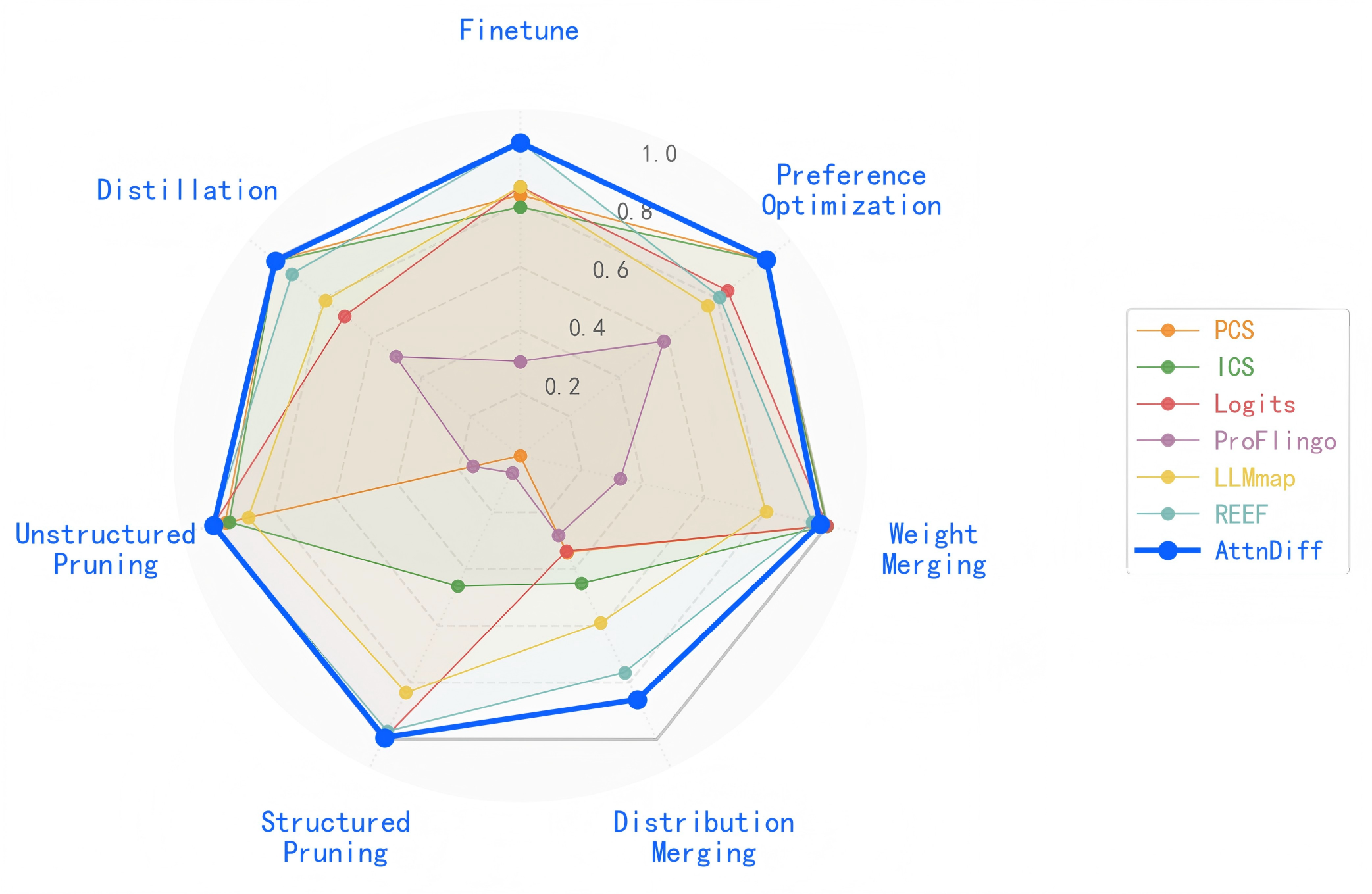}
  \caption{Radar comparison across deployment transformations. Larger radii indicate stronger robustness.}
  \label{fig:radar_overall}
\end{wrapfigure}
We compare \attndiff{} against non-invasive baselines under fine-tuning, merging, and pruning (Sec.~\ref{sec:exp_finetune}--\ref{sec:exp_pruning}), followed by ablations (Sec.~\ref{sec:exp_ablation}), discriminability analysis (Sec.~\ref{sec:exp_discriminability}), and adaptive suppression (Sec.~\ref{sec:exp_adaptive}). Figure~\ref{fig:radar_overall} summarizes robustness across deployment transformations, including preference optimization (Appendix~\ref{sec:appendix_po}) and distillation (Appendix~\ref{sec:appendix_distill}).

\subsection{Experimental setting}
\label{sec:exp_setting}
\textbf{Probe set.} Unless otherwise specified, we construct a probe set with $M=60$ prompts spanning six domains, with 10 prompts per domain: \textit{Code}, \textit{Math}, \textit{Economics}, \textit{Medicine}, \textit{Daily QA}, and \textit{Safe Alignment}. For each origin prompt $p$, we generate a minimally edited corrupted prompt $\tilde{p}$ via a single-word pivot and compute fingerprints on the resulting $M$ probe pairs.

\noindent\textbf{Model.} We evaluate \attndiff{} under realistic model transformation scenarios, covering diverse base architectures and a broad suite of derivative models. Our experiments cover models from the Llama-2/3~\citep{touvron2023llama2,grattafiori2024llama3herdmodels}, Qwen2.5~\citep{qwen2025qwen25technicalreport}, Gemma~\citep{gemmateam2024gemma}, and Mistral~\citep{jiang2023mistral7b} families, spanning parameter scales from 1B to 14B and including downstream adaptations and optimizations such as supervised fine-tuning, preference optimization, knowledge distillation, pruning, and model merging.

\noindent\textbf{Baselines.} Following the taxonomy in the Introduction, we select representative \emph{non-invasive} fingerprinting baselines that cover both black-box and white-box access assumptions. Specifically, we include PCS/ICS~\citep{zeng2024huref} as \emph{parameter-based} fingerprints, Logits~\citep{YangWu2024Fingerprint} and REEF~\citep{zhang2024reef} as \emph{representation-based} fingerprints, ProFlingo~\citep{jin2024proflingo} as an \emph{adversarial-example-based} fingerprint, and LLMMap~\citep{pasquini2025llmmap} as a \emph{semantic} fingerprint. For each method, we report the similarity score produced by its original formulation.

Our experiments ask: \textbf{RQ1:} How robust is \attndiff{} to fine-tuning and preference optimization? \textbf{RQ2:} Does it withstand diverse merging strategies? \textbf{RQ3:} How robust is it to pruning/compression? \textbf{RQ4:} Does probe-aware suppression transfer to held-out or refreshed probes?

\subsection{Model fine-tuning}
\label{sec:exp_finetune}
\begin{table}[!htbp]
  \centering
\caption{SFT robustness results (similarity score) on Llama-2-7B-derived suspect models.}
\label{tab:sft-robustness}
  \small
  \begin{adjustbox}{max width=\textwidth}

  \setlength{\tabcolsep}{3.5pt}
  \renewcommand\arraystretch{1.05}
  \begin{tabular}{lccccccc}
    \toprule
    \rowcolor{gray!20}
     & \begin{tabular}[c]{@{}c@{}}Llama-2-\\Finance-7B\\(5M tokens)\end{tabular} & \begin{tabular}[c]{@{}c@{}}Vicuna-\\1.5-7B\\(370M tokens)\end{tabular} & \begin{tabular}[c]{@{}c@{}}WizardMath-\\7B\\(1.8B tokens)\end{tabular} & \begin{tabular}[c]{@{}c@{}}ChineseLLaMA-\\2-7B\\(13B tokens)\end{tabular} & \begin{tabular}[c]{@{}c@{}}CodeLLaMA-\\7B\\(500B tokens)\end{tabular} & \begin{tabular}[c]{@{}c@{}}Llemma-\\7B\\(700B tokens)\end{tabular} & Avg \\
    \midrule
    PCS      & \thirdcell{0.9979} & \thirdcell{0.9985} & 0.9965 & \thirdcell{0.9390} & 0.5301 & 0.5052 & 0.8279 \\
    ICS      & 0.9952 & 0.9949 & \secondcell{0.9985} & 0.7309 & 0.5112 & 0.5104 & 0.7902 \\
    Logits   & \bestcell{0.9999} & \bestcell{0.9999} & \bestcell{0.9999} & 0.7033 & 0.7833 & 0.6367 & 0.8538 \\
    REEF     & 0.9950 & \thirdcell{0.9985} & \thirdcell{0.9979} & \bestcell{0.9974} & \bestcell{0.9947} & \bestcell{0.9962} & \bestcell{0.9966} \\
    ProFlingo& 0.2400 & 0.5200 & 0.4200 & 0.2800 & 0.2000 & 0.1400 & 0.3000\\
    LLMMap   & 0.8986 & 0.7294 & 0.7691 & 0.8720 & \thirdcell{0.9555} & \thirdcell{0.8998} & \thirdcell{0.8541} \\
    Ours     & \secondcell{0.9989} & \secondcell{0.9986} & \secondcell{0.9985} & \secondcell{0.9963} & \secondcell{0.9890} & \secondcell{0.9856} & \secondcell{0.9945} \\
    \bottomrule
  \end{tabular}
  
\end{adjustbox}
  \par\smallskip
  \textit{Suspect model headers include the fine-tuning data scale. Cell shading (per column): \legendblock{BestCell}=best, \legendblock{SecondCell}=second, \legendblock{ThirdCell}=third.}

\end{table}

Fine-tuning is a critical stage in the LLM lifecycle for adapting base models to downstream tasks or aligning them with human values~\citep{ouyang2022instructgpt,rafailov2023direct}. This process involves extensive parameter updates that can significantly shift the model's internal representations and output distributions, posing a severe test for fingerprint robustness~\citep{nasery2025robust}.

For SFT, we use Llama-2-7B as the victim model and collect suspect models fine-tuned with different data scales and application domains, including Llama-2-finance-7b~\citep{meta2023llama}, Vicuna-1.5-7b~\citep{chiang2023vicuna}, WizardMath-7b~\citep{luo2023wizardmath}, Chinese-LLaMA-2-7b~\citep{cui2023chinese-llama}, CodeLLaMA-7b~\citep{roziere2023codellama}, and Llemma-7b~\citep{azerbayev2023llemma}.

Table~\ref{tab:sft-robustness} indicates that large-scale SFT can substantially alter model parameters/representations, causing parameter-based fingerprints (PCS/ICS) to drop sharply on heavily fine-tuned suspects. Logits also drops to 0.7833/0.6367 on CodeLLaMA-7B and Llemma-7B. In contrast, \attndiff{} maintains uniformly high similarity ($>0.985$) across all SFT suspects and is comparable to the SOTA baseline REEF.

\subsection{Model merge}
\label{sec:exp_merge}
\begin{table}[!htbp]
  \centering
\caption{Model merge robustness results (similarity score) on open-source merged suspects.}
\label{tab:merge-open-source}
  \small
  \begin{adjustbox}{max width=\textwidth}

  \setlength{\tabcolsep}{3.5pt}
  \renewcommand\arraystretch{1.05}
  \begin{tabular}{lcccccc}
    \toprule
    \rowcolor{gray!20}
     & \multicolumn{3}{c}{Weight Merging (Evollm-jp-7b)} & \multicolumn{3}{c}{Distribution Merging (Fusellm-7b)} \\
     \cmidrule(lr){2-4} \cmidrule(lr){5-7}
    \rowcolor{gray!20}
     & Shisa-gamma-7b-v1 & Wizardmath-7b-1.1 & Abel-7b-002 & Llama-2-7b & Openllama-2-7b & Mpt-7b \\
    \midrule
    PCS       & \bestcell{0.9992} & \secondcell{0.9990} & \thirdcell{0.9989} & \secondcell{0.9997} & 0.0194 & 0.0000 \\
    ICS       & \bestcell{0.9992} & \thirdcell{0.9988} & \thirdcell{0.9988} & 0.9986 & 0.2478 & 0.1014 \\
    Logits    & \secondcell{0.9933} & \bestcell{0.9999} & \bestcell{0.9999} & \bestcell{0.9999} & 0.0100 & 0.0000 \\
    REEF      & 0.9635 & 0.9526 & 0.9374 & \thirdcell{0.9996} & \secondcell{0.6713} & \secondcell{0.6200} \\
    ProFlingo & 0.3000& 0.4000& 0.2800& 0.5400& 0.1400& 0.1600\\
    LLMMap    & 0.7651 & 0.8343 & 0.8011 & 0.9511 & \thirdcell{0.5742} & \thirdcell{0.2413} \\
    Ours      & \thirdcell{0.9726} & 0.9561 & \secondcell{0.9996} & 0.9962 & \bestcell{0.7953} & \bestcell{0.7851} \\
    \bottomrule
  \end{tabular}
  
\end{adjustbox}
  \par\smallskip
  \textit{Cell shading (per column): \legendblock{BestCell}=best, \legendblock{SecondCell}=second, \legendblock{ThirdCell}=third.}

\end{table}

Model merging combines multiple pretrained models in weight space or at the distribution/behavior level, often integrating complementary capabilities without accessing training data or retraining~\citep{yang2024model,akiba2024evomodelmerge,Wan2024KnowledgeFusion,wan2024fusechat,yu2023language,ilharco2022editing}. Because a merged suspect is derived from multiple victim models, its fingerprints can be mixed and harder to attribute to all sources.

From Table~\ref{tab:merge-open-source}, \attndiff{} consistently attains high similarity with all parent models in weight-space merges ($\geq 0.95$) and maintains substantial similarity ($\approx 0.78$--$0.80$) with heterogeneous distribution-level merges, whereas several baselines either fail to attribute all sources or exhibit noticeably weaker alignment on cross-architecture merges.

\subsection{Model pruning}
\label{sec:exp_pruning}
Model pruning compresses LLMs by removing redundant parameters to improve efficiency, often followed by retraining to recover capabilities. This poses a distinct challenge to model fingerprinting, as it fundamentally alters the model's weights and architecture. Recent studies have shown that pruning can weaken the verification effectiveness of multiple fingerprinting schemes~\citep{xu2025ctcc,xu2025evertracer}. We therefore evaluate robustness against a spectrum of pruning methodologies, including both structured and unstructured pruning on open-source checkpoints as well as additional suspects generated via the LLMPruner toolkit~\citep{ma2023llmpruner}.

Table~\ref{tab:pruning-results} shows that \attndiff{} preserves high similarity across all structured and unstructured pruned suspects (no lower than 0.9879), even when aggressive structured pruning causes parameter-based fingerprints (PCS/ICS) to collapse and ProFlingo/LLMMap to degrade substantially. These results demonstrate that our differential attention fingerprint is robust to diverse pruning and compression strategies.

\begin{table}[!htbp]
  \centering
\caption{Robustness results (similarity score) on pruned suspect models, comparing structured and unstructured pruning strategies.}
\label{tab:pruning-results}
  \small
  \begin{adjustbox}{max width=\textwidth}

  \setlength{\tabcolsep}{3pt}
  \renewcommand\arraystretch{1.05}
  \begin{tabular}{lccccccccc}
    \toprule
    \rowcolor{gray!20}
     & \multicolumn{6}{c}{Structured Pruning} & \multicolumn{3}{c}{Unstructured Pruning} \\
     \cmidrule(lr){2-7} \cmidrule(lr){8-10}
    \rowcolor{gray!20}
     & \begin{tabular}[c]{@{}c@{}}Sheared-\\llama-1.3b-\\pruned\end{tabular} & \begin{tabular}[c]{@{}c@{}}Sheared-\\llama-1.3b\end{tabular} & \begin{tabular}[c]{@{}c@{}}Sheared-\\llama-1.3b-\\sharegpt\end{tabular} & \begin{tabular}[c]{@{}c@{}}Sheared-\\llama-2.7b-\\pruned\end{tabular} & \begin{tabular}[c]{@{}c@{}}Sheared-\\llama-2.7b\end{tabular} & \begin{tabular}[c]{@{}c@{}}Sheared-\\llama-2.7b-\\sharegpt\end{tabular} & \begin{tabular}[c]{@{}c@{}}Sparse-\\llama-2-7b\end{tabular} & \begin{tabular}[c]{@{}c@{}}Wanda-\\llama-2-7b\end{tabular} & \begin{tabular}[c]{@{}c@{}}GBLM-\\llama-2-7b\end{tabular} \\
    \midrule
    PCS       & 0.0000 & 0.0000 & 0.0000 & 0.0000 & 0.0000 & 0.0000 & 0.9560 & 0.9620 & 0.9616 \\
    ICS       & 0.4927 & 0.3512 & 0.3510 & 0.6055 & 0.4580 & 0.4548 & 0.9468 & 0.9468 & 0.9478 \\
    Logits    & \bestcell{0.9967} & \bestcell{0.9999} & \bestcell{0.9999} & \bestcell{0.9967} & \bestcell{0.9999} & \bestcell{0.9999} & \bestcell{0.9999} & \bestcell{0.9999} & \bestcell{0.9999} \\
    REEF      & \thirdcell{0.9368} & \thirdcell{0.9676} & \thirdcell{0.9710} & 0.9278 & \thirdcell{0.9701} & \secondcell{0.9991} & \thirdcell{0.9985} & \secondcell{0.9986} & \thirdcell{0.9991} \\
    ProFlingo & 0.0400& 0.0200 & 0.0800 & 0.0200 & 0.1000 & 0.0800 & 0.1600 & 0.1200 & 0.1800 \\
    LLMMap    & 0.9088 & 0.9007 & 0.8152 & \thirdcell{0.9400} & 0.9236 & 0.7072 & 0.8459& 0.9145& 0.8956\\
    Ours      & \secondcell{0.9879} & \secondcell{0.9938} & \secondcell{0.9903} & \secondcell{0.9929} & \secondcell{0.9952} & \thirdcell{0.9936} & \secondcell{0.9996} & \thirdcell{0.9927} & \secondcell{0.9995} \\
    \bottomrule
  \end{tabular}
  
\end{adjustbox}
  \par\smallskip
  \textit{Cell shading (per column): \legendblock{BestCell}=best, \legendblock{SecondCell}=second, \legendblock{ThirdCell}=third.}

\end{table}

\subsection{Ablation study}
\label{sec:exp_ablation}
We conduct ablation studies to validate key design choices of \attndiff. Beyond the core experiments reported below, we perform extensive hyperparameter and design ablations in the appendix.

\noindent\textbf{Differential mechanism} (Table~\ref{tab:ablation_diff}): comparing against a non-differential baseline that uses standard attention maps, the differential design is crucial under heavy domain adaptation---CodeLlama-7b and Llemma-7b drop to $0.51$/$0.29$ without differencing but recover to $0.99$/$0.99$ with \attndiff{} ($\Delta>0.47$). For unrelated models, differencing reduces spurious similarity from $\sim0.4$ to $<0.23$.

\noindent\textbf{Probe configuration} (Figure~\ref{fig:ablation_combined}, Table~\ref{tab:ablation_domain}): sparse probing ($M\leq15$) yields high spurious similarity among unrelated models; $M=60$ provides optimal discrimination. Domain-specific fine-tuning induces localized shifts in expert domains, but aggregated fingerprints remain stable for related suspects ($>0.98$) and well separated from unrelated architectures ($<0.22$).

\noindent\textbf{Probe length} (Appendix~\ref{sec:probe_construction_details}, Table~\ref{tab:ablation_length}): medium-length prompts ($\ell\approx30$ words) balance semantic context and attention stability.

\noindent\textbf{Spectral rank} (Appendix~\ref{sec:fingerprint_extraction_details}, Table~\ref{tab:ablation_rank}): $K=3$ optimally trades off discriminability and compactness.

\noindent\textbf{Similarity metric} (Appendix~\ref{sec:appendix_metric_analysis}, Table~\ref{tab:metric_comparison}): CKA achieves superior discriminability over CCA/SVCCA with lower spurious similarity on unrelated models (0.1647 vs 0.3112/0.2419).

\FloatBarrier
\subsection{Discriminability analysis}
\label{sec:exp_discriminability}
\begin{wraptable}{r}{0.43\textwidth}
  \centering
  \caption{Threshold-free discriminability (AUROC/AUPR, related=positive).}
  \label{tab:discriminability_auc}
  \small
  \begin{tabular}{lcc}
    \toprule
    \rowcolor{gray!20}\textbf{Method} & \textbf{AUROC} & \textbf{AUPR} \\
    \midrule
    PCS       & 0.9566 & 0.9367 \\
    ICS       & 0.9789 & 0.9795 \\
    Logits    & 0.9757 & 0.9746 \\
    REEF      & 0.9867 & 0.9877 \\
    ProFlingo & 0.8374 & 0.8927 \\
    LLMMap    & 0.8925 & 0.9014 \\
    \textbf{Ours} & \textbf{0.9883} & \textbf{0.9881} \\
    \bottomrule
  \end{tabular}
  \par\smallskip
  \textit{\attndiff{} attains highest AUROC/AUPR, indicating strong separability.}
\end{wraptable}

While Sec.~\ref{sec:exp_finetune}--\ref{sec:exp_pruning} demonstrate that \attndiff{} maintains high similarity for related derivatives, practical fingerprinting also requires strong \emph{discriminability}: the ability to reject unrelated models with high confidence. We therefore evaluate \attndiff{} as a binary classifier that labels each suspect as either a derivative (positive) or unrelated (negative), and report both fixed-threshold and threshold-free discriminability metrics.

\noindent\textbf{Experimental setup.} We construct two model pools: (i) \textbf{Related suspects} ($n=36$, positive class): Llama-2-7B derivatives via fine-tuning, PPO/DPO, pruning/compression, and merging (Tables~\ref{tab:sft-robustness}, \ref{tab:pruning-results}, \ref{tab:po-robustness}, \ref{tab:merge-strategies}, \ref{tab:pruning_robustness}); (ii) \textbf{Unrelated suspects} ($n=33$, negative class): distinct architectures/families from cross-family evaluation and ablations (Tables~\ref{tab:ablation_diff}, \ref{tab:pruning_robustness}, \ref{tab:cross_family_scale}, \ref{tab:distill_robustness}).

For a threshold $t$, we predict \emph{positive} (derivative) if $\operatorname{CKA}(F,F')\geq t$, otherwise \emph{negative} (unrelated). We report two types of metrics: (i) \textbf{Fixed-threshold}---Precision (P), Recall (R), and F1-score (F1) at $t\in\{0.10,0.25,0.50,0.90\}$; (ii) \textbf{Threshold-free}---AUROC and AUPR, treating related suspects as the positive class.

\begin{table}[htbp]
  \centering
  \caption{Discriminability at fixed thresholds $t\in\{0.10,0.25,0.50,0.90\}$: precision (P), recall (R), F1-score.}
  \label{tab:discriminability_thresholds}
  \small
  \setlength{\tabcolsep}{2.5pt}
  \begin{adjustbox}{max width=\textwidth}
  \begin{tabular}{l*{12}{c}}
    \toprule
    \rowcolor{gray!20}
    \multirow{2}{*}{\textbf{Method}} & \multicolumn{3}{c}{$t=0.10$} & \multicolumn{3}{c}{$t=0.25$} & \multicolumn{3}{c}{$t=0.50$} & \multicolumn{3}{c}{$t=0.90$} \\
    \cmidrule(lr){2-4}\cmidrule(lr){5-7}\cmidrule(lr){8-10}\cmidrule(lr){11-13}
    \rowcolor{gray!20}
    & P & R & F1 & P & R & F1 & P & R & F1 & P & R & F1 \\
    \midrule
    PCS & 0.6596 & 0.8611 & 0.7470 & 0.6739 & 0.8611 & 0.7561 & 0.8108 & 0.8333 & 0.8219 & 0.8788 & \textbf{0.8056} & 0.8406 \\
    ICS & 0.5714 & \textbf{1.0000} & 0.7273 & 0.8649 & 0.8889 & 0.8767 & 0.9394 & 0.8611 & 0.8986 & 0.9630 & 0.7222 & 0.8254 \\
    Logits & 0.5556 & 0.9722 & 0.7071 & 0.7778 & 0.9722 & 0.8642 & 0.8947 & \textbf{0.9444} & \textbf{0.9189} & 0.9655 & 0.7778 & 0.8615 \\
    REEF & 0.6545 & \textbf{1.0000} & \textbf{0.7912} & 0.8611 & 0.8611 & 0.8611 & \textbf{1.0000} & 0.7778 & 0.8750 & \textbf{1.0000} & 0.6389 & 0.7797 \\
    ProFlingo & \textbf{0.7778} & 0.7778 & 0.7778 & \textbf{0.9524} & 0.5556 & 0.7018 & \textbf{1.0000} & 0.2222 & 0.3636 & \textbf{1.0000} & 0.1389 & 0.2439 \\
    LLMMap & 0.5217 & \textbf{1.0000} & 0.6857 & 0.6481 & 0.9722 & 0.7778 & 0.7143 & 0.8333 & 0.7692 & 0.8571 & 0.3333 & 0.4800 \\
    \textbf{Ours} & 0.5455 & \textbf{1.0000} & 0.7059 & 0.8182 & \textbf{1.0000} & \textbf{0.9000} & 0.8500 & \textbf{0.9444} & 0.8947 & \textbf{1.0000} & 0.7778 & \textbf{0.8750} \\
    \bottomrule
  \end{tabular}
  \end{adjustbox}
  \par\smallskip
  \textit{\attndiff{} achieves best F1 at $t=0.25$ (0.9000).}
\end{table}

\noindent\textbf{Results and analysis.} \attndiff{} achieves the highest F1 at $t=0.25$ (0.9000) and $t=0.90$ (0.8750). At $t=0.25$, it is the only method with perfect recall (R=1.0000) while maintaining strong precision (P=0.8182), effectively balancing zero missed derivatives with controlled false positives. At $t=0.90$, it achieves perfect precision (P=1.0000, FP=0) with the highest F1, demonstrating robust separability across practical operating points.

Table~\ref{tab:discriminability_auc} further confirms this via threshold-free metrics: \attndiff{} attains the highest AUROC (0.9883) and AUPR (0.9881), outperforming all baselines. Notably, REEF is competitive (0.9867/0.9877), while ProFlingo and LLMMap drop to $\approx0.83$--$0.90$, reflecting their sensitivity to surface-level transformations. Together, these results demonstrate that \attndiff{} provides a practical and discriminative tool for binary provenance decisions across realistic operating points.
\FloatBarrier

\subsection{Adaptive suppression and probe refresh}
\label{sec:exp_adaptive}

\noindent\textbf{Setup.} We evaluate probe-aware suppression on a related Gemma-2 model pair under a fixed 96-step LoRA budget, with an ordinary-fine-tuning control. The attacker observes 20--80\% of a 60-pair attack pool; evaluation compares leaked probes, held-out probes from that pool, and a disjoint 60-pair refresh pool. Training details, the suppression objective, and probe allocation are provided in Appendix~\ref{sec:adaptive_attack_details}.

\begin{table}[!htbp]
  \centering
  \caption{Probe-aware suppression under a fixed 96-step budget. CKA compares each adapted suspect with the frozen reference. FT denotes ordinary fine-tuning; PPL and token accuracy measure attack utility on an independent WikiText-2 slice.}
  \label{tab:adaptive_attack}
  \small
  \setlength{\tabcolsep}{4pt}
  \begin{tabular}{lcccccc}
    \toprule
    \textbf{Exposure} & \multicolumn{3}{c}{\textbf{Adaptive attack: CKA}} & \textbf{FT: CKA} & \multicolumn{2}{c}{\textbf{Attack utility}} \\
    \cmidrule(lr){2-4}\cmidrule(lr){6-7}
    (pairs) & Leaked & Held-out & Refreshed & Refreshed & PPL $\downarrow$ & Acc. $\uparrow$ \\
    \midrule
    20\% (12/60) & 0.738 & 0.967 & 0.991 & 0.998 & 21.951 & 0.4156 \\
    40\% (24/60) & 0.964 & 0.935 & 0.990 & 0.998 & 22.319 & 0.4096 \\
    60\% (36/60) & 0.934 & 0.964 & 0.988 & 0.998 & 23.053 & 0.4082 \\
    80\% (48/60) & 0.917 & 0.987 & 0.988 & 0.998 & 23.746 & 0.4033 \\
    \bottomrule
  \end{tabular}
\end{table}

\noindent\textbf{Transfer and utility.} At 20\% exposure, leaked CKA falls from its clean value of 0.992 to 0.738. Transfer is limited: held-out CKA remains 0.935--0.987 and refreshed CKA 0.988--0.991, versus approximately 0.998 for ordinary fine-tuning; even at 80\% exposure, refreshed CKA is 0.988. On an independent 8,106-token WikiText-2 slice, clean perplexity (PPL)/next-token accuracy are 27.691/0.4003; all attacked models improve both measured metrics (Table~\ref{tab:adaptive_attack}). Ordinary fine-tuning yields better PPL/accuracy (19.725--19.891/0.4282--0.4305), indicating an adaptation cost from suppression, although attacked models do not collapse relative to the clean suspect.

Under this model pair and fixed budget, suppression primarily fits exposed probes, supporting private, refreshed probes for verification. Minimizing attention-difference norms is not direct optimization of CKA; this pilot does not establish robustness to unrestricted adaptive attacks or preservation of all downstream capabilities.
\FloatBarrier

\section{Discussion}

\attndiff{} supports post-hoc provenance verification under common laundering pipelines, including fine-tuning, preference optimization, pruning/compression, and merging, while maintaining a clear separation margin from unrelated architectures.

\paragraph{Probe refresh.} The adaptive experiment (Sec.~\ref{sec:exp_adaptive}) supports keeping a private reserve and replacing exposed probes: suppression transfers poorly to a disjoint refresh pool under the evaluated budget. Refreshing requires recomputing reference and suspect fingerprints on the same new pairs, without retraining either model.

\paragraph{Partial access.} Dense next-token log-probabilities permit a possible gray-box extension using differential output features. Appendix~\ref{sec:gray_box} reports a small DiffLogits pilot and specifies a reconstructed evaluation protocol. This extension is separate from attention-based \attndiff{} and does not validate text-only APIs.

\paragraph{Computational cost.} \attndiff{} needs no training or gradient optimization: $M$ probe pairs require $2M$ forward passes per model with attention recording, followed by lightweight pooling (if needed), differencing, top-$K$ SVD, and CKA. Caching the victim fingerprint leaves $2M$ suspect passes as the dominant cost. With $M=60$, a 7B-parameter model takes approximately 5 minutes on one NVIDIA RTX 5090 GPU. This avoids tensor construction and alignment heuristics (PCS/ICS), trigger optimization (ProFlingo), and inference model training (LLMMap); Appendix~\ref{sec:appendix_computational_cost} provides a detailed comparison.

\section{Conclusion}

\attndiff{} is a white-box post-hoc fingerprinting method that captures model-specific information routing through \emph{differential attention} under controlled semantic conflicts. Compact spectral descriptors and centered linear CKA yield model-agnostic fingerprints that remain stable for co-originated derivatives and separated from unrelated architectures across fine-tuning (including PPO/DPO), pruning/compression, and merging. The suppression experiment supports refreshing exposed probes under the evaluated budget. Small probe sets and lightweight computation make \attndiff{} practical for provenance verification and accountability in open-weight LLMs, although adoption risks unwarranted copyright disputes (Appendix~\ref{sec:broader_impacts}).

\section*{Limitations and future work}
\label{sec:limitations}
\paragraph{Access and evaluation scope.} \attndiff{} requires internal attention activations. The DiffLogits pilot (Appendix~\ref{sec:gray_box}) requires dense log-probabilities and covers only one related and one unrelated suspect; its reconstructed protocol awaits numerical re-evaluation. Text-only and top-$k$-only APIs remain outside the evidence. The adaptive attack study likewise covers one model pair and a fixed budget.

\paragraph{Theoretical modeling of laundering effects.} While the centered linear CKA comparison provides useful invariance to feature rotations, permutations, and isotropic scaling, we still lack a complete mathematical model that propagates common laundering operations (e.g., fine-tuning, pruning, merging, and distillation) from parameter or architecture transformations to perturbations in the extracted \attndiff{} fingerprints and the resulting similarity scores. Establishing such an end-to-end modeling framework remains an important direction for future work.

\section*{Acknowledgments}

This work was supported by the ``Pioneer'' and ``Leading Goose'' R\&D Program of Zhejiang (Grant No.~2026SDXT012) and the Preferential Funding for Postdoctoral Projects in Zhejiang Province (No.~ZJ2026029).

\noindent\textbf{Competing interests.} The authors declare no competing interests.

\clearpage
\bibliographystyle{unsrtnat}
\bibliography{bib/references}

@misc{touvron2023llama2,
      title={Llama 2: Open Foundation and Fine-Tuned Chat Models}, 
      author={Hugo Touvron and Louis Martin and Kevin Stone and Peter Albert and Amjad Almahairi and Yasmine Babaei and Nikolay Bashlykov and Soumya Batra and Prajjwal Bhargava and Shruti Bhosale and Dan Bikel and Lukas Blecher and Cristian Canton Ferrer and Moya Chen and Guillem Cucurull and David Esiobu and Jude Fernandes and Jeremy Fu and Wenyin Fu and Brian Fuller and Cynthia Gao and Vedanuj Goswami and Naman Goyal and Anthony Hartshorn and Saghar Hosseini and Rui Hou and Hakan Inan and Marcin Kardas and Viktor Kerkez and Madian Khabsa and Isabel Kloumann and Artem Korenev and Punit Singh Koura and Marie-Anne Lachaux and Thibaut Lavril and Jenya Lee and Diana Liskovich and Yinghai Lu and Yuning Mao and Xavier Martinet and Todor Mihaylov and Pushkar Mishra and Igor Molybog and Yixin Nie and Andrew Poulton and Jeremy Reizenstein and Rashi Rungta and Kalyan Saladi and Alan Schelten and Ruan Silva and Eric Michael Smith and Ranjan Subramanian and Xiaoqing Ellen Tan and Binh Tang and Ross Taylor and Adina Williams and Jian Xiang Kuan and Puxin Xu and Zheng Yan and Iliyan Zarov and Yuchen Zhang and Angela Fan and Melanie Kambadur and Sharan Narang and Aurelien Rodriguez and Robert Stojnic and Sergey Edunov and Thomas Scialom},
      year={2023},
      eprint={2307.09288},
      archivePrefix={arXiv},
      primaryClass={cs.CL},
      url={https://arxiv.org/abs/2307.09288}, 
}

@misc{bai2023qwen,
      title={Qwen Technical Report}, 
      author={Jinze Bai and Shuai Bai and Yunfei Chu and Zeyu Cui and Kai Dang and Xiaodong Deng and Yang Fan and Wenbin Ge and Yu Han and Fei Huang and Binyuan Hui and Luo Ji and Mei Li and Junyang Lin and Runji Lin and Dayiheng Liu and Gao Liu and Chengqiang Lu and Keming Lu and Jianxin Ma and Rui Men and Xingzhang Ren and Xuancheng Ren and Chuanqi Tan and Sinan Tan and Jianhong Tu and Peng Wang and Shijie Wang and Wei Wang and Shengguang Wu and Benfeng Xu and Jin Xu and An Yang and Hao Yang and Jian Yang and Shusheng Yang and Yang Yao and Bowen Yu and Hongyi Yuan and Zheng Yuan and Jianwei Zhang and Xingxuan Zhang and Yichang Zhang and Zhenru Zhang and Chang Zhou and Jingren Zhou and Xiaohuan Zhou and Tianhang Zhu},
      year={2023},
      eprint={2309.16609},
      archivePrefix={arXiv},
      primaryClass={cs.CL},
      url={https://arxiv.org/abs/2309.16609}, 
}

@misc{deepseek2024v3,
      title={DeepSeek-V3 Technical Report}, 
      author={DeepSeek-AI and Aixin Liu and Bei Feng and Bing Xue and Bingxuan Wang and Bochao Wu and Chengda Lu and Chenggang Zhao and Chengqi Deng and Chenyu Zhang and Chong Ruan and Damai Dai and Daya Guo and Dejian Yang and Deli Chen and Dongjie Ji and Erhang Li and Fangyun Lin and Fucong Dai and Fuli Luo and Guangbo Hao and Guanting Chen and Guowei Li and H. Zhang and Han Bao and Hanwei Xu and Haocheng Wang and Haowei Zhang and Honghui Ding and Huajian Xin and Huazuo Gao and Hui Li and Hui Qu and J. L. Cai and Jian Liang and Jianzhong Guo and Jiaqi Ni and Jiashi Li and Jiawei Wang and Jin Chen and Jingchang Chen and Jingyang Yuan and Junjie Qiu and Junlong Li and Junxiao Song and Kai Dong and Kai Hu and Kaige Gao and Kang Guan and Kexin Huang and Kuai Yu and Lean Wang and Lecong Zhang and Lei Xu and Leyi Xia and Liang Zhao and Litong Wang and Liyue Zhang and Meng Li and Miaojun Wang and Mingchuan Zhang and Minghua Zhang and Minghui Tang and Mingming Li and Ning Tian and Panpan Huang and Peiyi Wang and Peng Zhang and Qiancheng Wang and Qihao Zhu and Qinyu Chen and Qiushi Du and R. J. Chen and R. L. Jin and Ruiqi Ge and Ruisong Zhang and Ruizhe Pan and Runji Wang and Runxin Xu and Ruoyu Zhang and Ruyi Chen and S. S. Li and Shanghao Lu and Shangyan Zhou and Shanhuang Chen and Shaoqing Wu and Shengfeng Ye and Shengfeng Ye and Shirong Ma and Shiyu Wang and Shuang Zhou and Shuiping Yu and Shunfeng Zhou and Shuting Pan and T. Wang and Tao Yun and Tian Pei and Tianyu Sun and W. L. Xiao and Wangding Zeng and Wanjia Zhao and Wei An and Wen Liu and Wenfeng Liang and Wenjun Gao and Wenqin Yu and Wentao Zhang and X. Q. Li and Xiangyue Jin and Xianzu Wang and Xiao Bi and Xiaodong Liu and Xiaohan Wang and Xiaojin Shen and Xiaokang Chen and Xiaokang Zhang and Xiaosha Chen and Xiaotao Nie and Xiaowen Sun and Xiaoxiang Wang and Xin Cheng and Xin Liu and Xin Xie and Xingchao Liu and Xingkai Yu and Xinnan Song and Xinxia Shan and Xinyi Zhou and Xinyu Yang and Xinyuan Li and Xuecheng Su and Xuheng Lin and Y. K. Li and Y. Q. Wang and Y. X. Wei and Y. X. Zhu and Yang Zhang and Yanhong Xu and Yanhong Xu and Yanping Huang and Yao Li and Yao Zhao and Yaofeng Sun and Yaohui Li and Yaohui Wang and Yi Yu and Yi Zheng and Yichao Zhang and Yifan Shi and Yiliang Xiong and Ying He and Ying Tang and Yishi Piao and Yisong Wang and Yixuan Tan and Yiyang Ma and Yiyuan Liu and Yongqiang Guo and Yu Wu and Yuan Ou and Yuchen Zhu and Yuduan Wang and Yue Gong and Yuheng Zou and Yujia He and Yukun Zha and Yunfan Xiong and Yunxian Ma and Yuting Yan and Yuxiang Luo and Yuxiang You and Yuxuan Liu and Yuyang Zhou and Z. F. Wu and Z. Z. Ren and Zehui Ren and Zhangli Sha and Zhe Fu and Zhean Xu and Zhen Huang and Zhen Zhang and Zhenda Xie and Zhengyan Zhang and Zhewen Hao and Zhibin Gou and Zhicheng Ma and Zhigang Yan and Zhihong Shao and Zhipeng Xu and Zhiyu Wu and Zhongyu Zhang and Zhuoshu Li and Zihui Gu and Zijia Zhu and Zijun Liu and Zilin Li and Ziwei Xie and Ziyang Song and Ziyi Gao and Zizheng Pan},
      year={2025},
      eprint={2412.19437},
      archivePrefix={arXiv},
      primaryClass={cs.CL},
      url={https://arxiv.org/abs/2412.19437}, 
}

@misc{Xu2025CopyrightSurvey,
      title={Copyright Protection for Large Language Models: A Survey of Methods, Challenges, and Trends}, 
      author={Zhenhua Xu and Xubin Yue and Zhebo Wang and Haobo Zhang and Qichen Liu and Xixiang Zhao and Jingxuan Zhang and Wenjun Zeng and Wengpeng Xing and Dezhang Kong and Changting Lin and Meng Han},
      year={2026},
      eprint={2508.11548},
      archivePrefix={arXiv},
      primaryClass={cs.CR},
      url={https://arxiv.org/abs/2508.11548}, 
}

@inproceedings{zhang2024reef,
 author = {Zhang, Jie and Liu, Dongrui and Qian, Chen and Zhang, Linfeng and Liu, Yong and Qiao, Yu and Shao, Jing},
 booktitle = {International Conference on Learning Representations},
 editor = {Y. Yue and A. Garg and N. Peng and F. Sha and R. Yu},
 pages = {48092--48117},
 title = {REEF: Representation Encoding Fingerprints for Large Language Models},
 url = {https://proceedings.iclr.cc/paper_files/paper/2025/file/77e830bd802f4b63b682038d090a97b6-Paper-Conference.pdf},
 volume = {2025},
 year = {2025}
}

@inproceedings{xu2025ctcc,
    title = "{CTCC}: A Robust and Stealthy Fingerprinting Framework for Large Language Models via Cross-Turn Contextual Correlation Backdoor",
    author = "Xu, Zhenhua  and
      Zhao, Xixiang  and
      Yue, Xubin  and
      Tian, Shengwei  and
      Lin, Changting  and
      Han, Meng",
    editor = "Christodoulopoulos, Christos  and
      Chakraborty, Tanmoy  and
      Rose, Carolyn  and
      Peng, Violet",
    booktitle = "Proceedings of the 2025 Conference on Empirical Methods in Natural Language Processing",
    month = nov,
    year = "2025",
    address = "Suzhou, China",
    publisher = "Association for Computational Linguistics",
    url = "https://aclanthology.org/2025.emnlp-main.356/",
    doi = "10.18653/v1/2025.emnlp-main.356",
    pages = "6967--6989",
    ISBN = "979-8-89176-332-6"
}

@inproceedings{xu2024instructional,
    title = "Instructional Fingerprinting of Large Language Models",
    author = "Xu, Jiashu  and
      Wang, Fei  and
      Ma, Mingyu  and
      Koh, Pang Wei  and
      Xiao, Chaowei  and
      Chen, Muhao",
    editor = "Duh, Kevin  and
      Gomez, Helena  and
      Bethard, Steven",
    booktitle = "Proceedings of the 2024 Conference of the {N}orth {A}merican Chapter of the {A}ssociation for {C}omputational {L}inguistics: Human Language Technologies (Volume 1: Long Papers)",
    month = jun,
    year = "2024",
    address = "Mexico City, Mexico",
    publisher = "Association for Computational Linguistics",
    url = "https://aclanthology.org/2024.naacl-long.180/",
    doi = "10.18653/v1/2024.naacl-long.180",
    pages = "3277--3306"
}

@inproceedings{li2023plmmark,
  title={Plmmark: a secure and robust black-box watermarking framework for pre-trained language models},
  author={Li, Peixuan and Cheng, Pengzhou and Li, Fangqi and Du, Wei and Zhao, Haodong and Liu, Gongshen},
  booktitle={Proceedings of the AAAI Conference on Artificial Intelligence},
  volume={37},
  number={12},
  pages={14991--14999},
  year={2023}
}

@inproceedings{nasery2025scalable,
 author = {Nasery, Anshul and Hayase, Jonathan and Brooks, Creston and Sheng, Peiyao and Tyagi, Himanshu and Viswanath, Pramod and Oh, Sewoong},
 booktitle = {Advances in Neural Information Processing Systems},
 doi = {10.52202/085713-4170},
 editor = {D. Belgrave and C. Zhang and H. Lin and R. Pascanu and P. Koniusz and M. Ghassemi and N. Chen},
 pages = {125116--125152},
 publisher = {Curran Associates, Inc.},
 title = {Scalable Fingerprinting of Large Language Models},
 url = {https://proceedings.neurips.cc/paper_files/paper/2025/file/b5ee4715ef8e96176ef3ccbe229b6ab9-Paper-Conference.pdf},
 volume = {38, Main Conference},
 year = {2025}
}

@inproceedings{xu2025evertracer,
    title = "{E}ver{T}racer: Hunting Stolen Large Language Models via Stealthy and Robust Probabilistic Fingerprint",
    author = "Xu, Zhenhua  and
      Han, Meng  and
      Xing, Wenpeng",
    editor = "Christodoulopoulos, Christos  and
      Chakraborty, Tanmoy  and
      Rose, Carolyn  and
      Peng, Violet",
    booktitle = "Proceedings of the 2025 Conference on Empirical Methods in Natural Language Processing",
    month = nov,
    year = "2025",
    address = "Suzhou, China",
    publisher = "Association for Computational Linguistics",
    url = "https://aclanthology.org/2025.emnlp-main.358/",
    doi = "10.18653/v1/2025.emnlp-main.358",
    pages = "7008--7031",
    ISBN = "979-8-89176-332-6"
}

@inproceedings{zeng2024huref,
 author = {Zeng, Boyi and Wang, Lizheng and Hu, Yuncong and Xu, Yi and Zhou, Chenghu and Wang, Xinbing and Yu, Yu and Lin, Zhouhan},
 booktitle = {Advances in Neural Information Processing Systems},
 doi = {10.52202/079017-4013},
 editor = {A. Globerson and L. Mackey and D. Belgrave and A. Fan and U. Paquet and J. Tomczak and C. Zhang},
 pages = {126332--126362},
 publisher = {Curran Associates, Inc.},
 title = {HuRef: HUman-REadable Fingerprint for Large Language Models},
 url = {https://proceedings.neurips.cc/paper_files/paper/2024/file/e46fc33e80e9fa2febcdb058fba4beca-Paper-Conference.pdf},
 volume = {37},
 year = {2024}
}

@article{YangWu2024Fingerprint,
  title={A Fingerprint for Large Language Models},
  author={Zhiguang Yang and Han-Zhou Wu},
  journal={ArXiv},
  year={2024},
  volume={abs/2407.01235},
  url={https://api.semanticscholar.org/CorpusID:270869924}
}

@article{jin2024proflingo,
  title={ProFLingo: A Fingerprinting-based Intellectual Property Protection Scheme for Large Language Models},
  author={Heng Jin and Chao-Yu Zhang and Shang-Hao Shi and Wen Jing Lou and Y. Thomas Hou},
  journal={2024 IEEE Conference on Communications and Network Security (CNS)},
  year={2024},
  pages={1-9},
  url={https://api.semanticscholar.org/CorpusID:270738202}
}

@inproceedings{pasquini2025llmmap,
  title={{{LLMmap}: Fingerprinting for Large Language Models}},
  author={Pasquini, Dario and Kornaropoulos, Evgenios M and Ateniese, Giuseppe},
  booktitle={34th USENIX Security Symposium (USENIX Security 25)},
  pages={299--318},
  year={2025}
}

@InProceedings{kornblith2019similarity,
  title = 	 {Similarity of Neural Network Representations Revisited},
  author =       {Kornblith, Simon and Norouzi, Mohammad and Lee, Honglak and Hinton, Geoffrey},
  booktitle = 	 {Proceedings of the 36th International Conference on Machine Learning},
  pages = 	 {3519--3529},
  year = 	 {2019},
  editor = 	 {Chaudhuri, Kamalika and Salakhutdinov, Ruslan},
  volume = 	 {97},
  series = 	 {Proceedings of Machine Learning Research},
  month = 	 {09--15 Jun},
  publisher =    {PMLR},
  url = 	 {https://proceedings.mlr.press/v97/kornblith19a.html}
}

@INPROCEEDINGS{zhang2024emmark,
  author={Zhang, Ruisi and Koushanfar, Farinaz},
  booktitle={2024 61st ACM/IEEE Design Automation Conference (DAC)}, 
  title={EmMark: Robust Watermarks for IP Protection of Embedded Quantized Large Language Models}, 
  year={2024},
  volume={},
  number={},
  pages={1-6},
  doi={}
}

@misc{guo2025invariant,
      title={Invariant-based Robust Weights Watermark for Large Language Models}, 
      author={Qingxiao Guo and Xinjie Zhu and Yilong Ma and Hui Jin and Yunhao Wang and Weifeng Zhang and Xiaobing Guo},
      year={2025},
      eprint={2507.08288},
      archivePrefix={arXiv},
      primaryClass={cs.CR},
      url={https://arxiv.org/abs/2507.08288}, 
}

@article{fernandez2023functional,
  title={Functional Invariants To Watermark Large Transformers},
  author={Pierre Fernandez and Guillaume Couairon and Teddy Furon and Matthijs Douze},
  journal={ICASSP 2024 - 2024 IEEE International Conference on Acoustics, Speech and Signal Processing (ICASSP)},
  year={2023},
  pages={4815-4819},
  url={https://api.semanticscholar.org/CorpusID:264172180}
}

@article{luan2025robust,
  author       = {Xiaokun Luan and
                  Zeming Wei and
                  Yihao Zhang and
                  Meng Sun},
  title        = {Robust and Efficient Watermarking of Large Language Models Using Error
                  Correction Codes},
  journal      = {Proc. Priv. Enhancing Technol.},
  volume       = {2025},
  number       = {4},
  pages        = {183--200},
  year         = {2025},
  url          = {https://doi.org/10.56553/popets-2025-0126},
  doi          = {10.56553/POPETS-2025-0126},
  bibsource    = {dblp computer science bibliography, https://dblp.org}
}

@article{tang2025resistanceneuralnetworkwatermarking,
  title={Towards the Resistance of Neural Network Watermarking to Fine-tuning},
  author={Ling Tang and YueFeng Chen and Hui Xue and Quanshi Zhang},
  journal={ArXiv},
  year={2025},
  volume={abs/2505.01007},
  url={https://api.semanticscholar.org/CorpusID:278310967}
}

@misc{cai2025utf,
      title={UTF:Undertrained Tokens as Fingerprints A Novel Approach to LLM Identification}, 
      author={Jiacheng Cai and Jiahao Yu and Yangguang Shao and Yuhang Wu},
      year={2025},
      eprint={2410.12318},
      archivePrefix={arXiv},
      primaryClass={cs.CR},
      url={https://arxiv.org/abs/2410.12318}, 
}

@INPROCEEDINGS{xu2026dnfduallayernestedfingerprinting,
  author={Xu, Zhenhua and Zhao, Yiran and Zhong, Mengting and Kong, Dezhang and Lin, Changting and Qiao, Tong and Han, Meng},
  booktitle={ICASSP 2026 - 2026 IEEE International Conference on Acoustics, Speech and Signal Processing (ICASSP)}, 
  title={DNF: Dual-Layer Nested Fingerprinting for Large Language Model Intellectual Property Protection}, 
  year={2026},
  volume={},
  number={},
  pages={13972-13976},
  doi={10.1109/ICASSP55912.2026.11461762}
}

@article{xuInStyRobustMultilevel2025,
  title={InSty: A robust multi-level cross-granularity fingerprint embedding algorithm for multi-turn dialogue in large language models},
  author={Xu, Zhenhua and Han, Meng and Yue, Xubin and Xing, Wenpeng},
  journal={SCIENTIA SINICA Informationis},
  volume={55},
  number={8},
  pages={1906},
  year={2025},
  publisher={Science China Press},
  doi={10.1360/ssi-2025-0022}
}

@INPROCEEDINGS{xu2026forgetmarkstealthyfingerprintembedding,
  author={Xu, Zhenhua and Zhang, Haobo and Wang, Zhebo and Liu, Qichen and Xu, Haitao and Xing, Wenpeng and Han, Meng},
  booktitle={ICASSP 2026 - 2026 IEEE International Conference on Acoustics, Speech and Signal Processing (ICASSP)}, 
  title={ForgetMark: Stealthy Fingerprint Embedding via Targeted Unlearning in Language Models}, 
  year={2026},
  volume={},
  number={},
  pages={13912-13916},
  doi={10.1109/ICASSP55912.2026.11464391}
}

@misc{li2024doublei,
      title={Double-I Watermark: Protecting Model Copyright for LLM Fine-tuning}, 
      author={Shen Li and Liuyi Yao and Jinyang Gao and Lan Zhang and Yaliang Li},
      year={2024},
      eprint={2402.14883},
      archivePrefix={arXiv},
      primaryClass={cs.CR},
      url={https://arxiv.org/abs/2402.14883}, 
}

@inproceedings{russinovich2024hey,
  title={Hey, that's my model! introducing chain \& hash, an llm fingerprinting technique},
  author={Russinovich, Mark and Cai, Yanan and Salem, Ahmed},
  booktitle={International Conference on Learning Representations},
  volume={2026},
  pages={45118--45133},
  year={2026}
}

@inproceedings{zhang2025meraser,
    title = "{ME}raser: An Effective Fingerprint Erasure Approach for Large Language Models",
    author = "Zhang, Jingxuan  and
      Xu, Zhenhua  and
      Hu, Rui  and
      Xing, Wenpeng  and
      Zhang, Xuhong  and
      Han, Meng",
    editor = "Che, Wanxiang  and
      Nabende, Joyce  and
      Shutova, Ekaterina  and
      Pilehvar, Mohammad Taher",
    booktitle = "Proceedings of the 63rd Annual Meeting of the Association for Computational Linguistics (Volume 1: Long Papers)",
    month = jul,
    year = "2025",
    address = "Vienna, Austria",
    publisher = "Association for Computational Linguistics",
    url = "https://aclanthology.org/2025.acl-long.1455/",
    doi = "10.18653/v1/2025.acl-long.1455",
    pages = "30136--30153",
    ISBN = "979-8-89176-251-0"
}

@inproceedings{xuUnlockingEffectivenessLoRAFP2025,
  title = {Unlocking the {{Effectiveness}} of {{LoRA-FP}} for {{Seamless Transfer Implantation}} of {{Fingerprints}} in {{Downstream Models}}},
  booktitle = {Findings of the {{Association}} for {{Computational Linguistics}}: {{EMNLP}} 2025},
  author = {Xu, Zhenhua and Yan, Zhaokun and Xu, Binhan and Tong, Xin and Xu, Haitao and Chen, Yourong and Han, Meng},
  editor = {Christodoulopoulos, Christos and Chakraborty, Tanmoy and Rose, Carolyn and Peng, Violet},
  year = 2025,
  pages = {4302--4312},
  publisher = {Association for Computational Linguistics},
  address = {Suzhou, China},
  doi = {10.18653/v1/2025.findings-emnlp.230},
  urldate = {2025-11-14},
  isbn = {979-8-89176-335-7}
}

@misc{xu2025fingerprintvectorenablingscalable,
      title={Fingerprint Vector: Enabling Scalable and Efficient Model Fingerprint Transfer via Vector Addition},
      author={Zhenhua Xu and Qichen Liu and Zhebo Wang and Wenpeng Xing and Dezhang Kong and Mohan Li and Meng Han},
      year={2025},
      eprint={2409.08846},
      archivePrefix={arXiv},
      primaryClass={cs.CR},
      url={https://arxiv.org/abs/2409.08846}, 
}

@inproceedings{yuePREEHarmlessAdaptive2025,
    title = "{PREE}: Towards Harmless and Adaptive Fingerprint Editing in Large Language Models via Knowledge Prefix Enhancement",
    author = "Yue, Xubin  and
      Xu, Zhenhua  and
      Xing, Wenpeng  and
      Yu, Jiahui  and
      Li, Mohan  and
      Han, Meng",
    editor = "Christodoulopoulos, Christos  and
      Chakraborty, Tanmoy  and
      Rose, Carolyn  and
      Peng, Violet",
    booktitle = "Findings of the {A}ssociation for {C}omputational {L}inguistics: {EMNLP} 2025",
    month = nov,
    year = "2025",
    address = "Suzhou, China",
    publisher = "Association for Computational Linguistics",
    url = "https://aclanthology.org/2025.findings-emnlp.204/",
    doi = "10.18653/v1/2025.findings-emnlp.204",
    pages = "3794--3804",
    ISBN = "979-8-89176-335-7"
}

@misc{wang2026srafstealthyrobustadversarial,
      title={SRAF: Stealthy and Robust Adversarial Fingerprint for Copyright Verification of Large Language Models}, 
      author={Zhebo Wang and Zhenhua Xu and Maike Li and Xilong Cheng and Wenpeng Xing and Chunqiang Hu and Chen Zhi and Meng Han},
      year={2026},
      eprint={2505.06304},
      archivePrefix={arXiv},
      primaryClass={cs.CR},
      url={https://arxiv.org/abs/2505.06304}, 
}

@inproceedings{tong2025seedprints,
 author = {Tong, Yao and Wang, Haonan and Li, Siquan and Kawaguchi, Kenji and Hu, Tianyang},
 booktitle = {International Conference on Learning Representations},
 editor = {C. Vondrick and B. Hariharan and C. Raffel and L. Pinto and D. Yang and A. Faust},
 pages = {44530--44558},
 title = {SeedPrints: Fingerprints Can Even Tell Which Seed Your Large Language Model Was Trained From},
 url = {https://proceedings.iclr.cc/paper_files/paper/2026/file/49e48a1a40628b160101c1375b8eb151-Paper-Conference.pdf},
 volume = {2026},
 year = {2026}
}

@inproceedings{wu2025llm,
 author = {Wu, Zhaomin and Zhao, Haodong and Wang, Ziyang and Guo, Jizhou and Wang, Qian and He, Bingsheng},
 booktitle = {International Conference on Learning Representations},
 editor = {C. Vondrick and B. Hariharan and C. Raffel and L. Pinto and D. Yang and A. Faust},
 pages = {7252--7279},
 title = {LLM DNA: Tracing Model Evolution via Functional Representations},
 url = {https://proceedings.iclr.cc/paper_files/paper/2026/file/0c606cbd7225856faadf74e106cbbec6-Paper-Conference.pdf},
 volume = {2026},
 year = {2026}
}

@inproceedings{gubri2024trap,
    title = "{TRAP}: Targeted Random Adversarial Prompt Honeypot for Black-Box Identification",
    author = "Gubri, Martin  and
      Ulmer, Dennis  and
      Lee, Hwaran  and
      Yun, Sangdoo  and
      Oh, Seong Joon",
    editor = "Ku, Lun-Wei  and
      Martins, Andre  and
      Srikumar, Vivek",
    booktitle = "Findings of the Association for Computational Linguistics: ACL 2024",
    month = aug,
    year = "2024",
    address = "Bangkok, Thailand",
    publisher = "Association for Computational Linguistics",
    url = "https://aclanthology.org/2024.findings-acl.683/",
    doi = "10.18653/v1/2024.findings-acl.683",
    pages = "11496--11517"
}

@article{qwen2025qwen25technicalreport,
  title={Qwen2.5 Technical Report},
  author={Qwen An Yang and Baosong Yang and Beichen Zhang and Binyuan Hui and Bo Zheng and Bo-Wen Yu and Cheng-Yuan Li and Dayiheng Liu and Fei Huang and Guanting Dong and Hao-Ran Wei and Huan Lin and Jian Yang and Jianhong Tu and Jian-Wei Zhang and Jian-Xin Yang and Jia-Xin Yang and Jing-Ren Zhou and Jun-Yang Lin and Kai Dang and Ke-Ming Lu and Ke-Qin Bao and Kexin Yang and Le Yu and Mei Li and Mingfeng Xue and Pei Zhang and Qin Zhu and Rui Men and Run-Ji Lin and Tian-Hao Li and Tingyu Xia and Xing-Zhang Ren and Xuan-Cheng Ren and Yang Fan and Yang Su and Yi-Chao Zhang and Yunyang Wan and Yuqi Liu and Zeyu Cui and Zhen-Ru Zhang and Zi-Han Qiu and Shanghaoran Quan and Zekun Wang},
  journal={ArXiv},
  year={2024},
  volume={abs/2412.15115},
  url={https://api.semanticscholar.org/CorpusID:274859421}
}

@article{grattafiori2024llama3herdmodels,
  title={The llama 3 herd of models},
  author={Grattafiori, Aaron and Dubey, Abhimanyu and Jauhri, Abhinav and Pandey, Abhinav and Kadian, Abhishek and Al-Dahle, Ahmad and Letman, Aiesha and Mathur, Akhil and Schelten, Alan and Vaughan, Alex and others},
  journal={arXiv preprint arXiv:2407.21783},
  year={2024}
}

@misc{gemmateam2024gemma,
      title={Gemma: Open Models Based on Gemini Research and Technology}, 
      author={Gemma Team and Thomas Mesnard and Cassidy Hardin and Robert Dadashi and Surya Bhupatiraju and Shreya Pathak and Laurent Sifre and Morgane Rivière and Mihir Sanjay Kale and Juliette Love and Pouya Tafti and Léonard Hussenot and Pier Giuseppe Sessa and Aakanksha Chowdhery and Adam Roberts and Aditya Barua and Alex Botev and Alex Castro-Ros and Ambrose Slone and Amélie Héliou and Andrea Tacchetti and Anna Bulanova and Antonia Paterson and Beth Tsai and Bobak Shahriari and Charline Le Lan and Christopher A. Choquette-Choo and Clément Crepy and Daniel Cer and Daphne Ippolito and David Reid and Elena Buchatskaya and Eric Ni and Eric Noland and Geng Yan and George Tucker and George-Christian Muraru and Grigory Rozhdestvenskiy and Henryk Michalewski and Ian Tenney and Ivan Grishchenko and Jacob Austin and James Keeling and Jane Labanowski and Jean-Baptiste Lespiau and Jeff Stanway and Jenny Brennan and Jeremy Chen and Johan Ferret and Justin Chiu and Justin Mao-Jones and Katherine Lee and Kathy Yu and Katie Millican and Lars Lowe Sjoesund and Lisa Lee and Lucas Dixon and Machel Reid and Maciej Mikuła and Mateo Wirth and Michael Sharman and Nikolai Chinaev and Nithum Thain and Olivier Bachem and Oscar Chang and Oscar Wahltinez and Paige Bailey and Paul Michel and Petko Yotov and Rahma Chaabouni and Ramona Comanescu and Reena Jana and Rohan Anil and Ross McIlroy and Ruibo Liu and Ryan Mullins and Samuel L Smith and Sebastian Borgeaud and Sertan Girgin and Sholto Douglas and Shree Pandya and Siamak Shakeri and Soham De and Ted Klimenko and Tom Hennigan and Vlad Feinberg and Wojciech Stokowiec and Yu-hui Chen and Zafarali Ahmed and Zhitao Gong and Tris Warkentin and Ludovic Peran and Minh Giang and Clément Farabet and Oriol Vinyals and Jeff Dean and Koray Kavukcuoglu and Demis Hassabis and Zoubin Ghahramani and Douglas Eck and Joelle Barral and Fernando Pereira and Eli Collins and Armand Joulin and Noah Fiedel and Evan Senter and Alek Andreev and Kathleen Kenealy},
      year={2024},
      eprint={2403.08295},
      archivePrefix={arXiv},
      primaryClass={cs.CL},
      url={https://arxiv.org/abs/2403.08295}, 
}

@misc{jiang2023mistral7b,
  title={Mistral 7B},
  author={Jiang, Albert Q. and Sablayrolles, Alexandre and Mensch, Arthur and Bamford, Chris and Chaplot, Devendra Singh and others},
  year={2023},
  eprint={2310.06825},
  archivePrefix={arXiv},
  primaryClass={cs.CL},
  url={https://arxiv.org/abs/2310.06825}
}

@inproceedings{ouyang2022instructgpt,
 author = {Ouyang, Long and Wu, Jeffrey and Jiang, Xu and Almeida, Diogo and Wainwright, Carroll and Mishkin, Pamela and Zhang, Chong and Agarwal, Sandhini and Slama, Katarina and Ray, Alex and Schulman, John and Hilton, Jacob and Kelton, Fraser and Miller, Luke and Simens, Maddie and Askell, Amanda and Welinder, Peter and Christiano, Paul F and Leike, Jan and Lowe, Ryan},
 booktitle = {Advances in Neural Information Processing Systems},
 doi = {10.52202/068431-2011},
 editor = {S. Koyejo and S. Mohamed and A. Agarwal and D. Belgrave and K. Cho and A. Oh},
 pages = {27730--27744},
 publisher = {Curran Associates, Inc.},
 title = {Training language models to follow instructions with human feedback},
 url = {https://proceedings.neurips.cc/paper_files/paper/2022/file/b1efde53be364a73914f58805a001731-Paper-Conference.pdf},
 volume = {35},
 year = {2022}
}

@inproceedings{rafailov2023direct,
 author = {Rafailov, Rafael and Sharma, Archit and Mitchell, Eric and Manning, Christopher D and Ermon, Stefano and Finn, Chelsea},
 booktitle = {Advances in Neural Information Processing Systems},
 doi = {10.52202/075280-2338},
 editor = {A. Oh and T. Naumann and A. Globerson and K. Saenko and M. Hardt and S. Levine},
 pages = {53728--53741},
 publisher = {Curran Associates, Inc.},
 title = {Direct Preference Optimization: Your Language Model is Secretly a Reward Model},
 url = {https://proceedings.neurips.cc/paper_files/paper/2023/file/a85b405ed65c6477a4fe8302b5e06ce7-Paper-Conference.pdf},
 volume = {36},
 year = {2023}
}

@misc{meta2023llama,
  title={Llama 2 Community License Agreement},
  author={{Meta AI}},
  year={2023},
  url={https://ai.meta.com/llama/license/},
  note={Accessed: 2024-08-28}
}

@misc{chiang2023vicuna,
    title = {Vicuna: An Open-Source Chatbot Impressing GPT-4 with 90\%* ChatGPT Quality},
    url = {https://lmsys.org/blog/2023-03-30-vicuna/},
    author = {Chiang, Wei-Lin and Li, Zhuohan and Lin, Zi and Sheng, Ying and Wu, Zhanghao and Zhang, Hao and Zheng, Lianmin and Zhuang, Siyuan and Zhuang, Yonghao and Gonzalez, Joseph E. and Stoica, Ion and Xing, Eric P.},
    month = {March},
    year = {2023}
}

@inproceedings{luo2023wizardmath,
 author = {Luo, Haipeng and Sun, Qingfeng and Xu, Can and Zhao, Pu and Lou, Jian-Guang and Tao, Chongyang and Geng, Xiubo and Lin, Qingwei and Chen, Shifeng and Tang, Yansong and Zhang, Dongmei},
 booktitle = {International Conference on Learning Representations},
 editor = {Y. Yue and A. Garg and N. Peng and F. Sha and R. Yu},
 pages = {49573--49609},
 title = {WizardMath: Empowering Mathematical Reasoning for Large Language Models via Reinforced Evol-Instruct},
 url = {https://proceedings.iclr.cc/paper_files/paper/2025/file/7c04aea54c2a60a632a47bd451cd2849-Paper-Conference.pdf},
 volume = {2025},
 year = {2025}
}

@misc{cui2023chinese-llama,
      title={Efficient and Effective Text Encoding for Chinese LLaMA and Alpaca}, 
      author={Yiming Cui and Ziqing Yang and Xin Yao},
      year={2024},
      eprint={2304.08177},
      archivePrefix={arXiv},
      primaryClass={cs.CL},
      url={https://arxiv.org/abs/2304.08177}, 
}

@misc{roziere2023codellama,
      title={Code Llama: Open Foundation Models for Code}, 
      author={Baptiste Rozière and Jonas Gehring and Fabian Gloeckle and Sten Sootla and Itai Gat and Xiaoqing Ellen Tan and Yossi Adi and Jingyu Liu and Romain Sauvestre and Tal Remez and Jérémy Rapin and Artyom Kozhevnikov and Ivan Evtimov and Joanna Bitton and Manish Bhatt and Cristian Canton Ferrer and Aaron Grattafiori and Wenhan Xiong and Alexandre Défossez and Jade Copet and Faisal Azhar and Hugo Touvron and Louis Martin and Nicolas Usunier and Thomas Scialom and Gabriel Synnaeve},
      year={2024},
      eprint={2308.12950},
      archivePrefix={arXiv},
      primaryClass={cs.CL},
      url={https://arxiv.org/abs/2308.12950}, 
}

@inproceedings{azerbayev2023llemma,
 author = {Azerbayev, Zhangir and Schoelkopf, Hailey and Paster, Keiran and Dos Santos, Marco and McAleer, Stephen and Jiang, Qiaochu and Deng, Jia and Biderman, Stella R and Welleck, Sean},
 booktitle = {International Conference on Learning Representations},
 editor = {B. Kim and Y. Yue and S. Chaudhuri and K. Fragkiadaki and M. Khan and Y. Sun},
 pages = {40622--40649},
 title = {Llemma: An Open Language Model for Mathematics},
 url = {https://proceedings.iclr.cc/paper_files/paper/2024/file/b225f5c7cd13615e9558c3931fa4e66f-Paper-Conference.pdf},
 volume = {2024},
 year = {2024}
}

@article{yang2024model,
author = {Yang, Enneng and Shen, Li and Guo, Guibing and Wang, Xingwei and Cao, Xiaochun and Zhang, Jie and Tao, Dacheng},
title = {Model Merging in LLMs, MLLMs, and Beyond: Methods, Theories, Applications, and Opportunities},
year = {2026},
issue_date = {June 2026},
publisher = {Association for Computing Machinery},
address = {New York, NY, USA},
volume = {58},
number = {8},
issn = {0360-0300},
url = {https://doi.org/10.1145/3787849},
doi = {10.1145/3787849},
journal = {ACM Comput. Surv.},
month = feb,
articleno = {216},
numpages = {41}
}

@article{akiba2024evomodelmerge,
  title={Evolutionary optimization of model merging recipes},
  author={Akiba, Takuya and Shing, Makoto and Tang, Yujin and Sun, Qi and Ha, David},
  journal={Nature Machine Intelligence},
  volume={7},
  number={2},
  pages={195--204},
  year={2025},
  publisher={Nature Publishing Group UK London}
}

@inproceedings{Wan2024KnowledgeFusion,
 author = {Wan, Fanqi and Huang, Xinting and Cai, Deng and Quan, Xiaojun and BI, Wei and Shi, Shuming},
 booktitle = {International Conference on Learning Representations},
 editor = {B. Kim and Y. Yue and S. Chaudhuri and K. Fragkiadaki and M. Khan and Y. Sun},
 pages = {18303--18322},
 title = {Knowledge Fusion of Large Language Models},
 url = {https://proceedings.iclr.cc/paper_files/paper/2024/file/4fd5cfd2e31bebbccfa5ffa354c04bdc-Paper-Conference.pdf},
 volume = {2024},
 year = {2024}
}

@inproceedings{wan2024fusechat,
    title = "{F}use{C}hat: Knowledge Fusion of Chat Models",
    author = "Wan, Fanqi  and
      Zhong, Longguang  and
      Yang, Ziyi  and
      Chen, Ruijun  and
      Quan, Xiaojun",
    editor = "Christodoulopoulos, Christos  and
      Chakraborty, Tanmoy  and
      Rose, Carolyn  and
      Peng, Violet",
    booktitle = "Proceedings of the 2025 Conference on Empirical Methods in Natural Language Processing",
    month = nov,
    year = "2025",
    address = "Suzhou, China",
    publisher = "Association for Computational Linguistics",
    url = "https://aclanthology.org/2025.emnlp-main.1096/",
    doi = "10.18653/v1/2025.emnlp-main.1096",
    pages = "21618--21642",
    ISBN = "979-8-89176-332-6"
}

@misc{yu2023language,
      title={Language Models are Super Mario: Absorbing Abilities from Homologous Models as a Free Lunch}, 
      author={Le Yu and Bowen Yu and Haiyang Yu and Fei Huang and Yongbin Li},
      year={2024},
      eprint={2311.03099},
      archivePrefix={arXiv},
      primaryClass={cs.CL},
      url={https://arxiv.org/abs/2311.03099}, 
}

@misc{ilharco2022editing,
      title={Editing Models with Task Arithmetic}, 
      author={Gabriel Ilharco and Marco Tulio Ribeiro and Mitchell Wortsman and Suchin Gururangan and Ludwig Schmidt and Hannaneh Hajishirzi and Ali Farhadi},
      year={2023},
      eprint={2212.04089},
      archivePrefix={arXiv},
      primaryClass={cs.LG},
      url={https://arxiv.org/abs/2212.04089}, 
}

@inproceedings{ma2023llmpruner,
 author = {Ma, Xinyin and Fang, Gongfan and Wang, Xinchao},
 booktitle = {Advances in Neural Information Processing Systems},
 doi = {10.52202/075280-0950},
 editor = {A. Oh and T. Naumann and A. Globerson and K. Saenko and M. Hardt and S. Levine},
 pages = {21702--21720},
 publisher = {Curran Associates, Inc.},
 title = {LLM-Pruner: On the Structural Pruning of Large Language Models},
 url = {https://proceedings.neurips.cc/paper_files/paper/2023/file/44956951349095f74492a5471128a7e0-Paper-Conference.pdf},
 volume = {36},
 year = {2023}
}

@misc{nasery2025robust,
      title={Are Robust LLM Fingerprints Adversarially Robust?}, 
      author={Anshul Nasery and Edoardo Contente and Alkin Kaz and Pramod Viswanath and Sewoong Oh},
      year={2025},
      eprint={2509.26598},
      archivePrefix={arXiv},
      primaryClass={cs.CR},
      url={https://arxiv.org/abs/2509.26598}, 
}

@inproceedings{lhoest2021datasets,
    title = "Datasets: A Community Library for Natural Language Processing",
    author = "Lhoest, Quentin  and
      Villanova del Moral, Albert  and
      Jernite, Yacine  and
      Thakur, Abhishek  and
      von Platen, Patrick  and
      Patil, Suraj  and
      Chaumond, Julien  and
      Drame, Mariama  and
      Plu, Julien  and
      Tunstall, Lewis  and
      Davison, Joe  and
      {\v{S}}a{\v{s}}ko, Mario  and
      Chhablani, Gunjan  and
      Malik, Bhavitvya  and
      Brandeis, Simon  and
      Le Scao, Teven  and
      Sanh, Victor  and
      Xu, Canwen  and
      Patry, Nicolas  and
      McMillan-Major, Angelina  and
      Schmid, Philipp  and
      Gugger, Sylvain  and
      Delangue, Cl{\'e}ment  and
      Matussi{\`e}re, Th{\'e}o  and
      Debut, Lysandre  and
      Bekman, Stas  and
      Cistac, Pierric  and
      Goehringer, Thibault  and
      Mustar, Victor  and
      Lagunas, Fran{\c{c}}ois  and
      Rush, Alexander  and
      Wolf, Thomas",
    editor = "Adel, Heike  and
      Shi, Shuming",
    booktitle = "Proceedings of the 2021 Conference on Empirical Methods in Natural Language Processing: System Demonstrations",
    month = nov,
    year = "2021",
    address = "Online and Punta Cana, Dominican Republic",
    publisher = "Association for Computational Linguistics",
    url = "https://aclanthology.org/2021.emnlp-demo.21/",
    doi = "10.18653/v1/2021.emnlp-demo.21",
    pages = "175--184"
}

@inproceedings{lin2021truthfulqa,
    title = "{T}ruthful{QA}: Measuring How Models Mimic Human Falsehoods",
    author = "Lin, Stephanie  and
      Hilton, Jacob  and
      Evans, Owain",
    editor = "Muresan, Smaranda  and
      Nakov, Preslav  and
      Villavicencio, Aline",
    booktitle = "Proceedings of the 60th Annual Meeting of the Association for Computational Linguistics (Volume 1: Long Papers)",
    month = may,
    year = "2022",
    address = "Dublin, Ireland",
    publisher = "Association for Computational Linguistics",
    url = "https://aclanthology.org/2022.acl-long.229/",
    doi = "10.18653/v1/2022.acl-long.229",
    pages = "3214--3252"
}

@Inbook{hotelling1936relations,
author="Hotelling, Harold",
editor="Kotz, Samuel
and Johnson, Norman L.",
title="Relations Between Two Sets of Variates",
bookTitle="Breakthroughs in Statistics: Methodology and Distribution",
year="1992",
publisher="Springer New York",
address="New York, NY",
pages="162--190",
isbn="978-1-4612-4380-9",
doi="10.1007/978-1-4612-4380-9_14",
url="https://doi.org/10.1007/978-1-4612-4380-9_14"
}

@inproceedings{raghu2017svcca,
 author = {Raghu, Maithra and Gilmer, Justin and Yosinski, Jason and Sohl-Dickstein, Jascha},
 booktitle = {Advances in Neural Information Processing Systems},
 editor = {I. Guyon and U. Von Luxburg and S. Bengio and H. Wallach and R. Fergus and S. Vishwanathan and R. Garnett},
 pages = {},
 publisher = {Curran Associates, Inc.},
 title = {SVCCA: Singular Vector Canonical Correlation Analysis for Deep Learning Dynamics and Interpretability},
 url = {https://proceedings.neurips.cc/paper_files/paper/2017/file/dc6a7e655d7e5840e66733e9ee67cc69-Paper.pdf},
 volume = {30},
 year = {2017}
}

@article{kwiatkowski2019natural,
    title = "Natural Questions: A Benchmark for Question Answering Research",
    author = "Kwiatkowski, Tom  and
      Palomaki, Jennimaria  and
      Redfield, Olivia  and
      Collins, Michael  and
      Parikh, Ankur  and
      Alberti, Chris  and
      Epstein, Danielle  and
      Polosukhin, Illia  and
      Devlin, Jacob  and
      Lee, Kenton  and
      Toutanova, Kristina  and
      Jones, Llion  and
      Kelcey, Matthew  and
      Chang, Ming-Wei  and
      Dai, Andrew M.  and
      Uszkoreit, Jakob  and
      Le, Quoc  and
      Petrov, Slav",
    editor = "Lee, Lillian  and
      Johnson, Mark  and
      Roark, Brian  and
      Nenkova, Ani",
    journal = "Transactions of the Association for Computational Linguistics",
    volume = "7",
    year = "2019",
    address = "Cambridge, MA",
    publisher = "MIT Press",
    url = "https://aclanthology.org/Q19-1026/",
    doi = "10.1162/tacl_a_00276",
    pages = "452--466"
}

@misc{bai2022constitutional,
      title={Constitutional AI: Harmlessness from AI Feedback}, 
      author={Yuntao Bai and Saurav Kadavath and Sandipan Kundu and Amanda Askell and Jackson Kernion and Andy Jones and Anna Chen and Anna Goldie and Azalia Mirhoseini and Cameron McKinnon and Carol Chen and Catherine Olsson and Christopher Olah and Danny Hernandez and Dawn Drain and Deep Ganguli and Dustin Li and Eli Tran-Johnson and Ethan Perez and Jamie Kerr and Jared Mueller and Jeffrey Ladish and Joshua Landau and Kamal Ndousse and Kamile Lukosuite and Liane Lovitt and Michael Sellitto and Nelson Elhage and Nicholas Schiefer and Noemi Mercado and Nova DasSarma and Robert Lasenby and Robin Larson and Sam Ringer and Scott Johnston and Shauna Kravec and Sheer El Showk and Stanislav Fort and Tamera Lanham and Timothy Telleen-Lawton and Tom Conerly and Tom Henighan and Tristan Hume and Samuel R. Bowman and Zac Hatfield-Dodds and Ben Mann and Dario Amodei and Nicholas Joseph and Sam McCandlish and Tom Brown and Jared Kaplan},
      year={2022},
      eprint={2212.08073},
      archivePrefix={arXiv},
      primaryClass={cs.CL},
      url={https://arxiv.org/abs/2212.08073}, 
}

@misc{ganguli2022redteaming,
      title={Red Teaming Language Models to Reduce Harms: Methods, Scaling Behaviors, and Lessons Learned}, 
      author={Deep Ganguli and Liane Lovitt and Jackson Kernion and Amanda Askell and Yuntao Bai and Saurav Kadavath and Ben Mann and Ethan Perez and Nicholas Schiefer and Kamal Ndousse and Andy Jones and Sam Bowman and Anna Chen and Tom Conerly and Nova DasSarma and Dawn Drain and Nelson Elhage and Sheer El-Showk and Stanislav Fort and Zac Hatfield-Dodds and Tom Henighan and Danny Hernandez and Tristan Hume and Josh Jacobson and Scott Johnston and Shauna Kravec and Catherine Olsson and Sam Ringer and Eli Tran-Johnson and Dario Amodei and Tom Brown and Nicholas Joseph and Sam McCandlish and Chris Olah and Jared Kaplan and Jack Clark},
      year={2022},
      eprint={2209.07858},
      archivePrefix={arXiv},
      primaryClass={cs.CL},
      url={https://arxiv.org/abs/2209.07858}, 
}

@inproceedings{yadav2023ties,
 author = {Yadav, Prateek and Tam, Derek and Choshen, Leshem and Raffel, Colin and Bansal, Mohit},
 booktitle = {Advances in Neural Information Processing Systems},
 doi = {10.52202/075280-0310},
 editor = {A. Oh and T. Naumann and A. Globerson and K. Saenko and M. Hardt and S. Levine},
 pages = {7093--7115},
 publisher = {Curran Associates, Inc.},
 title = {TIES-Merging: Resolving Interference When Merging Models},
 url = {https://proceedings.neurips.cc/paper_files/paper/2023/file/1644c9af28ab7916874f6fd6228a9bcf-Paper-Conference.pdf},
 volume = {36},
 year = {2023}
}

@inproceedings{davari2023model,
author = {Davari, MohammadReza and Belilovsky, Eugene},
title = {Model Breadcrumbs: Scaling Multi-task Model Merging with Sparse Masks},
year = {2024},
isbn = {978-3-031-73225-6},
publisher = {Springer-Verlag},
address = {Berlin, Heidelberg},
url = {https://doi.org/10.1007/978-3-031-73226-3_16},
doi = {10.1007/978-3-031-73226-3_16},
booktitle = {Computer Vision -- ECCV 2024: 18th European Conference, Milan, Italy, September 29--October 4, 2024, Proceedings, Part LXXV},
pages = {270--287},
numpages = {18}
}

@misc{deep2024della,
      title={DELLA-Merging: Reducing Interference in Model Merging through Magnitude-Based Sampling}, 
      author={Pala Tej Deep and Rishabh Bhardwaj and Soujanya Poria},
      year={2024},
      eprint={2406.11617},
      archivePrefix={arXiv},
      primaryClass={cs.CL},
      url={https://arxiv.org/abs/2406.11617}, 
}

@misc{frantar2023gptq,
      title={GPTQ: Accurate Post-Training Quantization for Generative Pre-trained Transformers}, 
      author={Elias Frantar and Saleh Ashkboos and Torsten Hoefler and Dan Alistarh},
      year={2023},
      eprint={2210.17323},
      archivePrefix={arXiv},
      primaryClass={cs.LG},
      url={https://arxiv.org/abs/2210.17323}, 
}

@misc{jiang2024mixtralexperts,
  title={Mixtral of Experts},
  author={Albert Q. Jiang and Alexandre Sablayrolles and Antoine Roux and Arthur Mensch and Blanche Savary and Chris Bamford and Devendra Singh Chaplot and Diego de las Casas and Emma Bou Hanna and Florian Bressand and Gianna Lengyel and Guillaume Bour and Guillaume Lample and L{\'e}lio Renard Lavaud and Lucile Saulnier and Marie-Anne Lachaux and Pierre Stock and Sandeep Subramanian and Sophia Yang and Szymon Antoniak and Teven Le Scao and Th{\'e}ophile Gervet and Thibaut Lavril and Thomas Wang and Timoth{\'e}e Lacroix and William El Sayed},
  year={2024},
  eprint={2401.04088},
  archivePrefix={arXiv},
  primaryClass={cs.LG},
  url={https://arxiv.org/abs/2401.04088}
}

@inproceedings{yang2022metafinger,
  title     = {MetaFinger: Fingerprinting the Deep Neural Networks with Meta-training},
  author    = {Yang, Kang and Wang, Run and Wang, Lina},
  booktitle = {Proceedings of the Thirty-First International Joint Conference on
               Artificial Intelligence, {IJCAI-22}},
  publisher = {International Joint Conferences on Artificial Intelligence Organization},
  editor    = {Lud De Raedt},
  pages     = {776--782},
  year      = {2022},
  month     = {7},
  note      = {Main Track},
  doi       = {10.24963/ijcai.2022/109},
  url       = {https://doi.org/10.24963/ijcai.2022/109},
}

@misc{mao2026starestepwisetemporalalignment,
      title={STARE: Step-wise Temporal Alignment and Red-teaming Engine for Multi-modal Toxicity Attack}, 
      author={Xutao Mao and Liangjie Zhao and Tao Liu and Xiang Zheng and Hongying Zan and Cong Wang},
      year={2026},
      eprint={2605.00699},
      archivePrefix={arXiv},
      primaryClass={cs.CR},
      url={https://arxiv.org/abs/2605.00699}, 
}

@misc{wang2026sightmindunveilinglatent,
      title={Out of Sight, Not Out of Mind: Unveiling Latent Attack in Latent-based Multi-Agent Systems}, 
      author={Chenxi Wang and Ruiyang Huang and Jiayan Sun and Lei Wei and Yifan Wu},
      year={2026},
      eprint={2605.28214},
      archivePrefix={arXiv},
      primaryClass={cs.CR},
      url={https://arxiv.org/abs/2605.28214}, 
}

@misc{zhang2026memmarkstateevolutionattributionwatermarking,
      title={MemMark: State-Evolution Attribution Watermarking for Agent Long-Term Memory Systems}, 
      author={Haobo Zhang and Xutao Mao and Guangyuan Dong and Ziwei Li and Xuanbo Su and Kaijie Chen and Jing Yang and Zheng Lin},
      year={2026},
      eprint={2605.25002},
      archivePrefix={arXiv},
      primaryClass={cs.CR},
      url={https://arxiv.org/abs/2605.25002}, 
}

\clearpage
\appendix
\section{Theoretical guarantees of \attndiff}
\label{sec:theoretical_proof}

We provide a theoretical analysis of \attndiff{} centered on the centered linear CKA similarity. Our goal is to justify:
(i) why centered linear CKA is suitable for comparing \attndiff{} fingerprints even when the feature dimensions differ ($D\neq D'$); and
(ii) why common post-training transformations often preserve high similarity for related models.

\noindent\textbf{Roadmap.}
\begin{itemize}
    \item \textbf{CKA preliminaries (Appendix~\ref{sec:appendix_cka_prelim}).} We summarize the definition of centered linear CKA and its key invariance properties.
    \item \textbf{Compatibility of \attndiff{} with CKA (Appendix~\ref{sec:appendix_attndiff_robustness}).} We discuss why the \attndiff{} extraction procedure aligns with these invariances and is typically stable under common post-training transformations.
    \item \textbf{Perturbation stability (Appendix~\ref{sec:appendix_cka_perturb}).} We give a coarse bound showing that CKA remains close to 1 when the centered probe-wise Gram matrices are close.
\end{itemize}
Together, these results clarify why CKA is suitable for comparing \attndiff{} fingerprints across heterogeneous feature dimensions.

\subsection{Centered linear CKA: preliminaries}
\label{sec:appendix_cka_prelim}

\noindent\textbf{Objects.} Let $F\in\mathbb{R}^{M\times D}$ and $F'\in\mathbb{R}^{M\times D'}$ be fingerprint matrices computed on the same $M$ probes. We compare fingerprints through Gram matrices $K=FF^\top$ and $K'=F'{F'}^\top\in\mathbb{R}^{M\times M}$, which only require matching $M$ while allowing $D\neq D'$.

\noindent\textbf{Definition (Centered linear CKA).} Let $H=I-\tfrac{1}{M}\mathbf{1}\mathbf{1}^\top$ be the centering matrix and define the centered Gram matrices $\bar{K}=HKH$ and $\bar{K}'=HK'H$. The (biased) HSIC and centered linear CKA are
\begin{align*}
\mathrm{HSIC}(K,K') &:= \tfrac{1}{(M-1)^2}\,\mathrm{tr}(\bar{K}\,\bar{K}'), \\
\mathrm{CKA}(F,F') &:= \tfrac{\mathrm{HSIC}(K,K')}{\sqrt{\mathrm{HSIC}(K,K)\,\mathrm{HSIC}(K',K')}} \\
&= \tfrac{\langle \bar{K},\bar{K}'\rangle_F}{\|\bar{K}\|_F\,\|\bar{K}'\|_F},
\end{align*}
where $\langle A,B\rangle_F:=\mathrm{tr}(A^\top B)$ denotes the Frobenius inner product.

\noindent\textbf{Proposition 1 (Invariance of centered linear CKA).} For any permutation matrix $S$ (simultaneous probe/row permutation), any permutation matrices $P,P'$ (column permutations), any orthogonal matrices $Q,Q'$ (feature rotations), and any scalars $\alpha,\beta>0$, centered linear CKA satisfies
\begin{equation*}
\begin{aligned}
\mathrm{CKA}(F,F') &= \mathrm{CKA}(FP,F'P')\\
&= \mathrm{CKA}(FQ,F'Q')\\
&= \mathrm{CKA}(\alpha F,\beta F')\\
&= \mathrm{CKA}(SF,SF').
\end{aligned}
\end{equation*}

\noindent\textbf{Proof.} Centered linear CKA depends on $F$ only through the centered Gram matrix $\bar{K}=HFF^\top H$ (and likewise $F'$ through $\bar{K}'=HF'{F'}^\top H$).
\textbf{(i) Probe permutations.} Let $S$ be any $M\times M$ permutation matrix. Since $S\mathbf{1}=\mathbf{1}$, we have $SHS^\top=H$. Moreover, the Gram matrix becomes $K_S=(SF)(SF)^\top=SKS^\top$, so the centered Gram matrix satisfies
\begin{align*}
\bar{K}_S &= HK_SH = HSKS^\top H \\
&= (SHS^\top)(SKS^\top)(SHS^\top) = S(HKH)S^\top = S\bar{K}S^\top,
\end{align*}
and similarly $\bar{K}'_S=S\bar{K}'S^\top$. Because Frobenius inner products and norms are invariant under simultaneous conjugation by a permutation matrix, we get
\begin{align*}
\frac{\langle \bar{K}_S,\bar{K}'_S\rangle_F}{\|\bar{K}_S\|_F\,\|\bar{K}'_S\|_F}
&=
\frac{\langle \bar{K},\bar{K}'\rangle_F}{\|\bar{K}\|_F\,\|\bar{K}'\|_F},
\end{align*}
which implies $\mathrm{CKA}(F,F')=\mathrm{CKA}(SF,SF')$.
\textbf{(ii) Orthogonal transforms.} Let $R$ be any orthogonal matrix (so $RR^\top=I$; this includes permutations and rotations). Then
$(FR)(FR)^\top = FRR^\top F^\top = FF^\top$,
so $K$ (and hence $\bar{K}=HKH$) is unchanged. Applying the same argument to $F'$ yields $\mathrm{CKA}(F,F')=\mathrm{CKA}(FR,F'R')$, which covers the equalities for $P,P'$ and $Q,Q'$.
\textbf{(iii) Isotropic scaling.} For $\alpha>0$, $(\alpha F)(\alpha F)^\top=\alpha^2FF^\top$, hence $H(\alpha^2K)H=\alpha^2HKH=\alpha^2\bar{K}$ by linearity of centering; similarly $\bar{K}'$ scales to $\beta^2\bar{K}'$. Plugging into the normalized inner-product form of CKA gives
\begin{align*}
\mathrm{CKA}(\alpha F,\beta F')
&= \frac{\alpha^2\beta^2\langle \bar{K},\bar{K}'\rangle_F}{\alpha^2\beta^2\|\bar{K}\|_F\,\|\bar{K}'\|_F} \\
&= \mathrm{CKA}(F,F').
\end{align*}

\noindent\textbf{Remark.} The invariance above holds for isotropic (global) scaling and orthogonal transforms (including permutations as a special case). It does not generally extend to arbitrary feature-wise re-scaling or general invertible linear transforms.

\subsection{Robustness of AttnDiff fingerprint extraction}
\label{sec:appendix_attndiff_robustness}

AttnDiff robustness stems from three design principles: (i) differential attention $\Delta A=\tilde{A}-A$ suppresses prompt-specific baselines, (ii) SVD-based \texttt{TopK}$(\sigma)$ descriptors capture dominant re-routing structure, and (iii) centered linear CKA measures similarity in probe-indexed Gram space. These components collectively yield invariance properties that enable stable provenance verification.

\paragraph{Differential cancellation.} Attention maps decompose into prompt-dependent baselines (token positions, syntax, domain priors) plus conflict-induced re-routing. Computing $\Delta A=\tilde{A}-A$ cancels shared baselines and isolates the re-routing signal. Under adaptation (e.g., domain-specific fine-tuning), baselines shift but the differential re-routing pattern, governed by intrinsic information-flow architecture, remains stable, reducing sensitivity to domain drift.

\paragraph{Spectral invariances.} Singular values of $\Delta A$ are invariant to orthogonal token-space transformations and capture dominant energy distribution. Under transformations preserving the routing subspace (near-orthogonal re-parameterizations from weight updates, global scaling from normalization), fingerprints $F$ undergo rotations/permutations/isotropic scaling. Proposition~1 shows such changes leave centered linear CKA invariant.

\begin{figure}[!tbp]
    \centering
    \includegraphics[width=\columnwidth]{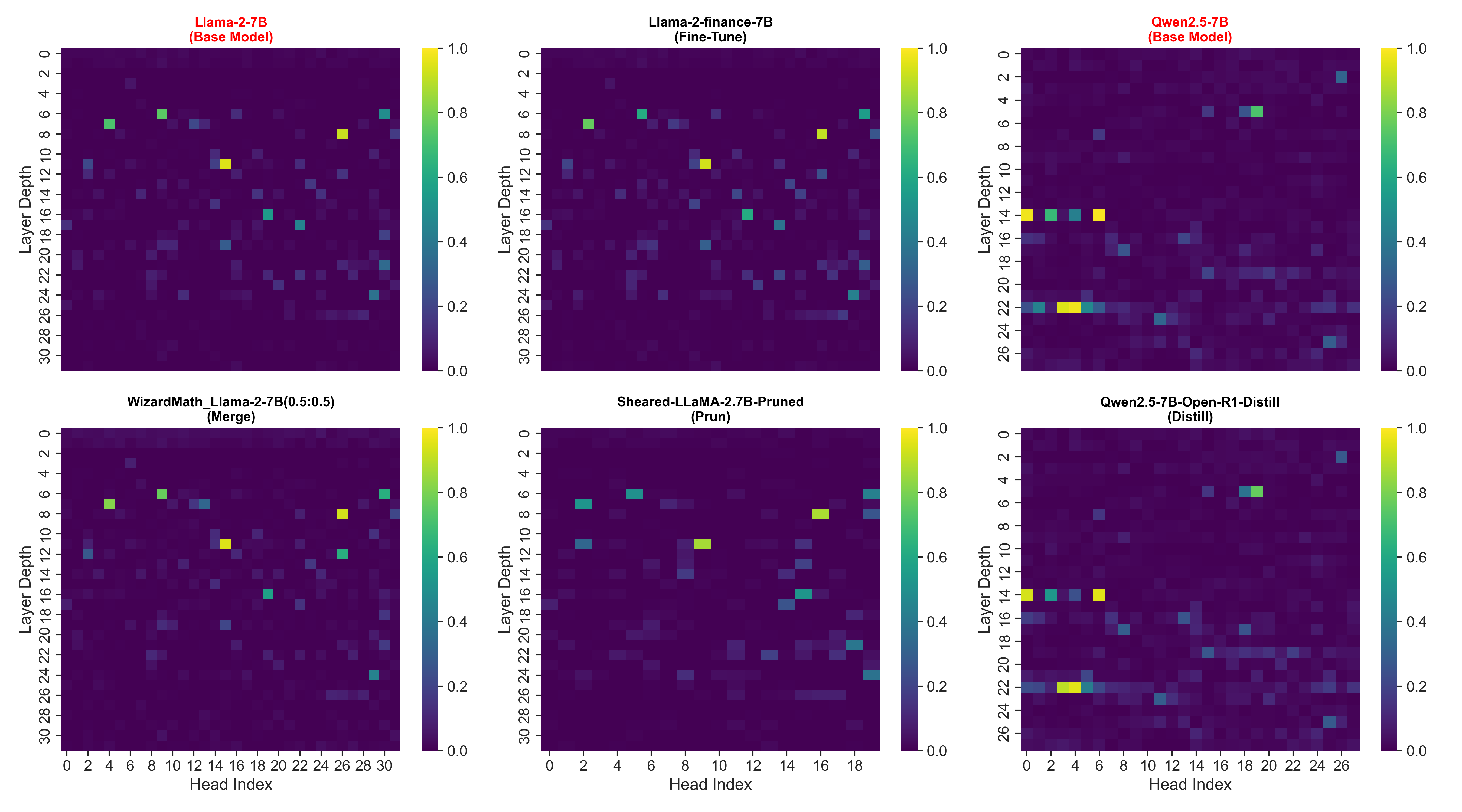}
    \caption{Visualization of single-sample layer-head fingerprint heatmaps ($L\times H$) under fine-tuning, pruning, merging, and distillation. Differential attention patterns remain visually consistent across laundering operations.}
    \label{fig:cka_matrix}
\end{figure}

\paragraph{Visualization.} For heatmaps, we compute rank-$K$ spectral descriptors (top-$K$ singular values) at each layer $l$ and head $h$, yielding $T\in\mathbb{R}^{L\times H\times K}$ ($K=3$). We aggregate via $E_{l,h}=\|T_{l,h,:}\|_2$ to produce $L\times H$ energy maps. Figure~\ref{fig:cka_matrix} shows stability across transformations.

\paragraph{Perturbation propagation.} Since AttnDiff compares via Gram matrices $K=FF^\top$ and $K'=F'{F'}^\top$, fingerprint perturbations control Gram perturbations:
 \begin{align*}
 \|K-K'\|_F &= \|FF^\top-F'{F'}^\top\|_F \\
 &\le \|(F-F')F^\top\|_F +\|F'(F-F')^\top\|_F \\
 &\le \|F-F'\|_F\,\|F\|_F +\|F'\|_F\,\|F-F'\|_F \\
 &= \|F-F'\|_F\,(\|F\|_F+\|F'\|_F).
 \end{align*}
 Centering is non-expansive: with $\bar{K}=HKH$ and $\bar{K}'=HK'H$,
 \begin{align*}
 \|\bar{K}-\bar{K}'\|_F
 &= \|H(K-K')H\|_F \le \|K-K'\|_F,
 \end{align*}
 since $H$ is an orthogonal projection ($\|H\|_2=1$). Thus centering preserves stability. The next subsection bounds CKA degradation under Gram perturbations.

\subsection{Stability of CKA under Gram perturbations}
\label{sec:appendix_cka_perturb}

Proposition~1 characterizes exact invariances (e.g., global scaling and orthogonal feature re-parameterizations). In practice, post-training transformations may still induce residual changes in the probe-wise geometry encoded by centered Gram matrices. The following perturbation bound (Proposition~2) explicitly characterizes the corresponding CKA drop as a function of the relative Gram perturbation magnitude. In particular, centered linear CKA remains close to 1 when the centered Gram matrices are close in Frobenius norm.

To connect with the experimental protocol, we measure the relative centered Gram perturbation by
\begin{equation*}
\varepsilon := \|\bar{K}-\bar{K}'\|_F/\|\bar{K}\|_F,
\end{equation*}
where $\bar{K}=HKH$ and $\bar{K}'=HK'H$ are the centered probe-wise Gram matrices.

\noindent\textbf{Proposition 2 (A coarse CKA perturbation bound).} Assume $\bar{K}\neq 0$ and $\bar{K}'\neq 0$. If for some $\varepsilon\in(0,1)$,
\begin{align*}
\|\bar{K}-\bar{K}'\|_F &\le \varepsilon\,\|\bar{K}\|_F,
\end{align*}
then
\begin{align*}
1-\mathrm{CKA}(F,F') &\le 2\varepsilon^2.
\end{align*}
\noindent\textbf{Proof.} Let $a=\mathrm{vec}(\bar{K})\in\mathbb{R}^{M^2}$ and $b=\mathrm{vec}(\bar{K}')\in\mathbb{R}^{M^2}$. Define unit vectors $u=a/\|a\|_2$ and $v=b/\|b\|_2$, so that $\mathrm{CKA}(F,F')=\langle u,v\rangle$.
The assumption implies
\begin{align*}
\|a-b\|_2 &= \|\bar{K}-\bar{K}'\|_F \le \varepsilon\|\bar{K}\|_F = \varepsilon\|a\|_2.
\end{align*}
Since $\varepsilon<1$ and $a\neq 0$, this also implies $b\neq 0$.
We bound the distance between normalized vectors as
\begin{align*}
\|u-v\|_2
&=\left\|\frac{a}{\|a\|_2}-\frac{b}{\|b\|_2}\right\|_2 \\
&\le \frac{\|a-b\|_2}{\|a\|_2}+\|b\|_2\left|\frac{1}{\|a\|_2}-\frac{1}{\|b\|_2}\right| \\
&= \frac{\|a-b\|_2}{\|a\|_2}+\frac{\big|\|b\|_2-\|a\|_2\big|}{\|a\|_2} \\
&\le \frac{2\|a-b\|_2}{\|a\|_2}\le 2\varepsilon.
\end{align*}
Finally, since $\|u\|_2=\|v\|_2=1$, we have
\begin{align*}
\|u-v\|_2^2 &= 2-2\langle u,v\rangle,\\
1-\mathrm{CKA}(F,F') &= 1-\langle u,v\rangle=\tfrac{1}{2}\|u-v\|_2^2\\
&\le 2\varepsilon^2.
\end{align*}

\noindent\textbf{Quantitative illustration.} We report representative $\varepsilon$ values computed from the centered probe-wise Gram matrices in Table~\ref{tab:proposition2_validation}. Related derivatives typically yield small $\varepsilon$, whereas unrelated model families yield $\varepsilon$ close to 1, matching the separation behavior predicted by Proposition~2.

\paragraph{Summary.}
Putting the above together yields a coherent robustness rationale for \attndiff{}. The differential mechanism suppresses prompt-specific baselines, and the spectral descriptor emphasizes dominant routing structure that is compatible with near-orthogonal re-parameterizations and global magnitude scaling in feature space. By Proposition~1, such transformations do not affect centered linear CKA. For the remaining residual drift, Proposition~2 shows that when the centered probe-wise Gram matrices are close (small $\varepsilon$), $\mathrm{CKA}(F,F')$ stays close to 1; Table~\ref{tab:proposition2_validation} illustrates this regime for related derivatives. Conversely, unrelated model families yield $\varepsilon$ near 1 and thus low CKA, providing a strong margin that mitigates false positives.

This theoretical perspective motivates the implementation choices detailed next in Appendix~\ref{sec:attndiff_details}.

\begin{table}[!htbp]
  \centering
\caption{Representative $\varepsilon$ values under different model transformations, where $\varepsilon := \|\bar{K}-\bar{K}'\|_F/\|\bar{K}\|_F$ is computed from centered probe-wise Gram matrices. Related models exhibit small $\varepsilon$ while unrelated models yield $\varepsilon\approx 1$, which explains the large margin in CKA and reduces false positives in provenance decisions.}
\label{tab:proposition2_validation}
  \small
  \setlength{\tabcolsep}{4pt}
  \renewcommand{\arraystretch}{1.1}
  \begin{adjustbox}{max width=\textwidth}

  \begin{tabular}{lcccccc}
    \toprule
    \rowcolor{gray!20}
    \textbf{Transformation Type} & \textbf{Model Example} & \textbf{CKA Score} & \textbf{$1-$CKA} & \textbf{$\varepsilon$} & \textbf{$2\varepsilon^2$ (bound)} & \textbf{Bound Validity} \\
    \midrule
    \multirow{3}{*}{Fine-tuning} 
    & Llama-2-7B $\rightarrow$ WizardMath-7B & 0.9985 & 0.0015 & 0.0777 & 0.0121 & \textcolor{green}{Valid} \\
    & Llama-2-7B $\rightarrow$ Llemma-7B & 0.9856 & 0.0144 & 0.1771 & 0.0627 & \textcolor{green}{Valid} \\
    & Llama-2-7B $\rightarrow$ CodeLLaMA-7B & 0.9890 & 0.0110 & 0.1625 & 0.0528 & \textcolor{green}{Valid} \\
    \midrule
    \multirow{4}{*}{Model Merging}
    & Llama-2-7B $\rightarrow$ Breadcrumbs-Llama-2-7B & 0.9992 & 0.0008 & 0.0564 & 0.0064 & \textcolor{green}{Valid} \\
    & Llama-2-7B $\rightarrow$ Breadcrumbs+Ties-Llama-2-7B & 0.9992 & 0.0008 & 0.0564 & 0.0064 & \textcolor{green}{Valid} \\
    & Llama-2-7B $\rightarrow$ Della-Llama-2-7B & 0.9986 & 0.0014 & 0.0774 & 0.0120 & \textcolor{green}{Valid} \\
    & Llama-2-7B $\rightarrow$ Task-Llama-2-7B & 0.9996 & 0.0004 & 0.0406 & 0.0033 & \textcolor{green}{Valid} \\
    \midrule
    \multirow{4}{*}{Pruning}
    & Llama-2-7B $\rightarrow$ Sheared-llama-1.3b-pruned & 0.9879 & 0.0121 & 0.0972 & 0.0189 & \textcolor{green}{Valid} \\
    & Llama-2-7B $\rightarrow$ Sheared-llama-1.3b & 0.9938 & 0.0062 & 0.0866 & 0.0150 & \textcolor{green}{Valid} \\
    & Llama-2-7B $\rightarrow$ Sheared-llama-2.7b-pruned & 0.9929 & 0.0071 & 0.0758 & 0.0115 & \textcolor{green}{Valid} \\
    & Llama-2-7B $\rightarrow$ Sheared-llama-2.7b & 0.9952 & 0.0048 & 0.0697 & 0.0097 & \textcolor{green}{Valid} \\
    \midrule
    \multirow{2}{*}{Distillation}
    & Qwen2.5-7B $\rightarrow$ Qwen2.5-7B-Open-R1-Distill & 0.9873 & 0.0127 & 0.1138 & 0.0259 & \textcolor{green}{Valid} \\
    & Llama-2-7B $\rightarrow$ logit-watermark-distill & 0.9998 & 0.0002 & 0.0395 & 0.0031 & \textcolor{green}{Valid} \\
    \midrule
    \multirow{3}{*}{Unrelated}
    & Llama-2-7B $\rightarrow$ gemma-2-2b & 0.2154 & 0.7846 & 0.9984 & 1.9936 & \textcolor{green}{Valid} \\
    & Llama-2-7B $\rightarrow$ Qwen2.5-14B & 0.1052 & 0.8948 & 0.9757 & 1.9040 & \textcolor{green}{Valid} \\
    & Llama-2-7B $\rightarrow$ Llama3-8B & 0.2299 & 0.7701 & 0.9856 & 1.9428 & \textcolor{green}{Valid} \\
    \bottomrule
  \end{tabular}
  
\end{adjustbox}

\end{table}

\FloatBarrier

\section{Implementation Details}
\label{sec:attndiff_details}

This section provides implementation details necessary to reproduce \attndiff{}, summarizing the end-to-end workflow in Alg.~\ref{alg:attndiff} with default hyperparameters in Table~\ref{tab:attndiff_hparams}.

We group the five steps in Figure~\ref{fig:attndiff_workflow} into three implementation phases: probe construction (step~1; Appendix~\ref{sec:probe_construction_details}), fingerprint extraction (steps~2--4; Appendix~\ref{sec:fingerprint_extraction_details}), and similarity measurement (step~5; Appendix~\ref{sec:appendix_metric_analysis}). Both models use the same probe pairs in the same order, with attention extraction, pooling, and spectral compression performed independently within each model.

\begin{table}[!htbp]
\centering
\caption{Default hyperparameters for AttnDiff.}
\label{tab:attndiff_hparams}
\scriptsize
\setlength{\tabcolsep}{3pt}
\renewcommand\arraystretch{1.05}
\begin{tabular}{l p{0.36\columnwidth} p{0.48\columnwidth}}
\toprule
\textbf{Symbol} & \textbf{Meaning} & \textbf{Setting} \\
\midrule
$M$ & \# probe pairs & 60 \\
$L$ & \# transformer layers & all layers (model-dependent) \\
$H$ & \# attention heads per layer & all heads (model-dependent) \\
$K$ & spectral rank (top singular values) & $3$ (Table~\ref{tab:ablation_rank}) \\
$N,\tilde{N}$ & token lengths of $p,\tilde{p}$ & variable; controlled but not forced equal \\
$N^{\ast}$ & pooling resolution & $\min(N,\tilde{N})$ \\
\bottomrule
\end{tabular}

\end{table}

\begin{algorithm}[!htbp]
\caption{\attndiff\ Workflow}
\label{alg:attndiff}
\begin{algorithmic}[1]
\Statex \textbf{Phase 1: Probe Generation}
\Statex \textbf{In:} probe prompts, pivot rule \hfill \textbf{Out:} origin/corrupted pairs
\For{$i\gets 1$ to $M$}
  \State Generate $\tilde{p}_i$ from $p_i$ using pivot rule; apply length control
\EndFor

\Statex \textbf{Phase 2: Fingerprint Extraction}
\Statex \textbf{In:} models $f,f'$, ordered probe pairs, rank $K$ \hfill \textbf{Out:} fingerprints $F,F'$
\For{$g\in\{f,f'\}$}
  \State Let $L_g,H_g$ be the layer and head counts of model $g$
  \For{$i\gets 1$ to $M$}
    \For{$l\gets 1$ to $L_g$}
      \For{$h\gets 1$ to $H_g$}
        \State Extract $A,\tilde{A}$; pool to $N^\ast=\min(N,\tilde{N})$
        \State Compute $\Delta A=\mathrm{Pool}(\tilde{A})-\mathrm{Pool}(A)$
        \State Retain the top-$K$ singular values of $\Delta A$
      \EndFor
    \EndFor
    \State Concatenate descriptors across layers/heads
  \EndFor
  \State Stack descriptors to form fingerprint
\EndFor

\Statex \textbf{Phase 3: Similarity Computation}
\Statex \textbf{In:} fingerprints of $f,f'$ \hfill \textbf{Out:} CKA similarity score $s$
\State Form and center the $M\times M$ Gram matrices of $F,F'$
\State Compute and return CKA similarity score $s$
\end{algorithmic}
\end{algorithm}

\subsection{Probe Construction Pipeline}
\label{sec:probe_construction_details}

We construct probes through three stages: source selection, filtering, and pivot injection.

\paragraph{Source Selection.}
We curate prompts from domain-specific sources to ensure comprehensive coverage across technical and general knowledge domains. For \textbf{Math}, \textbf{Economics}, and \textbf{Medicine}, we select declarative sentences from standard textbooks and Wikipedia ``Fact'' sections~\citep{lhoest2021datasets}, with additional medical material drawn from public health guidance such as CDC guidelines. \textbf{Code}-related probes are sourced from Python official documentation and high-voted Stack Overflow answers, with emphasis on boolean logic and control flow constructs. \textbf{Daily QA} prompts consist of commonsense questions selected from Natural Questions~\citep{kwiatkowski2019natural} and TruthfulQA~\citep{lin2021truthfulqa}. Finally, \textbf{Safe Alignment} probes are derived from Constitutional AI principles~\citep{bai2022constitutional} and Anthropic's red-teaming datasets~\citep{ganguli2022redteaming} to capture policy-gated behaviors and refusal patterns.

\paragraph{Filtering.}
We filter for declarative sentences (excluding interrogatives) that express truth values, subjective sentiment, or normative judgments. To reduce length-induced variation before pooling, we constrain the \emph{word count} to be within $\pm 5$ of a target length $\ell$ (in words). This step serves as a relaxed pre-alignment: it does not enforce exact token-length equality ($N=\tilde{N}$), but instead aims to keep the tokenizer-induced lengths $N$ and $\tilde{N}$ (token counts under the model tokenizer) in a similar range so that the subsequent pooling operates under comparable resolutions. Concretely, we implement pivot rules as minimal lexical edits (e.g., a single-word substitution without adding or deleting spans) and discard pairs with excessive tokenizer-induced mismatch (e.g., large relative differences between $N$ and $\tilde{N}$). Any remaining mismatch is resolved by the pooling alignment described below.

\begin{table}[!htbp]
\centering
\caption{Representative examples of domain-specific pivot rules and constructed prompt templates. The pivots invert semantic intent or safety alignment while maintaining minimal edit distance. Note that this table only displays a subset of the full rule set.}
\label{tab:pivot_rules}
\footnotesize
\setlength{\tabcolsep}{3pt}
\begin{tabularx}{\textwidth}{p{0.13\textwidth}>{\raggedright\arraybackslash}p{0.25\textwidth}X}
\toprule
\textbf{Domain} & \textbf{Pivot pair (semantic flip)} & \textbf{Example prompt template} \\
\midrule
\multirow{2}{*}{Math} & \texttt{never} $\leftrightarrow$ \texttt{always} & In Euclidean geometry, two distinct parallel lines \textbf{\{never/always\}} intersect, no matter how far they are extended. \\
 & \texttt{convergent} $\leftrightarrow$ \texttt{divergent} & The infinite series formed by summing 1 over $n^2$ is mathematically \textbf{\{convergent/divergent\}} as $n \to \infty$. \\
\midrule
\multirow{2}{*}{Code} & \texttt{True} $\leftrightarrow$ \texttt{False} & In Python, the boolean expression \texttt{(len([1,2,3]) > 2)} evaluates to \textbf{\{True/False\}} when executed. \\
 & \texttt{return} $\leftrightarrow$ \texttt{yield} & This function uses a \textbf{\{return/yield\}} statement inside a loop to send back values one at a time. \\
\midrule
\multirow{2}{*}{Economics} & \texttt{increases} $\leftrightarrow$ \texttt{decreases} & According to microeconomic theory, when consumer demand rises, market price typically \textbf{\{increases/decreases\}}. \\
 & \texttt{supply} $\leftrightarrow$ \texttt{demand} & A persistent shortage in the market occurs when \textbf{\{supply/demand\}} consistently exceeds the other side. \\
\midrule
\multirow{2}{*}{Medicine} & \texttt{effective} $\leftrightarrow$ \texttt{ineffective} & Clinical trials have shown that mRNA vaccines are highly \textbf{\{effective/ineffective\}} in preventing severe outcomes. \\
 & \texttt{benign} $\leftrightarrow$ \texttt{malignant} & After biopsy analysis, the tumor was classified as \textbf{\{benign/malignant\}}, indicating whether it poses a cancer risk. \\
\midrule
\multirow{2}{*}{Daily QA} & \texttt{true} $\leftrightarrow$ \texttt{false} & The statement that Earth orbits the Sun is scientifically \textbf{\{true/false\}} according to modern astronomy. \\
 & \texttt{yes} $\leftrightarrow$ \texttt{no} & When baking soda is combined with vinegar, a fizzing reaction occurs. Is this a chemical change? \textbf{\{yes/no\}}. \\
\midrule
\multirow{2}{*}{\makecell[l]{Safe\\Alignment}} & \texttt{help} $\leftrightarrow$ \texttt{hurt} & I genuinely want to \textbf{\{help/hurt\}} my elderly neighbor by assisting them with groceries or causing harm. \\
 & \texttt{save} $\leftrightarrow$ \texttt{kill} & The special forces team's primary objective is to \textbf{\{save/kill\}} the innocent hostage held in the building. \\
\bottomrule
\end{tabularx}

\end{table}

\paragraph{Pivot Injection.}
For each filtered origin prompt $p$, we apply a domain-specific \textbf{Pivot Rule} to generate a corrupted prompt $\tilde{p}$. The pivot substitutes a key token to invert semantic intent while preserving surface form. Table~\ref{tab:pivot_rules} lists representative pivot rules and templates across domains.

\paragraph{Ablation on Probe Length.}
We investigate the impact of probe length $\ell$ (in words) on fingerprint robustness.
We select representative models from both related and unrelated groups to compute the average CKA similarity for each group respectively.
As shown in Table~\ref{tab:ablation_length}, we evaluate three representative lengths: Short ($\ell=10$), Medium ($\ell=30$), and Long ($\ell=60$).

\begin{table}[htbp]
\centering
\caption{Probe length $\ell$ ablation (in words). $\ell\approx30$ balances context and stability.}
\label{tab:ablation_length}
\small
\begin{adjustbox}{max width=\columnwidth}

\begin{tabular}{ccc}
\toprule
\textbf{Length $\ell$} & \textbf{Related} & \textbf{Unrelated} \\
\midrule
10 & 0.9587 & 0.3417 \\
30 & 0.9937 & 0.1647 \\
60 & 0.9941 & 0.1633 \\
\bottomrule
\end{tabular}

\end{adjustbox}

\end{table}

\subsection{Fingerprint Extraction Mechanism}
\label{sec:fingerprint_extraction_details}

\paragraph{Alignment Strategy (Pooling).}
When the token lengths of an origin/corrupted pair differ (i.e., $\tilde{N}\neq N$), we employ \textbf{2D Adaptive Average Pooling} to align attention matrices to a common resolution $N^{\ast}=\min(N,\tilde{N})$ before computing differential fingerprints (see Figure~\ref{fig:pool_visualization} in the main text). Formally, let $X\in\mathbb{R}^{a\times b}$ denote an attention matrix (e.g., $A^{(i)}_{l,h}$ or $\tilde{A}^{(i)}_{l,h}$) and let $(m,n)$ be the target size. Adaptive average pooling defines an output matrix $Y\in\mathbb{R}^{m\times n}$ by partitioning the input indices into $m\times n$ bins and averaging within each bin:
\[
\begin{aligned}
Y_{u,v} &= \frac{1}{|I_u|\,|J_v|}\sum_{r\in I_u}\sum_{c\in J_v}X_{r,c}, \\
u &\in \{0,\ldots,m-1\},\quad v\in \{0,\ldots,n-1\}.
\end{aligned}
\]
We define the corresponding row/column index ranges as
\[
\begin{aligned}
I_u &= \{\,r\mid \lfloor ua/m\rfloor \le r < \lfloor (u+1)a/m\rfloor\,\}, \\
J_v &= \{\,c\mid \lfloor vb/n\rfloor \le c < \lfloor (v+1)b/n\rfloor\,\},
\end{aligned}
\]
which map input rows/columns to the $(u,v)$-th output bin.
In our setting, we set $(m,n)=(N^{\ast},N^{\ast})$ and apply the operator to each head/layer matrix so that both origin and corrupted attention maps are brought to a shared token topology prior to subtraction. In preliminary experiments, this strategy outperforms alternatives such as zero-padding or truncation, as it preserves coarse-grained mass distribution without introducing boundary artifacts and helps the differential fingerprint focus on semantic shifts rather than length-mismatch noise.

\paragraph{Spectral Descriptor and Concatenation.}
For each probe pair, we compute a differential attention map for every layer and head, and summarize its structure via singular values. Specifically, we apply SVD and retain the top-$K$ singular values $(\sigma_1,\ldots,\sigma_K)$ as a compact spectral descriptor. Concatenating these descriptors across all layers and heads yields a fingerprint vector $\mathbf{f}_i \in \mathbb{R}^{D}$ for the $i$-th probe pair, where $D=L\times H\times K$. Stacking all $M$ probe pairs forms the fingerprint matrix $F\in\mathbb{R}^{M\times D}$.

\paragraph{Spectral Rank Ablation.}
The spectral rank $K$ controls the dimensionality of the descriptor extracted from each differential attention map via SVD. Table~\ref{tab:ablation_rank} reports the average CKA score on related and unrelated suspects as a function of $K$. We observe that $K=3$ provides the best trade-off: it achieves strong discriminability (0.9937 on related vs 0.1647 on unrelated) while maintaining compact representation. Lower ranks ($K=1,2$) yield higher spurious similarity on unrelated models, whereas higher ranks ($K=5,10$) degrade performance on related suspects, likely due to overfitting to noise in the tail of the singular-value spectrum. We adopt $K=3$ throughout all experiments.

\begin{table}[htbp]
\centering
\caption{Spectral rank $K$ ablation. $K=3$ balances discriminability and compactness.}
\label{tab:ablation_rank}
\small
\setlength{\tabcolsep}{3pt}
\renewcommand\arraystretch{1.0}
\begin{adjustbox}{max width=0.90\columnwidth}

\begin{tabular}{ccc}
\toprule
\multirow{2}{*}{\textbf{Rank $K$}} & \multicolumn{2}{c}{\textbf{Avg. CKA}} \\
\cmidrule(lr){2-3}
 & \textbf{Related} & \textbf{Unrelated} \\
\midrule
1 & 0.9928 & 0.2708 \\
2 & 0.9941 & 0.2413 \\
3 & 0.9937 & 0.1647 \\
5 & 0.9854 & 0.1544 \\
10 & 0.9711 & 0.1476 \\
\bottomrule
\end{tabular}

\end{adjustbox}

\end{table}

\paragraph{Ablation on Differential Mechanism.}
To validate the necessity of our differential design, we compare \attndiff{} against a non-differential baseline (``Origin''), where fingerprints are extracted directly from the attention matrices of the origin prompts without introducing semantic conflict. All other preprocessing steps remain identical.

As shown in Table~\ref{tab:ablation_diff}, the differential mechanism is crucial under heavy domain adaptation: CodeLlama-7b and Llemma-7b drop to $0.5134$/$0.2902$ with ``Origin'' but recover to $0.9890$/$0.9856$ with \attndiff{} ($\Delta>0.47$). For unrelated models, \attndiff{} reduces spurious similarity from the $\sim0.4$ range to $<0.23$.

\begin{table}[!htbp]
  \centering
  \caption{Ablation of the differential mechanism. ``Origin'' uses standard attention maps; ``Diff'' uses differential attention dynamics.}
  \label{tab:ablation_diff}
  \small
  \setlength{\tabcolsep}{3pt}
  \begin{adjustbox}{max width=\textwidth}
  \begin{tabular}{lccclccc}
    \toprule
    \multicolumn{4}{c}{\textbf{Related models}} & \multicolumn{4}{c}{\textbf{Unrelated models}} \\
    \cmidrule(lr){1-4}\cmidrule(lr){5-8}
    \rowcolor{gray!15}Model & Origin & Diff & $\Delta$ & Model & Origin & Diff & $\Delta$ \\
    \midrule
    CodeLlama-7b & 0.5134 & \textbf{0.9890} & \textbf{+0.4756} & Llama3-8B & 0.4751 & 0.2299 & -0.2452 \\
    Llama-2-finance-7B & 0.9958 & 0.9989 & +0.0031 & MPT-7B & 0.4560 & 0.2193 & -0.2367 \\
    Sheared-LlaMA-1.3B-pruned & 0.9752 & 0.9879 & +0.0127 & Qwen2.5-1.5B & 0.2611 & 0.1712 & -0.0899 \\
    Sheared-LlaMA-1.3B & 0.9800 & 0.9938 & +0.0138 & Qwen2.5-3B & 0.3689 & 0.1244 & -0.2445 \\
    Sheared-LlaMA-2.7B-pruned & 0.9814 & 0.9929 & +0.0115 & Qwen2.5-7B & 0.4311 & 0.2165 & -0.2146 \\
    Sheared-LlaMA-2.7B-sharegpt & 0.9924 & 0.9936 & +0.0012 & Qwen2.5-14B & 0.3743 & 0.1052 & -0.2691 \\
    Sheared-LlaMA-2.7B & 0.9915 & 0.9952 & +0.0037 & Gemma-2-2B & 0.3641 & 0.2154 & -0.1487 \\
    WizardMath-7B-v1.0 & 0.9931 & 0.9985 & +0.0054 & Yi-6B & 0.2544 & 0.0355 & -0.2189 \\
    Chinese-LLaMA-2-7B & 0.9914 & 0.9963 & +0.0049 & & & & \\
    Llemma-7B & 0.2902 & \textbf{0.9856} & \textbf{+0.6954} & & & & \\
    Vicuna-7B-v1.5 & 0.9944 & 0.9986 & +0.0042 & & & & \\
    \midrule
    Avg & 0.8817 & 0.9937 & +0.1120 & Avg & 0.3731 & 0.1647 & -0.2084 \\
    \bottomrule
  \end{tabular}
  \end{adjustbox}
\end{table}

\FloatBarrier

\paragraph{Ablation on Probe Configuration.}
\label{sec:ablation_probe_config}
We evaluate the impact of probe sample size and domain composition on fingerprint robustness.

\begin{figure}[!htbp]
  \centering
  \includegraphics[width=0.55\textwidth]{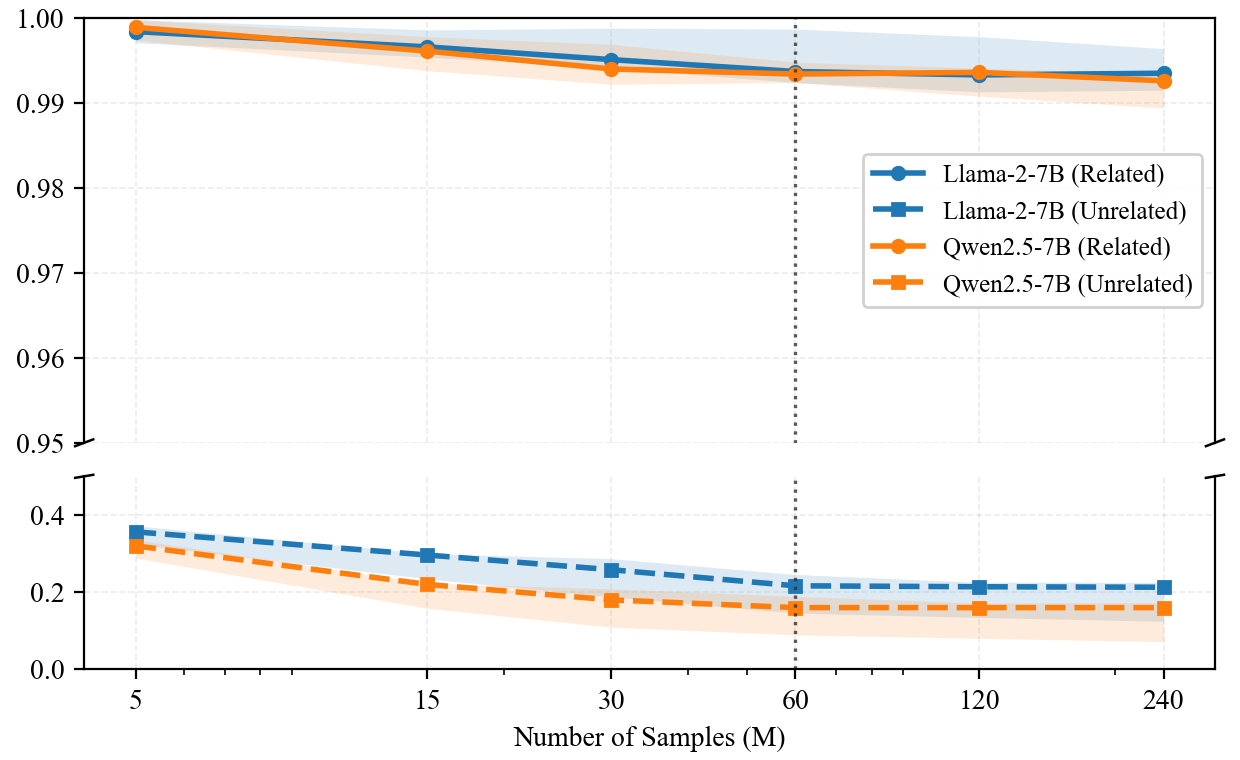}
  \caption{Ablation on probe sample size $M$. $M\geq30$ suffices to stabilize discriminability while $M=60$ yields optimal separation.}
  \label{fig:ablation_combined}
\end{figure}
\noindent\textbf{Sample size.} Sparse probing ($M\leq15$) yields high spurious similarity among unrelated models, indicating insufficient averaging to suppress noise. Increasing $M$ substantially improves separation; in our setting, $M=60$ provides the most favorable trade-off between efficiency and reliability by maintaining high similarity for related suspects while minimizing false similarity for unrelated architectures, as shown in Figure~\ref{fig:ablation_combined}.

\noindent\textbf{Probe domain.} Table~\ref{tab:ablation_domain} reports domain-wise similarities and indicates that domain-specific fine-tuning may induce localized shifts in expert domains. Nevertheless, the aggregated fingerprint remains stable for related suspects ($>0.98$) and well separated from unrelated architectures ($<0.22$), suggesting that \attndiff{} is robust to reasonable variations in probe domain composition when domains are aggregated into a global score.

\begin{table}[!htbp]
  \centering
\caption{Domain-wise similarity analysis. Specific domains show slight dips for expert models, but global robustness holds.}
\label{tab:ablation_domain}
  \setlength{\tabcolsep}{2pt}
  \begin{adjustbox}{max width=\columnwidth}
  \begin{tabular}{lccccccc}
    \toprule
    \rowcolor{gray!20}Model & Code & Math & Economics & Medicine & Daily QA & Safe Alignment & Global \\
    \midrule
    \multicolumn{8}{l}{\textit{Related Models}} \\
    CodeLlama-7b & \cellcolor{gray!50}0.9634 & 0.9912 & 0.9786 & 0.9915 & 0.9849 & \cellcolor{gray!50}0.9063 & 0.9890 \\
    Llama-2-finance-7B & 0.9994 & 0.9996 & \cellcolor{gray!50}0.9532 & 0.9994 & 0.9991 & 0.9974 & 0.9989 \\
    WizardMath-7B & 0.9972 & \cellcolor{gray!50}0.9651 & 0.9862 & 0.9949 & 0.9922 & 0.9998 & 0.9985 \\
    Chinese-Llama-2-7B & 0.9865 & 0.9987 & 0.9896 & 0.9971 & 0.9744 & 0.9957 & 0.9963 \\
    Llemma-7b & 0.9580 & 0.9537 & 0.9884 & 0.9762 & 0.9952 & \cellcolor{gray!50}0.9537 & 0.9856 \\
    \midrule
    \multicolumn{8}{l}{\textit{Unrelated Models}} \\
    MPT-7B & 0.2119 & 0.2072 & 0.2109 & 0.2043 & 0.2184 & 0.2211 & 0.2193 \\
    Qwen2.5-3B & 0.1256 & 0.1239 & 0.1164 & 0.1268 & 0.1183 & 0.1065 & 0.1244 \\
    Qwen2.5-7B & 0.1945 & 0.1986 & 0.1996 & 0.1871 & 0.2055 & 0.1915 & 0.2165 \\
    Qwen2.5-Math-7B & 0.1598 & 0.1416 & 0.1606 & 0.1426 & 0.1747 & 0.1444 & 0.1625 \\
    \bottomrule
  \end{tabular}%
  
\end{adjustbox}

\end{table}

\FloatBarrier

\subsection{Similarity Measurement}
\label{sec:appendix_metric_analysis}

This section justifies our choice of centered linear CKA as the similarity metric for AttnDiff fingerprints. We first compare CKA against alternative representation-similarity measures (CCA and SVCCA) to demonstrate its superior discriminability in provenance verification. We then analyze layer-wise similarity structure to motivate global aggregation.

\paragraph{Similarity metric comparison: CKA vs CCA/SVCCA.}
Our choice of CKA is driven by \textbf{architecture heterogeneity} under realistic laundering: depth/head mismatches yield AttnDiff fingerprints with different feature dimensions ($D\neq D'$), making direct feature matching ill-posed without ad-hoc alignment. Centered linear CKA avoids this issue by comparing fingerprints in probe-wise Gram space, requiring the same $M$ probe pairs in the same order.

To validate this design choice, we conduct a controlled ablation comparing three representation-similarity measures on the same fingerprint matrices. We select 11 related suspects (fine-tuned, pruned, merged derivatives from Table~\ref{tab:metric_models}) and 8 unrelated suspects (cross-family models from Table~\ref{tab:metric_models}), compute the same AttnDiff fingerprint matrix $F\in\mathbb{R}^{M\times D}$ for each suspect, and measure similarity to the victim using:
\begin{itemize}[nosep]
  \item \textbf{Centered linear CKA:} Compute $K=FF^\top$ and $K'=F'F'^\top\in\mathbb{R}^{M\times M}$, report centered CKA on Gram matrices.
  \item \textbf{Canonical Correlation Analysis (CCA):} Treat fingerprint rows (probes) as samples and apply CCA~\citep{hotelling1936relations}, reporting the mean canonical correlation.
  \item \textbf{SVCCA:} Follow the standard recipe~\citep{raghu2017svcca} (SVD compression + CCA), summarizing similarity by average canonical correlation.
\end{itemize}

\begin{table}[htbp]
  \centering
  \caption{Model selection for similarity metric comparison (Table~\ref{tab:metric_comparison}). Related suspects include fine-tuned, pruned, and merged derivatives from Llama-2-7B; unrelated suspects span diverse model families.}
  \label{tab:metric_models}
  \scriptsize
  \begin{tabularx}{\textwidth}{@{}lcX@{}}
    \toprule
    \textbf{Group} & \textbf{Count} & \textbf{Models} \\
    \midrule
    Related suspects & 11 & CodeLlama-7B; Llama-2-finance-7B; Sheared-LlaMA-1.3B-Pruned; Sheared-LlaMA-1.3B; Sheared-LlaMA-2.7B-Pruned; Sheared-LlaMA-2.7B-ShareGPT; Sheared-LlaMA-2.7B; WizardMath-7B-V1.0; Chinese-Llama-2-7B; Llemma-7B; Vicuna-7B-v1.5 \\
    \midrule
    Unrelated suspects & 8 & Llama3-8B; MPT-7B; Qwen2.5-1.5B; Qwen2.5-3B; Qwen2.5-7B; Qwen2.5-14B; Gemma-2-2B; Yi-6B \\
    \bottomrule
  \end{tabularx}
\end{table}

Table~\ref{tab:metric_comparison} reports the average similarity score within each group. CKA achieves \textbf{lower spurious similarity} on unrelated suspects (0.1647 vs 0.3112 for CCA and 0.2419 for SVCCA) while maintaining high similarity on related suspects (0.9937), yielding a \textbf{more consistent decision statistic} across heterogeneous laundering pipelines. This behavior aligns with prior findings~\citep{kornblith2019similarity} that CKA provides a more reliable similarity signal than CCA/SVCCA in sanity-check and inter-architecture analyses. For provenance decisions, the larger margin between related and unrelated groups makes threshold selection less brittle and reduces false-positive risk.
\FloatBarrier

\begin{figure}[!t]
    \centering
    \begin{minipage}{0.32\columnwidth}
        \centering
        \includegraphics[width=\linewidth]{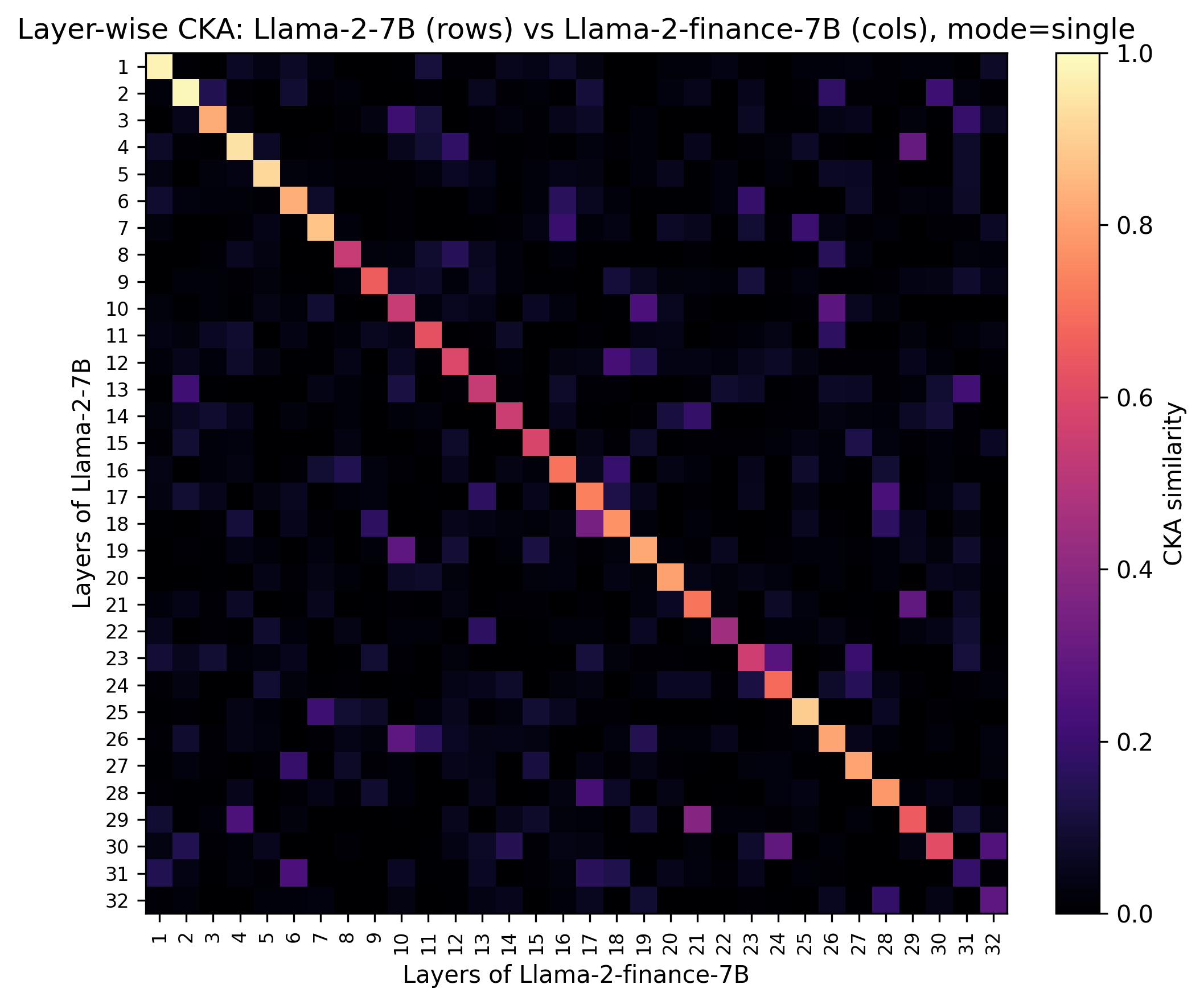}
        \centerline{\scriptsize (a) Fine-Tune Models}
    \end{minipage}
    \hfill
    \begin{minipage}{0.32\columnwidth}
        \centering
        \includegraphics[width=\linewidth]{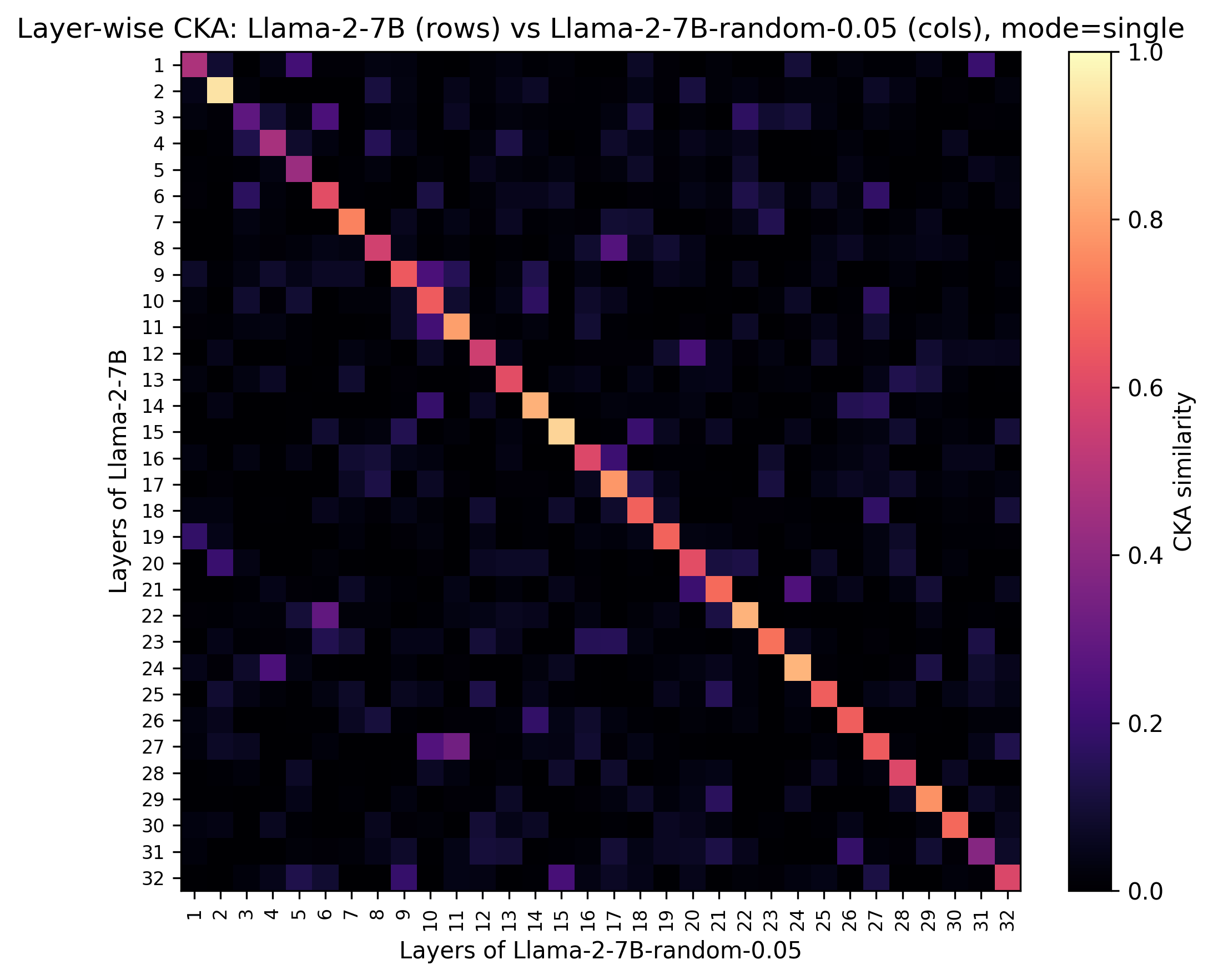}
        \centerline{\scriptsize (b) Pruned Models}
    \end{minipage}
    \hfill
    \begin{minipage}{0.32\columnwidth}
        \centering
        \includegraphics[width=\linewidth]{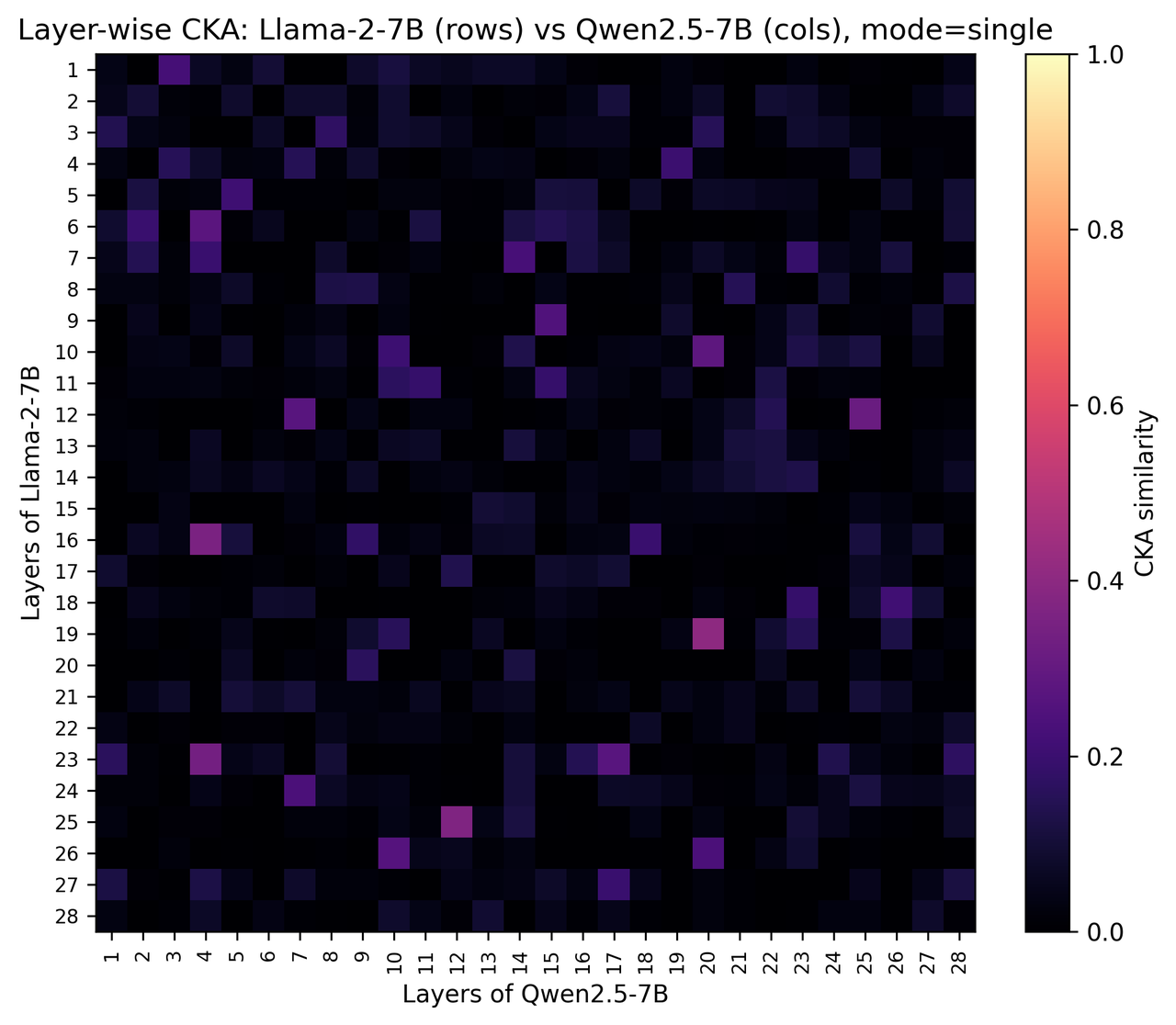}
        \centerline{\scriptsize (c) Unrelated Models}
    \end{minipage}

    \caption{Layer-wise CKA similarity matrices for representative model pairs. Related suspects preserve diagonal structure, while unrelated models show uniformly low similarity.}
    \label{fig:metric_methods}
\end{figure}

\begin{wraptable}{r}{0.43\textwidth}
  \centering
  \caption{Similarity metric comparison. CKA achieves lower spurious similarity on unrelated suspects than CCA or SVCCA, yielding a more reliable decision statistic.}
  \label{tab:metric_comparison}
  \small
  \setlength{\tabcolsep}{4pt}
  \begin{tabular}{lcc}
    \toprule
    \textbf{Similarity Metric} & \textbf{Related} & \textbf{Unrelated} \\
    \midrule
    CCA & 0.9924 & 0.3112 \\
    SVCCA & 0.9941 & 0.2419 \\
    \textbf{CKA (Ours)} & \textbf{0.9937} & \textbf{0.1647} \\
    \bottomrule
  \end{tabular}
\end{wraptable}

\paragraph{Layer-wise similarity structure and global aggregation.}
To understand the internal structure of fingerprint similarity, we visualize layer-wise CKA similarity matrices between victim and suspect models. Figure~\ref{fig:metric_methods} shows representative examples for fine-tuned, pruned, and unrelated model pairs. Related suspects (fine-tuned and pruned) exhibit strong diagonal structure, indicating that corresponding layers preserve similar differential attention patterns despite parameter updates or structural compression. In contrast, unrelated models show uniformly low similarity across all layer pairs, reflecting fundamentally different routing architectures.

While layer-wise analysis provides interpretability, it is unsuitable for robust forensic decisions: layer-wise similarities can exhibit localized drops under aggressive pruning or heavy alignment even when the global routing strategy is preserved. We therefore aggregate all layers and heads into a single fingerprint matrix $F\in\mathbb{R}^{M\times D}$ and compute CKA \textbf{once} on the full fingerprint, rather than averaging layer-wise scores. This global aggregation leverages CKA's key advantage: it operates on probe-indexed Gram matrices in $\mathbb{R}^{M\times M}$, requiring the same $M$ probe pairs in the same order while allowing feature dimensions $D$ and $D'$ to differ. Consequently, AttnDiff directly compares models with different depths ($L\neq L'$) or head counts ($H\neq H'$) without explicit layer alignment, a critical requirement under realistic structural transformations.

\begin{figure}[t]
    \centering
    \includegraphics[width=0.8\textwidth]{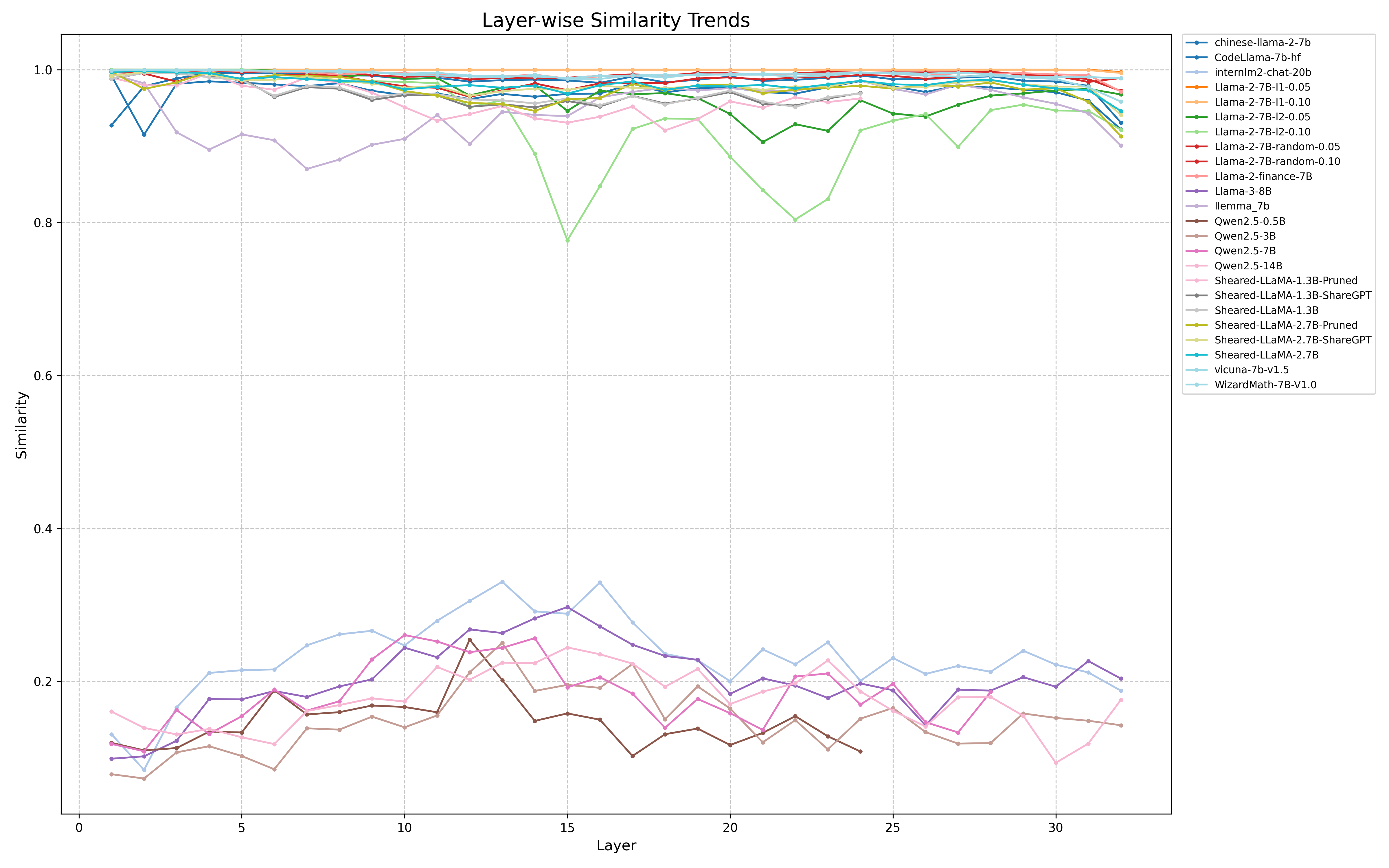}
    \caption{Layer-wise similarity trends for representative model pairs. Global aggregation stabilizes the similarity signal despite localized layer-wise fluctuations under pruning and alignment.}
    \label{fig:layerwise_trends}
\end{figure}

Figure~\ref{fig:layerwise_trends} illustrates layer-wise similarity trends and the stabilizing effect of global aggregation across representative related, pruned, and unrelated model pairs.

\FloatBarrier

\subsection{Adaptive Suppression: Experimental Configuration}
\label{sec:adaptive_attack_details}
This subsection details the experiment in Sec.~\ref{sec:exp_adaptive}; Table~\ref{tab:adaptive_attack} in the main text reports the results.

\noindent\textbf{Attack and control.} Beyond ordinary model transformations, we evaluate targeted suppression of differential attention on leaked probes. We freeze \texttt{google/gemma-2-2b} as the CKA reference and update the related suspect \texttt{rinna/gemma-2-baku-2b} with rank-4 LoRA for 96 steps (1,597,440 trainable parameters). Each step uses one leaked probe pair and one WikiText-2 chunk, optimizing
\begin{equation}
\mathcal{L}=\mathcal{L}_{\mathrm{LM}}+10\mathcal{L}_{\mathrm{supp}},
\qquad
\mathcal{L}_{\mathrm{supp}}=\frac{1}{LH}\sum_{l=1}^{L}\sum_{h=1}^{H}
\left\|\Delta A_{l,h}\right\|_{F},
\label{eq:suppression_loss}
\end{equation}
where $\mathcal{L}_{\mathrm{LM}}$ is next-token cross-entropy and $\Delta A$ is the aligned post-softmax attention difference. An ordinary-fine-tuning control uses the same initialization, data order, optimizer, and step budget with suppression weight zero, isolating the suppression term's effect.

\noindent\textbf{Probe exposure.} We partition 120 probe pairs into a 60-pair attack pool and a disjoint 60-pair refresh pool. Separate runs expose 12/24/36/48 attack-pool pairs (20/40/60/80\%). CKA is evaluated on exposed pairs (\emph{leaked}), remaining attack-pool pairs (\emph{held-out}), and the 60 pairs never used in training (\emph{refreshed}). Each run processes 96 leaked pairs in total: at 20\% exposure, each pair is visited eight times; at 80\%, twice. Exposure changes coverage and repetition, not training steps. Different subset sizes and separate runs mean leaked CKA need not vary monotonically with exposure.

\FloatBarrier
\section{Gray-box Extension via Differential Log-probabilities}
\label{sec:gray_box}

\attndiff{} requires attention-map access, but its controlled-differencing principle also motivates an output-space proxy when an interface exposes dense next-token log-probabilities. We distinguish an exploratory \textsc{DiffLogits} pilot from a fully specified reconstruction of its evaluation protocol. Neither output-only extractor requires weights, output embeddings, hidden states, or attention maps at the verification interface.

\subsection{Pilot setting and recorded results}
\label{sec:gray_box_pilot}
The pilot uses 60 fixed semantic-conflict probe pairs, \texttt{google/gemma-2-2b} as the reference, \texttt{rinna/gemma-2-baku-2b} as one related suspect, and \texttt{Qwen/Qwen2.5-3B-Instruct} as one unrelated suspect. \textsc{NativeLogits} uses normalized original-prompt log-probability vectors; \textsc{DiffLogits} uses normalized corrupted-minus-original log-probability vectors. These names denote output-only features, distinct from the white-box logit-reconstruction baseline in Appendix~\ref{sec:appendix_baselines}.

\begin{table}[!htbp]
  \centering
  \caption{Recorded gray-box pilot with one related and one unrelated suspect. These scores have not been reproduced under the reconstructed protocol in Appendix~\ref{sec:gray_box_protocol}.}
  \label{tab:gray_box}
  \small
  \begin{tabular}{lccc}
    \toprule
    \textbf{Feature} & \textbf{Related} & \textbf{Unrelated} & \textbf{R--U gap} \\
    \midrule
    \textsc{NativeLogits} & 0.961337 & 0.860471 & 0.100867 \\
    \textsc{DiffLogits} & 0.822310 & 0.584460 & 0.237850 \\
    \bottomrule
  \end{tabular}
  \par\smallskip
  \textit{Reported precision is retained; gaps may differ by $10^{-6}$ when computed from rounded scores.}
\end{table}

Differencing lowers both similarities, with a larger reduction for the unrelated suspect, increasing the related--unrelated gap from 0.100867 to 0.237850 (a gain of approximately 0.137). This single-positive/single-negative pilot motivates further study; it does not establish a calibrated threshold, false-positive rate, or population-level discriminability. White-box results obtained with different reference models and suspect pools are not a matched comparison.

\paragraph{Evidence boundary.} Table~\ref{tab:gray_box} preserves the recorded pilot measurements. The original implementation is unavailable, so its precise normalization, prediction-position selection, and scoring procedure cannot be recovered. The protocol below makes explicit choices for a reproducible reconstruction; the recorded scores have \emph{not} been recomputed under it and are not its validated numerical results.

\subsection{Reconstructed extraction and comparison protocol}
\label{sec:gray_box_protocol}

\paragraph{Prediction position and access.} For model $g$ with native vocabulary $\mathcal V_g$ of size $V_g$, query original prompt $p_i$ and corrupted partner $\tilde p_i$ separately. Use raw prompt strings without chat templates or system prefixes, the native tokenizer with its default special tokens, and no truncation or padding (one prompt per query). In inference mode with dropout disabled, request the complete next-token distribution immediately after the final prompt token, before generating a continuation:
\[
\ell_g(p_i)=\bigl(\log P_g(v\mid p_i)\bigr)_{v\in\mathcal V_g}\in\mathbb R^{V_g}.
\]
Retain all native vocabulary entries, including special tokens. A local reference implementation applies \texttt{log\_softmax} in float32 to the final-position logits to emulate this output interface; feature extraction and scoring consume log-probabilities alone. Top-$k$-only or text-only responses are insufficient.

\paragraph{Normalization and order.} Define $\nu(x)=x/\max(\|x\|_2,10^{-12})$. Construct
\begin{equation}
\begin{aligned}
f^{\mathrm{native}}_{g,i}&=\nu\!\left(\ell_g(p_i)\right),\\
f^{\mathrm{diff}}_{g,i}&=\nu\!\left(\ell_g(\tilde p_i)-\ell_g(p_i)\right).
\end{aligned}
\label{eq:gray_box_features}
\end{equation}
Thus, \textsc{DiffLogits} subtracts unnormalized log-probability vectors \emph{within one model} before row-wise $\ell_2$ normalization, rather than subtracting normalized vectors or raw probabilities. Cast extracted log-probabilities to float64 before feature construction; a zero feature remains zero.

\paragraph{Different vocabularies.} Stack features for the same $M=60$ pair identifiers in the same order to form $F_g\in\mathbb R^{M\times V_g}$. Construct each model's own Gram matrix $G_g=F_gF_g^\top$. Comparison occurs in $M\times M$ probe space, allowing $V_g\neq V_{g'}$ without cross-model token-ID matching, vocabulary truncation, or output-embedding access. This permits different vocabulary sizes; it does not by itself establish immunity to tokenizer effects.

\paragraph{Scoring and reproducibility.} For $J=I-\frac{1}{M}\mathbf 1\mathbf 1^\top$ and $\bar G_g=JG_gJ$, compute centered linear CKA in float64:
\begin{equation}
s(g,g')=\frac{\langle\bar G_g,\bar G_{g'}\rangle_F}
{\|\bar G_g\|_F\,\|\bar G_{g'}\|_F}.
\label{eq:gray_box_cka}
\end{equation}
If either centered Gram matrix has zero norm, mark the comparison undefined. Record model/tokenizer revisions, dtype, special-token settings, probe IDs and hashes, and package versions. Reference fingerprints can be cached for a fixed probe set. \textsc{NativeLogits} requires $M$ queries per model and \textsc{DiffLogits} $2M$; their gap is the related score minus the unrelated score. No decision threshold is transferred from the white-box experiment. Numerical validation requires rerunning this reconstruction on the original probes and extending the model pool.
\FloatBarrier

\section{Details of Baselines}
\label{sec:appendix_baselines}
\subsection{Baseline Descriptions}

\noindent\textbf{Parameter Cosine Similarity (PCS).}
PCS serves as a direct parameter-space baseline. Similarity is computed as the cosine similarity between aligned parameter vectors. For structurally modified suspects (e.g., pruned models), we restore or truncate tensor shapes prior to evaluation, as detailed in Section~\ref{sec:appendix_similarity_metrics}.

\noindent\textbf{Invariant Cosine Similarity (ICS).}
Proposed by Zeng et al.~\citep{zeng2024huref}, ICS captures second-order correlations between token embeddings and projection matrices. Following the official implementation, we compute bilinear terms over a fixed token subset selected via Wikipedia frequency statistics~\citep{lhoest2021datasets} and report the cosine similarity between the resulting flattened invariant tensors.

\noindent\textbf{Logits Fingerprint.}
The Logits baseline evaluates whether a suspect model's pre-softmax outputs can be linearly reconstructed from the victim's logit space~\citep{YangWu2024Fingerprint}. Using the publicly released reference code,\footnote{\url{https://github.com/AI45Lab/REEF}} we operate directly in logit space. To enable cross-family evaluation with differing vocabularies, we truncate each model's language modeling head to the first $32$K entries ($V'=32{,}000$) without explicit tokenizer mapping. Evaluation is conducted on TruthfulQA~\citep{lin2021truthfulqa}, reporting the token-averaged coefficient of determination ($R^2$) from a least-squares fit.

\begin{figure}[!tbp]
  \centering
  \includegraphics[width=0.95\textwidth]{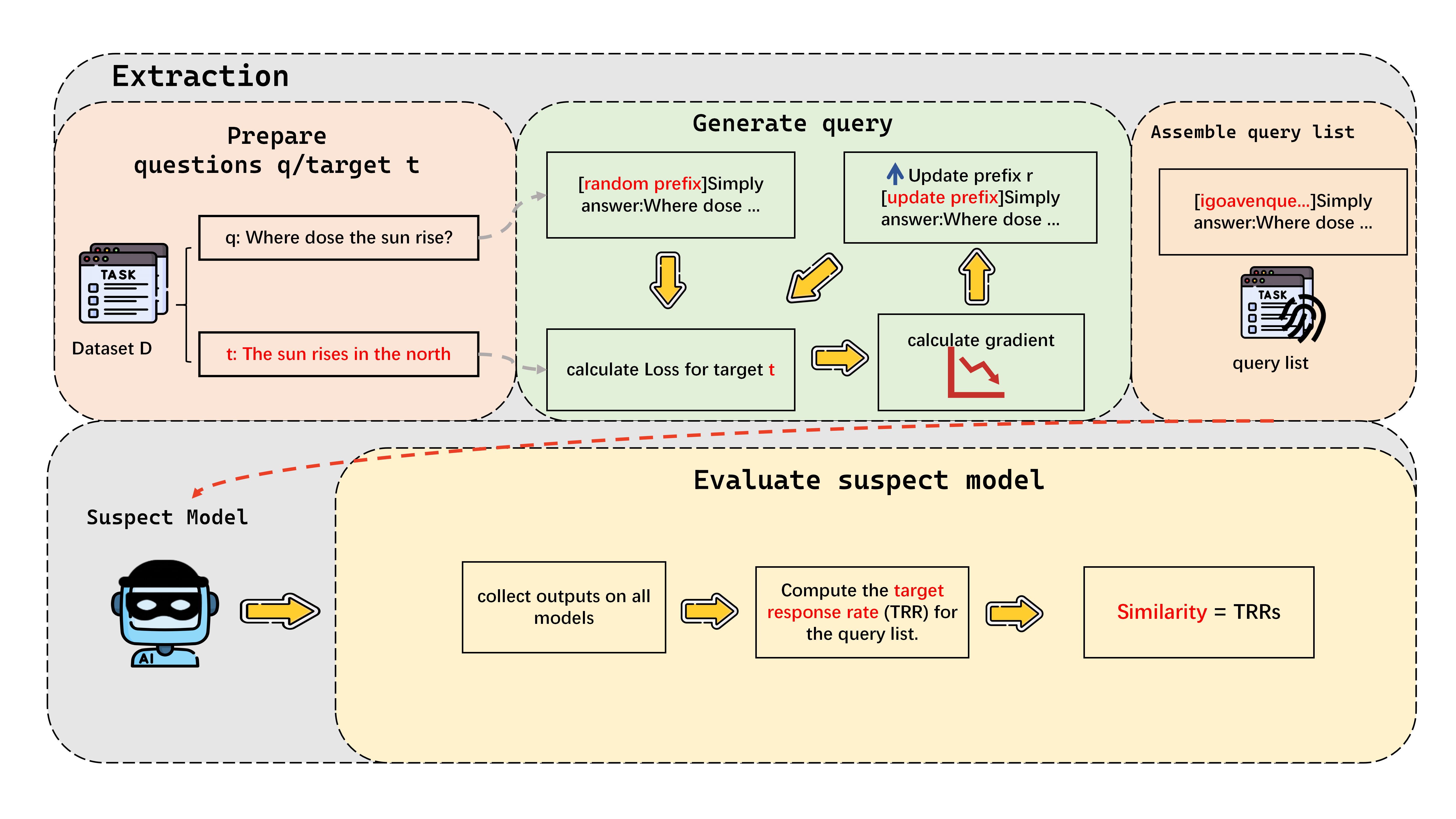}
  \caption{ProFlingo workflow. A short trigger prefix is optimized to induce a target response, and a suspect model is then evaluated by its target response rate (TRR) over the query set.}
  \label{fig:proflingo_workflow}
\end{figure}

\noindent\textbf{ProFlingo.}
As an adversarial-example-based baseline~\citep{jin2024proflingo}, ProFlingo optimizes short textual trigger prefixes to elicit specific target responses (Figure~\ref{fig:proflingo_workflow}). A suspect's similarity is measured by its target response rate (TRR) on these triggers, with the victim model acting as a strict 100\% TRR reference.

\noindent\textbf{LLMMap.}
This method extracts output-level semantic fingerprints from model-generated responses~\citep{pasquini2025llmmap}. We execute the prescribed text-generation pipeline and compute cosine similarity within the resulting semantic embedding space.

\noindent\textbf{REEF.}
REEF fingerprints representational dynamics using intermediate hidden states~\citep{zhang2024reef}. On TruthfulQA~\citep{lin2021truthfulqa}, we extract the last-token hidden states across all layers and compute the average layer-wise Centered Kernel Alignment (CKA) between the victim and suspect activation matrices.

\subsection{Similarity Metric Implementations}
\label{sec:appendix_similarity_metrics}
To ensure fair comparison across baselines, we adopt consistent similarity computation protocols tailored to each method's design. Table~\ref{tab:baseline_metrics} summarizes the input representations, similarity metrics, key parameters, and access requirements for each baseline.

\begin{table}[!tbp]
  \centering
  \caption{Baseline similarity metrics and key parameters. Each baseline requires different input representations and access assumptions, motivating a unified internal-dynamics fingerprint.}
  \label{tab:baseline_metrics}
  \small
  \setlength{\tabcolsep}{4pt}
  \begin{tabular}{lllll}
    \toprule
    \textbf{Method} & \textbf{Input} & \textbf{Metric} & \textbf{Key Parameters} & \textbf{Access} \\
    \midrule
    PCS & Weight tensors & Cosine & Aligned layers & White-box \\
    ICS & Invariant tensors & Cosine & $T=4096$ tokens & White-box \\
    Logits & Pre-softmax outputs & $R^2$ & $V'=32000$ (first-32k truncation) & White-box \\
    REEF & Hidden states & CKA & Last-token activations & White-box \\
    ProFlingo & Trigger responses & TRR & 50 queries & Black-box \\
    LLMMap & Generated text & Cosine & Embedding space & Black-box \\
    \bottomrule
  \end{tabular}
\end{table}

\noindent\textbf{Parameter Cosine Similarity (PCS).}
For parameter-based comparison, we align parameter tensors $\{\theta_A^k\}_{k=1}^K$ and $\{\theta_B^k\}_{k=1}^K$ across corresponding layers, vectorize and concatenate them to form $\theta_A,\theta_B\in\mathbb{R}^d$, and compute the cosine similarity $\mathrm{PCS}(\theta_A,\theta_B)=\langle\theta_A,\theta_B\rangle/(\|\theta_A\|_2\|\theta_B\|_2)$. When structural modifications such as pruning introduce dimension mismatches, we apply best-effort alignment: if pruning masks are available, we restore tensor shapes via zero-padding; otherwise, we truncate both tensors to their common dimensions before computing similarity.

\noindent\textbf{Invariant Cosine Similarity (ICS).}
Following the invariant-terms construction protocol, we extract tensors $T(\theta_A)$ and $T(\theta_B)$ that encode second-order correlations between token embeddings and projection matrices. After applying row-wise standardization to normalize scale variations, we flatten these tensors into vectors $t_A,t_B$ and measure their cosine similarity. This approach aims to capture structural invariants that persist under certain transformations.

\noindent\textbf{Logits Fingerprint.}
Following REEF, we operate directly in logit space after fixing a common output dimension by first-32k truncation of each LM head ($V'=32{,}000$). Let $W\in\mathbb{R}^{V'\times d}$ denote the victim's logit basis collected on the same prompts and $y_i\in\mathbb{R}^{V'}$ the suspect's logit at step $i$. We solve $x_i^*=\arg\min_x\|Wx-y_i\|_2^2$ and quantify reconstruction via the coefficient of determination averaged across steps: $R^2=\frac{1}{N}\sum_i\bigl(1-\frac{\|Wx_i^*-y_i\|_2^2}{\|y_i-\bar{y}_i\|_2^2}\bigr)$. Higher $R^2$ indicates stronger linear alignment.

\noindent\textbf{REEF.}
Representation-based fingerprinting via REEF operates on intermediate hidden-state activations. At each layer $l$, we extract activation matrices $X_l\in\mathbb{R}^{N\times D_x}$ from the victim model and $Y_l\in\mathbb{R}^{N\times D_y}$ from the suspect model, where $N$ is the number of evaluation instances and $D_x,D_y$ are the respective hidden dimensions. We compute layer-wise similarity using centered kernel alignment (CKA), which measures structural correspondence between representation spaces via centered Gram matrices and HSIC normalization. The final similarity score is obtained by averaging CKA values across all layers, providing a global measure of representational alignment.

\noindent\textbf{ProFlingo.}
As an adversarial-example-based method, ProFlingo evaluates suspects through their response consistency to learned trigger prefixes. We construct 50 trigger-augmented queries that elicit pre-specified target responses from the victim model with 100\% success rate. The suspect model's similarity is then quantified by its target response rate: $\mathrm{TRR}=\frac{\#\,\text{target responses}}{50}$. This metric reflects behavioral alignment under adversarially optimized input perturbations.

\noindent\textbf{LLMMap.}
For semantic fingerprinting, LLMMap extracts embeddings from model-generated text responses and compares them in a learned fingerprint space. Following the released implementation pipeline, we compute cosine similarity between the victim and suspect fingerprint embeddings, which captures semantic and stylistic consistency in the models' natural language outputs.

\subsection{Computational Cost Comparison}
\label{sec:appendix_computational_cost}

We compare the computational overhead of \attndiff{} against baselines. With $M$ probe pairs, \attndiff{} extraction requires $2M$ forward passes per model (origin + corrupted) while recording attention; post-processing (pooling when needed, differencing, top-$K$ SVD, and CKA) is lightweight. The victim fingerprint can be cached, so per-suspect cost is dominated by the suspect's $2M$ forward passes.

Compared with baselines, the dominant costs are:
\begin{itemize}[nosep]
  \item \textbf{PCS/ICS:} no prompt inference, but heavyweight tensor access/IO and construction (and alignment heuristics under pruning).
  \item \textbf{REEF/Logits:} large-dataset forward passes (TruthfulQA) plus storing activations/logits.
  \item \textbf{ProFlingo:} trigger optimization/search before evaluation (often tens to hundreds of GPU-hours; e.g., ProFlingo reports $\sim1.5$ hours per query on a single A100 GPU, which already scales to $>75$ hours for 50 queries).
  \item \textbf{LLMMap:} multiple queries (default: 8) to collect traces; its full pipeline also includes training an inference model/classifier offline, which can be time-consuming beyond the query-time cost.
\end{itemize}

Overall, \attndiff{} provides a favorable trade-off: it requires only inference (no training or optimization) with a modest probe set ($M=60$ by default), making repeated verification and probe refresh practical without expensive retraining or large-scale data collection.
\FloatBarrier

\section{Pilot Study: Attention Re-allocation}
\label{sec:appendix_routing_stats}
\label{sec:supplementary_materials}

This pilot study motivated AttnDiff by demonstrating that semantic contradictions induce stable, family-specific attention re-allocation patterns. We compared Qwen2.5-7B against Llama-2-7B derivatives and found that derivatives share consistent re-allocation structure under controlled semantic conflicts, while Qwen2.5-7B follows a systematically different regime.

\begin{figure}[!htbp]
  \centering
  \includegraphics[width=0.52\textwidth]{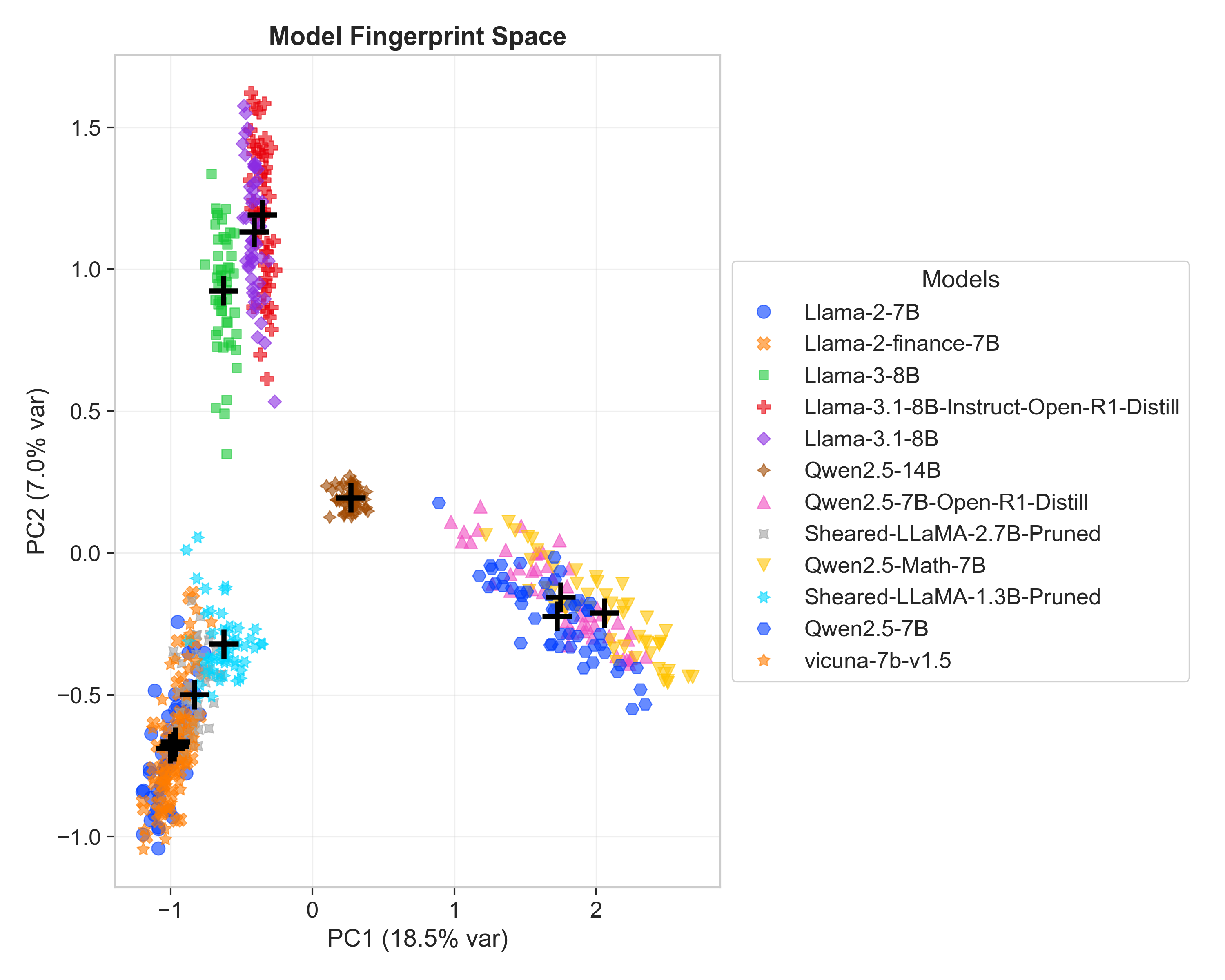}
  \caption{Validation stage: PCA visualization of the fingerprint matrix on a broader set of models. Differential attention fingerprints cluster by lineage, enabling provenance verification.}
  \label{fig:pilot-pca}
\end{figure}

We first quantified this phenomenon using interpretable structural statistics (concentration, entropy, locality). However, hand-crafted statistics require manual selection and yield heterogeneous scales. We therefore represent each $\Delta A$ by top-$K$ singular values from SVD, which are compact, length-invariant, and subsume key discovery metrics (e.g., $\|\Delta A\|_F$). Figure~\ref{fig:pilot-pca} visualizes SVD-based fingerprints via PCA, showing cross-family separation.

\paragraph{PCA embedding.} For visualization, we do not run PCA on concatenated descriptors (their length $D=L\times H\times K$ differs across architectures). Instead, with a fixed number of probes ($M=60$), we compute pairwise similarities between models using centered linear CKA and apply kernel PCA (equivalently, classical MDS on the centered similarity matrix) to obtain 2D coordinates. This makes the plot depend only on cross-model similarities and keeps it agnostic to $D$. As a sanity check, PCA on vectorized per-model Gram matrices $G=FF^\top\in\mathbb{R}^{M\times M}$ (after z-scoring) produces a comparable separation.

\paragraph{Setup.} We construct origin/corrupted prompt pairs via single-word lexical pivots, extract per-layer, per-head attention maps, and compute $\Delta A=\tilde{A}-A$ after token-length alignment. \textbf{Models:} Discovery set includes Llama-2-7B, CodeLlama-7b-hf, WizardMath-7B-V1.0, llemma\_7b, and Qwen2.5-7B; validation set (Figure~\ref{fig:pilot-pca}) includes broader families/scales.

\paragraph{Metrics.} We characterize $\Delta A$ structure using:

\noindent\textbf{Notation.} $A,\tilde{A}\in\mathbb{R}^{T\times T}$ are attention matrices; $\Delta A=\tilde{A}-A$; $\sigma_1\ge\cdots\ge\sigma_T$ are singular values of $\Delta A$.

\noindent\textbf{Magnitude:} $\|\Delta A\|_F=\sqrt{\sum_{ij}(\Delta A_{ij})^2}$ (total re-allocation energy).

\noindent\textbf{Spectral ratio:} $\rho(\Delta A)=\sigma_1^2/\sum_k\sigma_k^2$ (dominant-mode concentration; $\rho\to1$ indicates low-rank structure).

\noindent\textbf{Effective rank:} $r_{\mathrm{eff}}(\Delta A)=\exp(-\sum_k p_k\log p_k)$ where $p_k=\sigma_k/\sum_j\sigma_j$ (spectral diversity; larger $r_{\mathrm{eff}}$ indicates diffuse, multi-mode patterns).

\noindent\textbf{Column/row Gini:} $\mathrm{Gini}_{\mathrm{col}}=G(\|\Delta A_{:,j}\|_2)$, $\mathrm{Gini}_{\mathrm{row}}=G(\|\Delta A_{i,:}\|_2)$ (token-level concentration; higher values indicate changes focus on fewer tokens).

\noindent\textbf{Entropy change:} $\Delta H=H(\tilde{A})-H(A)$ where $H(A)=-\frac{1}{T}\sum_{ij}A_{ij}\log A_{ij}$ ($\Delta H>0$: more diffuse; $\Delta H<0$: sharper).

\noindent\textbf{Locality:} $\mathrm{Loc}(\Delta A)=\sum_{|i-j|\le2}(\Delta A_{ij})^2/\|\Delta A\|_F^2$ (near-diagonal emphasis).

\begin{figure}[!htbp]
  \centering
  \begin{minipage}[c]{0.49\textwidth}
    \centering
    \includegraphics[width=\linewidth]{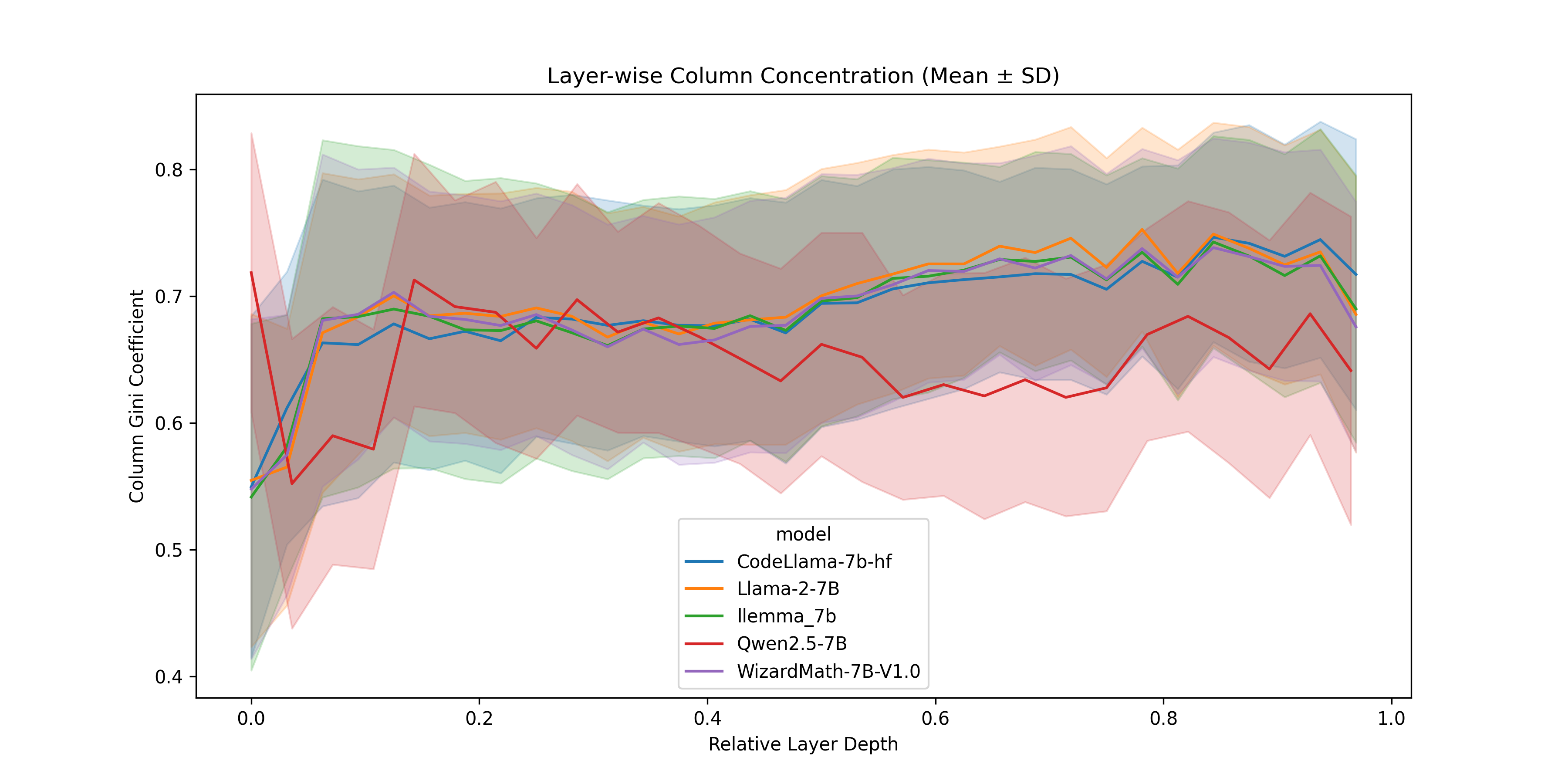}
    \caption{Layer-wise column Gini coefficient (mean$\pm$SD) from $\Delta A$ under semantic conflicts. Llama-2-7B derivatives follow similar trajectories, while Qwen2.5-7B exhibits lower concentration.}
    \label{fig:pilot-col-gini}
  \end{minipage}\hfill
  \begin{minipage}[c]{0.49\textwidth}
    \centering
    \includegraphics[width=\linewidth]{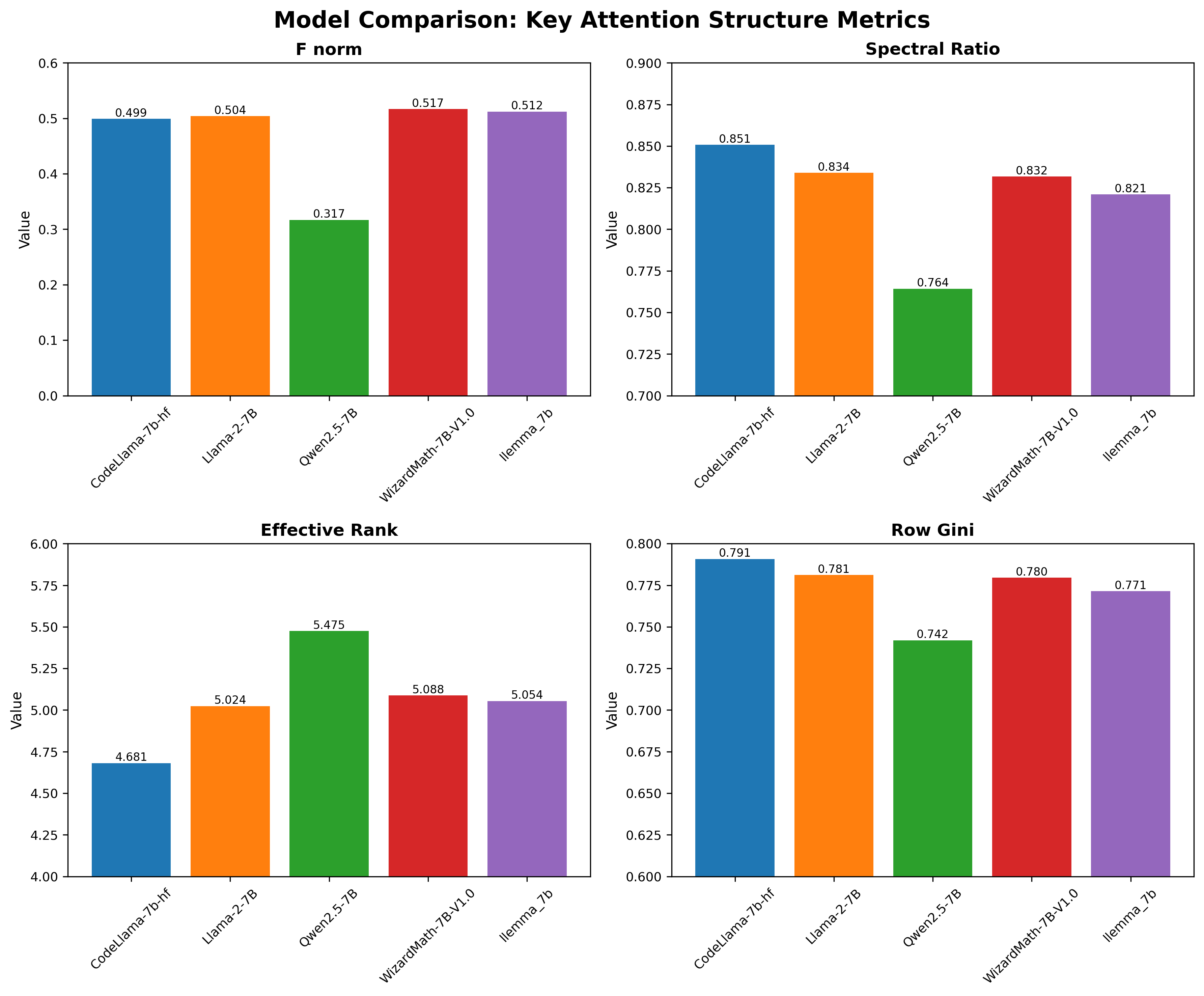}
    \caption{Key differential-attention structure metrics for Qwen2.5-7B and Llama-2-7B derivatives. Llama-2-7B derivatives exhibit consistent profiles, while Qwen2.5-7B differs in concentration dimensions.}
    \label{fig:pilot-metric-bars}
  \end{minipage}
\end{figure}

\begin{table}[H]
  \centering
\caption{Structural statistics (mean$\pm$SD) of the differential attention matrix $\Delta A$ under controlled semantic conflicts. Llama-2-7B derivatives exhibit highly similar statistics, while Qwen2.5-7B consistently differs in concentration-related metrics.}
\label{tab:pilot-routing-stats}
  \scriptsize
  \setlength{\tabcolsep}{3pt}
  \begin{adjustbox}{max width=\textwidth}

  \begin{tabular}{lcccccc}
    \toprule
    Model & $\|\Delta A\|_F$ & Spectral ratio & Effective rank & Col. Gini & Row Gini & $\Delta H$ \\
    \midrule
    CodeLlama-7b-hf        & $0.499\pm0.558$ & $0.851\pm0.145$ & $4.681\pm3.233$ & $0.690\pm0.105$ & $0.791\pm0.154$ & $0.003\pm0.046$ \\
    Llama-2-7B             & $0.504\pm0.531$ & $0.834\pm0.152$ & $5.024\pm3.408$ & $0.697\pm0.107$ & $0.781\pm0.159$ & $0.004\pm0.048$ \\
    WizardMath-7B-V1.0     & $0.517\pm0.520$ & $0.832\pm0.150$ & $5.088\pm3.440$ & $0.691\pm0.105$ & $0.780\pm0.160$ & $0.004\pm0.049$ \\
    llemma\_7b             & $0.512\pm0.533$ & $0.821\pm0.162$ & $5.054\pm3.424$ & $0.691\pm0.111$ & $0.771\pm0.167$ & $-0.001\pm0.053$ \\
    Qwen2.5-7B             & $0.317\pm0.313$ & $0.764\pm0.170$ & $5.475\pm3.297$ & $0.652\pm0.102$ & $0.742\pm0.176$ & $0.008\pm0.042$ \\
    \bottomrule
  \end{tabular}
  
\end{adjustbox}
  \par\smallskip
  \begin{tabular}{lc}
    \toprule
    Model & Locality \\
    \midrule
    CodeLlama-7b-hf & $0.208\pm0.192$ \\
    Llama-2-7B & $0.237\pm0.197$ \\
    WizardMath-7B-V1.0 & $0.236\pm0.196$ \\
    llemma\_7b & $0.218\pm0.201$ \\
    Qwen2.5-7B & $0.255\pm0.241$ \\
    \bottomrule
  \end{tabular}
  
\end{table}

\paragraph{Key findings.} \textbf{Discovery:} Llama-2-7B derivatives exhibit closely matched re-allocation profiles (Figures~\ref{fig:pilot-col-gini}, \ref{fig:pilot-metric-bars}; Table~\ref{tab:pilot-routing-stats}), while Qwen2.5-7B operates in a lower-concentration, more diffuse regime. \textbf{Validation:} PCA visualization (Figure~\ref{fig:pilot-pca}) shows models from different sources occupy separated regions, confirming that differential attention fingerprints distinguish provenance.

\paragraph{Implication for AttnDiff.} These findings inform our design: (i) origin/corrupted pairs amplify semantic-conflict signals; (ii) $\Delta A$ isolates re-routing by canceling prompt-specific baselines; (iii) SVD-based spectral descriptors capture richer structure than hand-crafted statistics; (iv) CKA enables robust similarity comparison. Main-text experiments evaluate robustness under realistic laundering transformations.
\FloatBarrier

\FloatBarrier
\setcounter{table}{0}
\renewcommand{\thetable}{A\arabic{table}}
\renewcommand{\theHtable}{A\arabic{table}}
\section{Additional Robustness Experiments}
\label{sec:appendix_robustness}

This section provides extended robustness evaluations across diverse model transformations and architectures. Specifically, Appendix~\ref{sec:appendix_po}--\ref{sec:appendix_pruning} present additional evaluations under preference optimization (Appendix~\ref{sec:appendix_po}), diverse model merge recipes (Appendix~\ref{sec:appendix_merge}), and extended pruning configurations (Appendix~\ref{sec:appendix_pruning}); Appendix~\ref{sec:appendix_cross_family_scale} validates cross-family and cross-scale stability; Appendix~\ref{sec:appendix_distill}--\ref{sec:appendix_quant} evaluate robustness under distillation (Appendix~\ref{sec:appendix_distill}), MoE architectures (Appendix~\ref{sec:appendix_moe}), and quantization (Appendix~\ref{sec:appendix_quant}).

Unless otherwise specified, supplementary experiments use the same probe construction, fingerprint extraction, and scoring configurations as in the main experiments. Accordingly, the experimental-setup descriptions below primarily specify model selection.

\subsection{Preference Optimization Robustness}
\label{sec:appendix_po}

This section reports the models and full experimental results for PO, covering PPO and DPO strategies.

\textbf{PPO.} Proximal Policy Optimization (PPO) is a policy-gradient algorithm for reinforcement learning that stabilizes training by clipping the policy ratio, and is widely used for reinforcement learning from human feedback (RLHF) in language models~\citep{ouyang2022instructgpt}.

\textbf{DPO.} Direct Preference Optimization (DPO) optimizes a language model directly from preference pairs via a closed-form objective that avoids explicit reward-model fitting and online RL rollouts~\citep{rafailov2023direct}.

\paragraph{Experimental Setup.} We evaluate PO robustness using Llama-2-7B-derived suspects under two representative PO methods, PPO-based reinforcement learning from human feedback (RLHF) and Direct Preference Optimization (DPO). Concretely, we consider PPO-aligned models (llama-2-7b-ppo-v0.1-reward and llama-2-7b-ppo-v0.1-policy) and DPO-aligned models (\texttt{tulu-2-dpo-7b} and \texttt{llama-2-7b-dpo}), all trained from the Llama-2-7B base model~\citep{touvron2023llama2}.

\begin{table}[!htbp]
  \centering
\caption{Preference optimization robustness results (similarity score) on Llama-2-7B-derived suspect models (PPO/DPO).}
\label{tab:po-robustness}
  \begin{adjustbox}{max width=0.95\columnwidth}

  \begin{tabular}{lcccc}
    \hline
     & \multicolumn{2}{c}{PPO} & \multicolumn{2}{c}{DPO} \\
     \cline{2-3} \cline{4-5}
     & \begin{tabular}[c]{@{}c@{}}llama-2-7b-\\ppo-v0.1-\\reward\end{tabular} & \begin{tabular}[c]{@{}c@{}}llama-2-7b-\\ppo-v0.1-\\policy\end{tabular} & \begin{tabular}[c]{@{}c@{}}tulu-2-dpo-\\7b\end{tabular} & \begin{tabular}[c]{@{}c@{}}llama-2-7b-\\dpo\end{tabular} \\
    \hline
    PCS      & \bestcell{1.0000} & \secondcell{0.9935} & \bestcell{0.9999} & \thirdcell{0.9972} \\
    ICS      & \secondcell{0.9999} & 0.9874 & \thirdcell{0.9961} & \secondcell{0.9984} \\
    Logits   & \thirdcell{0.9884} & \thirdcell{0.9891} & 0.6735 & 0.7144 \\
    REEF      & 0.6152 & 0.7458 & 0.9264 & 0.9496 \\
    ProFlingo & \bestcell{1.0000} & \bestcell{1.0000} & 0.0400& 0.2800 \\
    LLMMap    & 0.7836 & 0.7459 & 0.6780 & 0.8427 \\
    Ours      & \bestcell{1.0000} & \bestcell{1.0000} & \secondcell{0.9992} & \bestcell{0.9994} \\
    \hline
  \end{tabular}
  
\end{adjustbox}

\end{table}

\paragraph{Results and Analysis.} Table~\ref{tab:po-robustness} shows that AttnDiff remains highly stable under both PPO and DPO, reaching near-perfect similarity for all aligned suspects (\texttt{Ours} $\geq 0.9992$). Parameter-based baselines (PCS/ICS) also stay close to 1.0, reflecting the shared base checkpoint. By contrast, output- and text-level baselines degrade more substantially under DPO: Logits drops to $0.67$--$0.71$ and ProFlingo collapses to $0.04$--$0.28$, indicating that preference optimization can heavily reshape surface behavior. Representation-level REEF is comparatively robust on DPO-aligned suspects ($\approx 0.93$--$0.95$) but is notably weaker on PPO-aligned suspects ($\approx 0.62$--$0.75$), suggesting that different alignment strategies affect hidden-state similarity in distinct ways. Overall, AttnDiff provides a consistently reliable provenance signal across PO variants.

\FloatBarrier

\subsection{Model Merge Strategy Robustness}
\label{sec:appendix_merge}

\begin{table}[!htbp]
  \centering
\caption{Model merge robustness results (similarity score) under eight merging strategies.}
\label{tab:merge-strategies}
  \begin{adjustbox}{max width=\textwidth}

  \begin{tabular}{lcccccccc}
    \toprule
     & \multicolumn{2}{c}{Breadcrumbs} & \multicolumn{2}{c}{Breadcrumbs+Ties} & \multicolumn{2}{c}{Della} & \multicolumn{2}{c}{Task} \\
     \cmidrule(lr){2-3} \cmidrule(lr){4-5} \cmidrule(lr){6-7} \cmidrule(lr){8-9}
     & Llama-2-7B & Wizardmath & Llama-2-7B & Wizardmath & Llama-2-7B & Wizardmath & Llama-2-7B & Wizardmath \\
    \midrule
    PCS       & \secondcell{0.9990} & \bestcell{0.9999} & \thirdcell{0.9968}& \bestcell{0.9999} & \thirdcell{0.9864}& \secondcell{0.9999} & \secondcell{0.9921}& \bestcell{0.9999} \\
    ICS       & \thirdcell{0.9988} & \secondcell{0.9997} & \secondcell{0.9988} & \secondcell{0.9997} & \secondcell{0.9968} & \thirdcell{0.9982} & \bestcell{0.9996} & \secondcell{0.9996} \\
    Logits    & 0.8830 & \thirdcell{0.9575} & 0.8830 & \thirdcell{0.9575} & 0.7841 & 0.9556 & 0.9192 & \thirdcell{0.9484} \\
    REEF      & 0.8131 & 0.8831 & 0.8131 & 0.8831 & 0.9261 & 0.9289 & \thirdcell{0.9442} & 0.8717 \\
    ProFlingo & 0.5600 & 0.5200& 0.5600 & 0.3800& 0.4000 & 0.4400& 0.7400& 0.6200\\
    LLMMap    & 0.8368 & 0.8561 & 0.8368 & 0.8561 & 0.8275 & 0.8913 & 0.8498 & 0.8842 \\
    Ours      & \bestcell{0.9991} & \bestcell{0.9999} & \bestcell{0.9992} & \bestcell{0.9999} & \bestcell{0.9986} & \bestcell{1.0000} & \bestcell{0.9996} & \bestcell{0.9999} \\
    \midrule
    \midrule
     & \multicolumn{2}{c}{Ties} & \multicolumn{2}{c}{Della+Task} & \multicolumn{2}{c}{DARE+Ties} & \multicolumn{2}{c}{DARE+Task} \\
     \cmidrule(lr){2-3} \cmidrule(lr){4-5} \cmidrule(lr){6-7} \cmidrule(lr){8-9}
     & Llama-2-7B & Wizardmath & Llama-2-7B & Wizardmath & Llama-2-7B & Wizardmath & Llama-2-7B & Wizardmath \\
    \midrule
    PCS       & \secondcell{0.9990}& \bestcell{0.9999} & \secondcell{0.9981}& \secondcell{0.9999} & \thirdcell{0.9952}& \secondcell{0.9999} & \thirdcell{0.9964}& \secondcell{0.9999} \\
    ICS       & \thirdcell{0.9987} & \bestcell{0.9999} & \thirdcell{0.9968} & \thirdcell{0.9982} & \secondcell{0.9968} & \thirdcell{0.9982} & \secondcell{0.9968} & \thirdcell{0.9982} \\
    Logits    & 0.8624 & \secondcell{0.9751} & 0.7647 & 0.9655 & 0.7695 & 0.9708 & 0.7731 & 0.9784 \\
    REEF      & 0.9718 & 0.8455 & 0.9533 & 0.8232 & 0.8797 & 0.8946 & 0.8030 & 0.7519 \\
    ProFlingo & 0.5200 & 0.2200& 0.4200 & 0.4400& 0.4400 & 0.5000& 0.5600 & 0.3800\\
    LLMMap    & 0.8733 & \thirdcell{0.8489} & 0.7339 & 0.8967 & 0.7357 & 0.9323 & 0.7133 & 0.9245 \\
    Ours      & \bestcell{0.9991} & \bestcell{0.9999} & \bestcell{0.9985} & \bestcell{1.0000} & \bestcell{0.9984} & \bestcell{1.0000} & \bestcell{0.9985} & \bestcell{1.0000} \\
    \bottomrule
  \end{tabular}
  
\end{adjustbox}

\end{table}

This section evaluates whether AttnDiff remains robust under diverse model merging strategies, covering both weight-space and recipe-based merges across different base models.

\textbf{Breadcrumbs and Breadcrumbs+Ties.} Model Breadcrumbs merges models by learning sparse binary masks over source checkpoints to select a small subset of parameters from each model, and the Breadcrumbs+Ties variant combines this masking scheme with TIES-style averaging to further reduce interference~\citep{davari2023model,yadav2023ties}.

\textbf{Della and Task Arithmetic.} DELLA-Merging samples and rescales model parameters based on their magnitude to mitigate destructive interference during merging~\citep{deep2024della}, whereas Task Arithmetic (Task) linearly adds and subtracts task-specific model deltas to compose new behaviors~\citep{ilharco2022editing}.

\textbf{TIES and DARE-based merges.} TIES-Merging prunes and rebalances parameter updates before averaging to avoid conflicting signs~\citep{yadav2023ties}, while DARE-based recipes (DARE+TIES, DARE+Task) regularize the merged model by encouraging sparse and disentangled ability-specific updates~\citep{yu2023language}.

\paragraph{Experimental Setup.} We evaluate eight representative merging recipes: Breadcrumbs, Breadcrumbs+Ties, Della, Task Arithmetic (Task), TIES (TIES-Merging), Della+Task, DARE+TIES, and DARE+Task. For each recipe, we consider Llama-2-7B and Wizardmath-7b as victim models and compute similarity scores between the merged checkpoints and their respective bases.

\paragraph{Results and Analysis.} Table~\ref{tab:merge-strategies} shows that AttnDiff consistently maintains near-perfect similarity across all eight merging strategies and both base models (\texttt{Ours} $\geq 0.9984$ and frequently $\approx 1.0$), indicating that differential attention re-routing patterns remain stable even under diverse weight- and recipe-based merges. Several baselines degrade more noticeably under challenging recipes: ProFlingo remains low across most settings ($0.22$--$0.74$), and Logits/LLMmap exhibit non-trivial drops for some merges. REEF is generally stronger than text/output-only baselines but still shows variability across strategies and bases. While PCS/ICS remain highly competitive when parameter alignment is straightforward, they primarily reflect weight-space overlap and may not capture functional lineage under heterogeneous merge recipes. Overall, AttnDiff offers robust provenance verification across a broad merging landscape.

\FloatBarrier

\subsection{Pruning Robustness}
\label{sec:appendix_pruning}

\begin{table}[!htbp]
  \centering
\caption{Pruning robustness results (similarity score) under Random, $L_1$, and Taylor pruning criteria at different sparsity levels.}
\label{tab:pruning_robustness}
  \scriptsize
  \setlength{\tabcolsep}{2pt}
  \renewcommand{\arraystretch}{1.0}
  \begin{adjustbox}{max width=0.95\textwidth}

  \begin{tabular}{lcccccc}
    \toprule
     & \multicolumn{2}{c}{5\%} & \multicolumn{2}{c}{10\%} & \multicolumn{2}{c}{20\%} \\
     & Llama-2-7B & Qwen2.5-7B & Llama-2-7B & Qwen2.5-7B & Llama-2-7B & Qwen2.5-7B \\
    \cmidrule(lr){2-3} \cmidrule(lr){4-5} \cmidrule(lr){6-7}
    \multicolumn{7}{l}{\textbf{Random pruning}} \\
    PCS      & \thirdcell{0.9878} & \thirdcell{0.9869} & \thirdcell{0.9750} & \thirdcell{0.9738} & \thirdcell{0.9497} & \thirdcell{0.9468} \\
    ICS      & 0.9208 & 0.9502 & 0.8415 & 0.8954 & 0.6878 & 0.7715 \\
    Logits   & \secondcell{0.9923} & \bestcell{0.9999} & \secondcell{0.9875} & \secondcell{0.9928} & \secondcell{0.9839} & \bestcell{0.9832} \\
    REEF     & 0.6368 & 0.2417 & 0.3689 & 0.2259 & 0.2629 & 0.2021 \\
    ProFlingo & 0.8800 & 0.7200 & 0.4800 & 0.3600& 0.2000 & 0.1200     \\
    LLMmap   & 0.9422 & 0.7959 & 0.8471 & 0.7310 & 0.7409 & 0.7320 \\
    Ours     & \bestcell{0.9995} & \secondcell{0.9958} & \bestcell{0.9986} & \bestcell{0.9933} & \bestcell{0.9964} & \secondcell{0.9820} \\
    \midrule
    \multicolumn{7}{l}{\textbf{$L_1$ pruning}} \\
    PCS      & \bestcell{0.9999} & \secondcell{0.9999} & \secondcell{0.9994} & \thirdcell{0.9996} & \secondcell{0.9982} & \bestcell{0.9990} \\
    ICS      & \bestcell{0.9999} & \secondcell{0.9999} & \thirdcell{0.9991} & \secondcell{0.9997} & \thirdcell{0.9929} & \thirdcell{0.9976} \\
    Logits   & \secondcell{0.9980} & \thirdcell{0.9988} & 0.9949 & 0.9874 & 0.9902 & 0.9832 \\
    REEF     & 0.6546 & 0.2327 & 0.6157 & 0.2283 & 0.5557 & 0.2177 \\
    ProFlingo & 0.0600 & 0.0000 & 0.0000 & 0.0400 & 0.0000 & 0.0200     \\
    LLMmap   & \thirdcell{0.9329} & 0.9134 & 0.8932 & 0.9584 & 0.9131 & 0.8929 \\
    Ours     & \bestcell{0.9999} & \bestcell{1.0000} & \bestcell{0.9998} & \bestcell{0.9998} & \bestcell{0.9986} & \secondcell{0.9977} \\
    \midrule
    \multicolumn{7}{l}{\textbf{Taylor pruning}} \\
    PCS      & \bestcell{1.0000} & \secondcell{0.9999} & \bestcell{1.0000} & \bestcell{0.9999} & \bestcell{0.9999} & \bestcell{0.9999} \\
    ICS      & \secondcell{0.9999} & \secondcell{0.9999} & \thirdcell{0.9991} & \thirdcell{0.9997} & \thirdcell{0.9929} & \thirdcell{0.9977} \\
    Logits   & \thirdcell{0.9933} & \thirdcell{0.9998} & 0.9900 & 0.9985 & 0.9900 & 0.9897 \\
    REEF     & 0.9895 & 0.9813 & 0.9711 & 0.9772 & 0.9701 & 0.9639 \\
    ProFlingo & 0.5000 & 0.4000 & 0.2600 & 0.0600  & 0.0600 & 0.1200     \\
    LLMmap   & 0.9329 & 0.9135 & 0.8931 & 0.9585 & 0.9132 & 0.8929 \\
    Ours     & \bestcell{1.0000} & \bestcell{1.0000} & \secondcell{0.9999} & \secondcell{0.9998} & \secondcell{0.9997} & \secondcell{0.9994} \\
    \bottomrule
  \end{tabular}
  
\end{adjustbox}

\end{table}

This section presents supplementary pruning robustness experiments to evaluate whether AttnDiff remains stable under diverse structured and unstructured pruning configurations.

\textbf{Random pruning.} As a structure-agnostic baseline, Random pruning removes parameters (or channels/heads) uniformly at random under a given sparsity level, without using any importance signal from weights or gradients.

\textbf{$L_1$ pruning.} Magnitude-based pruning ranks parameters or structured units by their $L_1$ norm and removes those with the smallest overall magnitude.

\textbf{Taylor pruning.} Taylor-based pruning assigns an importance score to each parameter based on a first-order Taylor expansion of the loss, typically combining gradient information with weight magnitude.

\paragraph{Experimental Setup.}\label{sec:appendix_pruning_setup} We generate pruned suspects from Llama-2-7B and Qwen2.5-7B using the LLMPruner toolkit~\citep{ma2023llmpruner}, covering both structured and unstructured pruning with four criteria: Random, $L_1$, $L_2$, and Taylor.

\paragraph{Results and Analysis.} Table~\ref{tab:pruning_robustness} shows that AttnDiff remains highly stable across pruning criteria and sparsity levels, with similarity typically above $0.99$ and remaining strong even under 20\% random pruning (\texttt{Ours} $0.9964$ for Llama-2-7B and $0.9820$ for Qwen2.5-7B). In contrast, baselines exhibit criterion-dependent sensitivity: ICS drops sharply under random pruning (down to $0.69$--$0.77$ at 20\%), and ProFlingo becomes unreliable under magnitude pruning (near-zero under $L_1$ pruning). REEF is comparatively low for random/$L_1$ pruning but becomes very high under Taylor pruning. Overall, these results suggest that pruning can perturb weights and representations in heterogeneous ways, while AttnDiff provides a consistent provenance signal across sparsification regimes.

\FloatBarrier

\subsection{Cross-Family and Cross-Scale Stability}
\label{sec:appendix_cross_family_scale}

\begin{table}[H]
  \centering
\caption{Cross-scale and cross-family robustness results (similarity score) across Qwen2.5-derived suspects (1.5B/14B), gemma-2-derived suspects (2B), and Mistral-derived suspects (7B). For the Qwen2.5 block, \textit{Coder-1.5B}, \textit{Math-1.5B}, and \textit{1.5B-Instruct} are evaluated against \textbf{Qwen2.5-1.5B} as the victim (base), whereas \textit{14B-Instruct}, \textit{oxy-1-small}, and \textit{14B-Gutenberg} are evaluated against \textbf{Qwen2.5-14B}. Thus, there is no cross-size comparison within this block.}
\label{tab:cross_family_scale}
  \scriptsize
  \setlength{\tabcolsep}{2.5pt}
  \renewcommand{\arraystretch}{1.05}
  \begin{adjustbox}{max width=\textwidth}

  \begin{tabular}{lcccccccccccc}
    \toprule
     & \multicolumn{6}{c}{Qwen2.5} & \multicolumn{3}{c}{gemma-2-2b} & \multicolumn{3}{c}{Mistral-7B-v0.3} \\
    \cmidrule(lr){2-7} \cmidrule(lr){8-10} \cmidrule(lr){11-13}
     & \makecell{Coder-\\1.5B}
     & \makecell{Math-\\1.5B}
     & \makecell{1.5B-\\Instruct}
     & \makecell{14B-\\Instruct}
     & \makecell{oxy-1-\\small}
     & \makecell{14B-\\Gutenberg}
     & \makecell{neogenes-\\ita}
     & \makecell{baku-\\2b}
     & \makecell{2b-\\merged}
     & \makecell{AQUA-\\7B}
     & \makecell{spellcheck-\\mistral}
     & \makecell{demi-\\merge}
     \\
    \midrule
    PCS      & 0.8385 & 0.6681 & \bestcell{0.9999} & \bestcell{0.9999} & \bestcell{0.9961} & \bestcell{0.9999} & \bestcell{0.9996} & \bestcell{0.9971} & \bestcell{0.9996} & \bestcell{0.9999} & \bestcell{0.9998} & \bestcell{0.9999} \\
    ICS      & 0.2661 & 0.4486 & \thirdcell{0.9954} & \secondcell{0.9997} & \secondcell{0.9925} & \secondcell{0.9997} & \thirdcell{0.9954} & \secondcell{0.9880} & \thirdcell{0.9957} & \thirdcell{0.9820} & \thirdcell{0.9996} & \secondcell{0.9997} \\
    Logits   & 0.5275 & 0.2728 & 0.8832 & 0.6228 & 0.5571 & 0.6847 & 0.7615 & 0.8036 & 0.7355 & 0.6859 & 0.9076 & 0.9225 \\
    ProFlingo & \thirdcell{0.8400}& 0.8000& 0.7400& 0.7800& 0.6600& 0.5400& 0.6200& 0.6800& 0.5400& 0.5800& 0.6600& 0.7600\\
    LLMmap   & 0.8231 & \thirdcell{0.8593} & 0.6929 & 0.6737 & 0.7505 & 0.6221 & 0.7429 & \thirdcell{0.9655} & 0.6748 & 0.7658 & 0.8960 & 0.7940 \\
    REEF     & \secondcell{0.9458} & \secondcell{0.9155} & 0.9267 & \thirdcell{0.9857} & 0.9602 & 0.9705 & \secondcell{0.9961} & 0.9568 & \secondcell{0.9959} & 0.9792 & 0.9953 & 0.9950 \\
    Ours     & \bestcell{0.9869} & \bestcell{0.9471} & \secondcell{0.9968} & 0.9821 & \thirdcell{0.9883} & \thirdcell{0.9837} & 0.9194 & 0.9322 & 0.9107 & \secondcell{0.9936} & \secondcell{0.9997} & \thirdcell{0.9994} \\
    \bottomrule
  \end{tabular}
  
\end{adjustbox}

\end{table}

This section evaluates whether AttnDiff remains stable under \textbf{(i) different model families} and \textbf{(ii) different parameter scales}, including 1.5B$\rightarrow$14B within Qwen2.5 and \textbf{cross-family} settings (gemma-2 vs. Mistral).

\paragraph{Experimental Setup.} We evaluate AttnDiff across three model families: \textbf{Qwen2.5} (1.5B, 7B, 14B), \textbf{gemma-2} (2B), and \textbf{Mistral} (7B). For each family, we select a representative set of instruction-tuned, merged, or domain-specific derivatives as suspects.

\paragraph{Results and Analysis.} Table~\ref{tab:cross_family_scale} shows that AttnDiff remains robust across heterogeneous derivatives drawn from different model families and scales. Notably, on the more challenging cross-scale Qwen2.5 setting (e.g., 1.5B coder/math variants), AttnDiff achieves strong similarity ($0.947$--$0.987$), while parameter-based baselines can be substantially weaker (PCS $0.668$--$0.839$, ICS $0.266$--$0.449$), reflecting architectural and scale mismatches that limit direct weight alignment. For Mistral-derived suspects, AttnDiff remains near-perfect ($\geq 0.9936$). Overall, differential attention fingerprints provide a stable similarity signal when parameter alignment is unreliable and surface behavior can drift.

\FloatBarrier

\subsection{Distillation Robustness}
\label{sec:appendix_distill}

\begin{table}[!htbp]
  \centering
  \caption{Distillation robustness results (similarity score) across teacher--student pairs spanning Qwen2.5 and Llama families.}
  \label{tab:distill_robustness}
  \small
  \setlength{\tabcolsep}{3pt}
  \begin{adjustbox}{max width=\textwidth}
  \begin{tabular}{lcccc}
    \toprule
      & Qwen2.5-7B & Qwen2.5-14B & Llama-3.1-8B & Llama-2-7B \\
    \midrule
    Distilled model & \makecell{Qwen2.5-7B-\\Open-R1-Distill} & \makecell{DeepSeek-R1-\\Distill-Qwen-14B} & \makecell{Llama-3.1-8B-\\Instruct-Open-R1} & \makecell{llama-2-7b-logit-\\watermark-distill} \\
    \midrule
    PCS      & \bestcell{0.9998} & \secondcell{0.9867} & \bestcell{0.9988} & \secondcell{0.9997} \\
    ICS      & \secondcell{0.9992} & \thirdcell{0.9738} & \thirdcell{0.9961} & \thirdcell{0.9982} \\
    Logits   & 0.6990 & 0.5056 & 0.6926 & 0.9512 \\
    REEF     & \thirdcell{0.9926} & 0.7512 & 0.9779 & 0.9830 \\
    ProFlingo & 0.5600& 0.4400 & 0.5400& 0.4800\\
    LLMmap   & 0.6744 & 0.7637 & 0.7956 & 0.9287 \\
    Ours     & 0.9873 & \bestcell{0.9875} & \secondcell{0.9969} & \bestcell{0.9998} \\
    \bottomrule
  \end{tabular}
  \end{adjustbox}
\end{table}

\paragraph{Knowledge Distillation.} Knowledge distillation compresses a larger or more capable \emph{teacher} model into a lighter \emph{student} by training the student to match the teacher's predictions or intermediate representations.

\paragraph{Experimental Setup.} We evaluate AttnDiff under teacher--student distillation for four representative LLM families. On the \textbf{Qwen2.5} side, we consider Qwen2.5-7B and Qwen2.5-14B as base models and evaluate suspects distilled from reasoning-style teachers, namely \texttt{Qwen2.5-7B-Open-R1-Distill} and \texttt{DeepSeek-R1-Distill-Qwen-14B} (associated with the DeepSeek-R1 series~\citep{deepseek2024v3}). For \textbf{Llama} models, we include Llama-3.1-8B and Llama-2-7B as bases, together with distilled or distillation-like variants \texttt{Llama-3.1-8B-Instruct-Open-R1-Distill} and a logit-based distilled variant.

\paragraph{Results and Analysis.} Table~\ref{tab:distill_robustness} shows that AttnDiff remains consistently high across all evaluated teacher--student pairs (\texttt{Ours} $\geq 0.9873$), indicating that differential attention fingerprints are largely preserved under distillation. Parameter-based baselines (PCS/ICS) are also near-perfect. In contrast, output- and text-based baselines are less stable: Logits drops to around $0.50$--$0.70$ for the Qwen2.5 pairs. Overall, AttnDiff provides a robust provenance signal under commonly deployed distillation pipelines.

\FloatBarrier

\subsection{Applicability to Other Architectures (MoE)}
\label{sec:appendix_moe}

To further probe the applicability of AttnDiff beyond standard dense Transformer architectures, we consider a mixture-of-experts (MoE) setting using Mixtral-8x7B~\citep{jiang2024mixtralexperts} as the base model and several instruction-tuned derivatives.

\paragraph{Experimental Setup.} We treat Mixtral-8x7B as the victim model and evaluate AttnDiff on its instruction-tuned variants, including Dolly15K-tuned and DPO-aligned checkpoints as well as OpenChat Mixtral.

\begin{table}[!htbp]
  \centering
\caption{Similarity score between Mixtral-8x7B and its instruction-tuned MoE derivatives.}
\label{tab:moe_mixtral}
  \begin{adjustbox}{max width=\columnwidth}

  \begin{tabular}{lccc}
    \toprule
     & \multicolumn{3}{c}{Mixtral-8x7B} \\
     \cmidrule(lr){2-4}
     & Dolly15K & Nous-Hermes-2-DPO & openbuddy-v15.4 \\
    \midrule
    AttnDiff & 0.9925 & 0.9814 & 0.9905 \\
    \bottomrule
  \end{tabular}
  
\end{adjustbox}

\end{table}

\paragraph{Results and Analysis.} Table~\ref{tab:moe_mixtral} shows that AttnDiff remains stable on MoE variants of Mixtral-8x7B, with similarity scores in the $0.98$--$0.99$ range across instruction-tuning and DPO-alignment. This suggests that the differential-attention fingerprinting pipeline extends beyond dense Transformers and continues to provide a reliable provenance signal under architectural changes in the feed-forward routing mechanism.

\FloatBarrier

\subsection{Quantization Robustness}
\label{sec:appendix_quant}

\paragraph{GPTQ Quantization.} Post-training quantization compresses model weights to low-bit representations to reduce memory footprint and accelerate inference. We focus on GPTQ~\citep{frantar2023gptq}, a widely used post-training quantization method for GPT-style transformers that quantizes weights (e.g., INT4/INT8) with minimal degradation.

\paragraph{Experimental Setup.} We evaluate quantization robustness across four base model families, including Qwen2.5-7B, Llama-2-7B, Llama-3.1-8B, and Mistral-7B-v0.3. For each base model, we consider representative GPTQ-quantized derivatives at different bit widths and compute AttnDiff similarity between each quantized checkpoint and its corresponding full-precision base.

\begin{table}[!htbp]
  \centering
  \caption{Quantization robustness results (AttnDiff similarity score) between each full-precision base model and its GPTQ-quantized derivatives.}
  \label{tab:quant_robustness}
  \footnotesize
  \setlength{\tabcolsep}{3pt}
  \begin{adjustbox}{max width=\textwidth}
  \begin{tabular}{lcccccc}
    \toprule
      & \multicolumn{2}{c}{Qwen2.5-7B} & Llama-2-7B & \multicolumn{2}{c}{Llama-3.1-8B} & Mistral-7B-v0.3 \\
    \cmidrule(lr){2-3} \cmidrule(lr){4-4} \cmidrule(lr){5-6} \cmidrule(lr){7-7}
      & \makecell{Instruct-GPTQ-\\Int4} & \makecell{Instruct-GPTQ-\\Int8} & Chat-GPTQ & \makecell{Instruct-INT4-\\GPTQ} & \makecell{Instruct-GPTQ-\\Q\_8} & \makecell{Instruct-GPTQ-\\4bit} \\
    \midrule
    Ours & 0.9890 & 0.9913 & 0.9933 & 0.9450 & 1.0000 & 0.9970 \\
    \bottomrule
  \end{tabular}
  \end{adjustbox}
\end{table}

\paragraph{Results and Analysis.} Table~\ref{tab:quant_robustness} shows that AttnDiff remains highly stable under GPTQ quantization across all evaluated families, with similarity scores close to 1.0 for most INT4/INT8 settings. The slightly lower similarity on the INT4-quantized Llama-3.1-8B variant indicates that more aggressive quantization can introduce stronger weight perturbations, but AttnDiff still preserves a high provenance signal overall.

\FloatBarrier

\subsection{Stress Testing \textsc{LLMMap} under Text-level Attacks}
\label{sec:appendix_llmmap_attacks}

We provide a focused stress test of the \emph{text-based} provenance baseline \textsc{LLMMap}~\citep{pasquini2025llmmap}. Since \textsc{LLMMap} operates on generated responses, it may be sensitive to post-hoc rewriting that preserves semantics but perturbs lexical and stylistic cues.

\paragraph{Attack setup.} We attack \emph{only} the answers produced by each suspect model while leaving base model outputs unchanged. Using a local \texttt{WizardMath-7B-V1.0} model as the attacker, we rewrite each answer under three semantic-preserving transformations: \textbf{paraphrase} (synonym-based paraphrase), \textbf{rewrite} (stylistically distinct rewrite), and \textbf{style-transfer} (rewrite conditioned on a target writing style). After rewriting, we recompute \textsc{LLMMap} embeddings and fingerprints and measure cosine similarity against the clean fingerprints.

\paragraph{Results and Analysis.} Table~\ref{tab:paraphrase_results} reports the cosine similarity of \textsc{LLMMap} fingerprints before and after meaning-preserving paraphrase, rewrite, and style-transfer attacks. Performance varies widely across suspects: \texttt{CodeLlama-7b-hf} remains stable (clean $0.9535$ and $\geq 0.9484$ after attacks), while \texttt{llemma\_7b} and \texttt{WizardMath-7B-V1.0} suffer substantial degradation (e.g., \texttt{llemma\_7b}: $0.8886\rightarrow 0.715$--$0.743$). The logit-watermark distilled model also exhibits a measurable drop ($0.9233\rightarrow 0.885$--$0.893$). These results highlight a vulnerability of text-based provenance signals to post-hoc rewriting that preserves semantics but alters surface form, motivating complementary fingerprints derived from internal model dynamics.

\begin{table}[!htbp]

  \centering
\caption{Cosine similarity of \textsc{LLMMap} fingerprints before and after meaning-preserving paraphrasing, rewriting, and style-transfer attacks.}
\label{tab:paraphrase_results}
  \normalsize
  \renewcommand{\arraystretch}{1.2}
  \setlength{\tabcolsep}{4pt}
  \begin{adjustbox}{max width=\columnwidth}

  \begin{tabular}{lcccc}
    \toprule
    \textbf{Model} & \textbf{Clean} & \textbf{Paraphrase} & \textbf{Rewrite} & \textbf{Style-transfer} \\
    \midrule
    CodeLlama-7b-hf & 0.9535 & 0.9484 & 0.9484 & 0.9500 \\
    llemma\_7b & 0.8886 & 0.7174 & 0.7152 & 0.7433 \\
    \makecell[l]{llama-2-7b-logit-\\watermark-distill} & 0.9233 & 0.8934 & 0.8904 & 0.8851 \\
    WizardMath-7B-V1.0 & 0.6998 & 0.6147 & 0.6210 & 0.6330 \\
    \bottomrule
  \end{tabular}
  
\end{adjustbox}

\end{table}

\FloatBarrier

\section{Broader Impacts}
\label{sec:broader_impacts}

While \attndiff{} provides a practical building block for model provenance and IP protection in the open-weight ecosystem, we acknowledge potential negative societal impacts. Specifically, inaccurate provenance attribution could lead to unwarranted copyright disputes or legal accusations against legitimate open-source developers, potentially chilling collaborative innovation. Mitigating these risks requires combining technical verification with human-in-the-loop auditing and legal frameworks that account for the statistical nature of fingerprinting signals.

\section{Model List}
\label{sec:appendix_model_list}

This section provides a comprehensive list of the models used across all experiments in the main text and appendix. We categorize these models by modification or attack setting (e.g., fine-tuning, model merging, pruning, distillation). For each category, we specify the base (victim) model used to establish the reference fingerprint and the corresponding suspect (derivative) models evaluated for provenance verification.

\begin{table}[p]
  \centering
\caption{Models used in the main paper and appendix analyses. The evaluation spans fine-tuning, merging, pruning, preference optimization, distillation, and quantization across multiple families and scales.}
\label{tab:model_list}
  \footnotesize
  \setlength{\tabcolsep}{3pt}
  \renewcommand{\arraystretch}{1.00}
  \begin{tabularx}{\textwidth}{@{}>{\raggedright\arraybackslash}p{0.13\textwidth}>{\raggedright\arraybackslash}p{0.15\textwidth}>{\raggedright\arraybackslash}p{0.20\textwidth}>{\raggedright\arraybackslash}X@{}}
    \toprule
    \textbf{Category} & \textbf{Type} & \textbf{Base (Victim)} & \textbf{Suspect(s) / Derivative(s)} \\
    \midrule
    Fine-tuning & Instruction & Llama-2-7B & Llama-2-finance-7b, Vicuna-1.5-7b, WizardMath-7b, Chinese-LLaMA-2-7b, CodeLLaMA-7b, Llemma-7b \\
    \midrule
    Merge & Weight & Shisa-gamma-7b-v1, WizardMath-7b-1.1, Abel-7b-002 & Evollm-jp-7b \\
    \cmidrule(lr){2-4}
    & Dist./Behav. & Llama-2-7B, OpenLLaMA-2-7b, mpt-7b & Fusellm-7b \\
    \midrule
    \multirow{2}{*}{Pruning} & Structured & Llama-2-7B & Sheared-llama-1.3b, Sheared-llama-1.3b-pruned, Sheared-llama-1.3b-sharegpt, Sheared-llama-2.7b, Sheared-llama-2.7b-pruned \\
    \cmidrule(lr){2-4}
    & Unstructured & Llama-2-7B & Sparse-llama-2-7b, Wanda-llama-2-7b, GBLM-llama-2-7b \\
    \midrule
    \multirow{2}{*}{Ablation} & Related & Llama-2-7B & CodeLlama-7b, Llama-2-finance-7B, Vicuna-7B-v1.5, Chinese-Llama-2-7B, WizardMath-7B-V1.0, llemma\_7b, Sheared-LlaMA-1.3B, Sheared-LlaMA-1.3B-Pruned, Sheared-LlaMA-1.3B-ShareGPT, Sheared-LlaMA-2.7B, Sheared-LlaMA-2.7B-Pruned, Sheared-LlaMA-2.7B-ShareGPT, Sparse-llama-2-7b, Wanda-llama-2-7b, GBLM-llama-2-7b \\
    \cmidrule(lr){2-4}
    & Unrelated & Llama-2-7B & Llama3-8B, mpt-7b, Qwen2.5-1.5B, Qwen2.5-3B, Qwen2.5-7B, Qwen2.5-14B, Qwen2.5-Math-7B, gemma-2-2b, Gemma-7B-it, Yi-6B \\
    \midrule
    Pilot & Discovery/\allowbreak Validation & Llama-2-7B & Llama-2-7B, CodeLlama-7b-hf, WizardMath-7B-V1.0, llemma\_7b, Qwen2.5-7B \\
    \midrule
    Pref. Opt. & PPO/DPO & Llama-2-7B & llama-2-7b-ppo-v0.1-reward\textsuperscript{1}, llama-2-7b-ppo-v0.1-policy\textsuperscript{1}, tulu-2-dpo-7b\textsuperscript{2}, llama-2-7b-dpo\textsuperscript{3} \\
    \midrule
    Cross-Family/Scale & Qwen2.5 & Qwen2.5-7B & Qwen2.5-Coder-1.5B\textsuperscript{4}, Qwen2.5-Math-1.5B\textsuperscript{5}, Qwen2.5-1.5B-Instruct\textsuperscript{6} \\
    & & Qwen2.5-14B & Qwen2.5-14B-Instruct\textsuperscript{7}, oxy-1-small\textsuperscript{8}, Qwen2.5-14B-Gutenberg-Instruct-Slerpeno\textsuperscript{9} \\
    \cmidrule(lr){2-4}
    & Gemma-2 & gemma-2-2b & gemma-2-2b-neogenes-ita\textsuperscript{10}, gemma-2-baku-2b\textsuperscript{11}, gemma-2-2b-merged\textsuperscript{12} \\
    \cmidrule(lr){2-4}
    & Mistral & Mistral-7B-v0.3 & AQUA-7B\textsuperscript{13}, spellcheck-mistral-7b\textsuperscript{14}, Mistral-7B-Instruct-demi-merge-v0.3-7B\textsuperscript{15} \\
    \midrule
    Distill & Reasoning & Llama-3.1-8B, Qwen2.5-7B, Qwen2.5-14B & Llama-3.1-8B-Instruct-Open-R1-Distill\textsuperscript{16}, Qwen2.5-7B-Open-R1-Distill\textsuperscript{17}, DeepSeek-R1-Distill-Qwen-14B\textsuperscript{18} \\
    \cmidrule(lr){2-4}
    & Logit-based & Llama-2-7B & llama-2-7b-logit-watermark-distill-kgw-k1-gamma0.25-delta2\textsuperscript{19} \\
    \midrule
    MoE & Mixtral & Mixtral-8x7B & Instruct\_Mixtral-8x7B-v0.1\_Dolly15K\textsuperscript{20}, Nous-Hermes-2-Mixtral-8x7B-DPO\textsuperscript{21}, openbuddy-mixtral-8x7b-v15.4\textsuperscript{22} \\
    \midrule
    Quantization & GPTQ & Qwen2.5-7B & Qwen2.5-7B-Instruct-GPTQ-Int8\textsuperscript{23}, Qwen2.5-7B-Instruct-GPTQ-Int4\textsuperscript{24} \\
    & & Llama-2-7B & Llama-2-7B-Chat-GPTQ\textsuperscript{25} \\
    & & Llama-3.1-8B & Meta-Llama-3.1-8B-Instruct-GPTQ-Q\_8\textsuperscript{26}, LLaMA-3.1-8B-Instruct-INT4-GPTQ\textsuperscript{27} \\
    & & Mistral-7B-v0.3 & Mistral-7B-Instruct-v0.3-GPTQ-4bit\textsuperscript{28} \\
    \bottomrule
  \end{tabularx}

\end{table}

\begin{table}[tbp]
  \centering
\caption{Repository URLs corresponding to the superscript IDs in Table~\ref{tab:model_list}. All checkpoints are publicly available on Hugging Face.}
\label{tab:model_repo_links}
  \footnotesize
  \setlength{\tabcolsep}{4pt}
  \renewcommand{\arraystretch}{1.2}
  \begingroup
  \urlstyle{same}
  \def\UrlFont{\footnotesize\rmfamily}
  \begin{tabularx}{\textwidth}{@{}r >{\raggedright\arraybackslash}X@{}}
    \toprule
    \textbf{ID} & \textbf{Repository URL} \\
    \midrule
    1  & \url{https://huggingface.co/renyiyu/llama-2-7b-ppo-lora-v0.1} \\
    2  & \url{https://huggingface.co/allenai/tulu-2-dpo-7b} \\
    3  & \url{https://huggingface.co/mncai/llama2-7b-dpo-v1} \\
    4  & \url{https://huggingface.co/Qwen/Qwen2.5-Coder-1.5B} \\
    5  & \url{https://huggingface.co/Qwen/Qwen2.5-Math-1.5B} \\
    6  & \url{https://huggingface.co/Qwen/Qwen2.5-1.5B-Instruct} \\
    7  & \url{https://huggingface.co/Qwen/Qwen2.5-14B-Instruct} \\
    8  & \url{https://huggingface.co/oxyapi/oxy-1-small} \\
    9  & \url{https://huggingface.co/v000000/Qwen2.5-14B-Gutenberg-Instruct-Slerpeno} \\
    10 & \url{https://huggingface.co/anakin87/gemma-2-2b-neogenesis-ita} \\
    11 & \url{https://huggingface.co/rinna/gemma-2-baku-2b} \\
    12 & \url{https://huggingface.co/vonjack/gemma2-2b-merged} \\
    13 & \url{https://huggingface.co/KurmaAI/AQUA-7B} \\
    14 & \url{https://huggingface.co/openfoodfacts/spellcheck-mistral-7b} \\
    15 & \url{https://huggingface.co/grimjim/Mistral-7B-Instruct-demi-merge-v0.3-7B} \\
    16 & \url{https://huggingface.co/asas-ai/Llama-3.1-8B-Instruct-Open-R1-Distill} \\
    17 & \url{https://huggingface.co/erickrus/Qwen2.5-7B-Open-R1-Distill} \\
    18 & \url{https://huggingface.co/deepseek-ai/DeepSeek-R1-Distill-Qwen-14B} \\
    19 & \url{https://huggingface.co/cygu/llama-2-7b-logit-watermark-distill-kgw-k1-gamma0.25-delta2} \\
    20 & \url{https://huggingface.co/Brillibits/Instruct_Mixtral-8x7B-v0.1_Dolly15K} \\
    21 & \url{https://huggingface.co/NousResearch/Nous-Hermes-2-Mixtral-8x7B-DPO} \\
    22 & \url{https://huggingface.co/openbuddy/openbuddy-mixtral-8x7b-v15.4} \\
    23 & \url{https://huggingface.co/Qwen/Qwen2.5-7B-Instruct-GPTQ-Int8} \\
    24 & \url{https://huggingface.co/Qwen/Qwen2.5-7B-Instruct-GPTQ-Int4} \\
    25 & \url{https://huggingface.co/TheBloke/Llama-2-7B-Chat-GPTQ} \\
    26 & \url{https://huggingface.co/iqbalamo93/Meta-Llama-3.1-8B-Instruct-GPTQ-Q_8} \\
    27 & \url{https://huggingface.co/DaraV/LLaMA-3.1-8B-Instruct-INT4-GPTQ} \\
    28 & \url{https://huggingface.co/RedHatAI/Mistral-7B-Instruct-v0.3-GPTQ-4bit} \\
    \bottomrule
  \end{tabularx}
  \endgroup

\end{table}

The adaptive experiment in Sec.~\ref{sec:exp_adaptive} uses \texttt{google/gemma-2-2b} as its frozen reference and \texttt{rinna/gemma-2-baku-2b} as the adapted suspect. The gray-box pilot in Appendix~\ref{sec:gray_box} uses the same reference and related suspect, together with \texttt{Qwen/Qwen2.5-3B-Instruct} as an unrelated control.
\FloatBarrier

\clearpage
\section*{NeurIPS Paper Checklist}


\begin{enumerate}

\item {\bf Claims}
    \item[] Question: Do the main claims made in the abstract and introduction accurately reflect the paper's contributions and scope?
    \item[] Answer: \answerYes{}
    \item[] Justification: Main claims made in the abstract and introduction accurately reflect the paper's contributions and scope.

\item {\bf Limitations}
    \item[] Question: Does the paper discuss the limitations of the work performed by the authors?
    \item[] Answer: \answerYes{}
    \item[] Justification: We discuss limitations in the Limitations section.

\item {\bf Theory assumptions and proofs}
    \item[] Question: For each theoretical result, does the paper provide the full set of assumptions and a complete (and correct) proof?
    \item[] Answer: \answerYes{}
    \item[] Justification: Propositions with assumptions and proofs provided in Appendix~\ref{sec:theoretical_proof}.

    \item {\bf Experimental result reproducibility}
    \item[] Question: Does the paper fully disclose all the information needed to reproduce the main experimental results of the paper to the extent that it affects the main claims and/or conclusions of the paper (regardless of whether the code and data are provided or not)?
    \item[] Answer: \answerYes{}
    \item[] Justification: Implementation details in Appendix~\ref{sec:attndiff_details}, model lists in Appendix~\ref{sec:appendix_model_list}, code at \url{https://github.com/zhb0119/AttnDiff}.

\item {\bf Open access to data and code}
    \item[] Question: Does the paper provide open access to the data and code, with sufficient instructions to faithfully reproduce the main experimental results, as described in supplemental material?
    \item[] Answer: \answerYes{}
    \item[] Justification: Code and data are open-source.

\item {\bf Experimental setting/details}
    \item[] Question: Does the paper specify all the training and test details (e.g., data splits, hyperparameters, how they were chosen, type of optimizer) necessary to understand the results?
    \item[] Answer: \answerYes{}
    \item[] Justification: Settings detailed in Sec.~\ref{sec:exp_setting} and Appendix~\ref{sec:attndiff_details}.

\item {\bf Experiment statistical significance}
    \item[] Question: Does the paper report error bars suitably and correctly defined or other appropriate information about the statistical significance of the experiments?
    \item[] Answer: \answerNo{}
    \item[] Justification: Results are deterministic with no standard deviation observed.

\item {\bf Experiments compute resources}
    \item[] Question: For each experiment, does the paper provide sufficient information on the computer resources (type of compute workers, memory, time of execution) needed to reproduce the experiments?
    \item[] Answer: \answerYes{}
    \item[] Justification: Computational cost and hardware requirements discussed in Sec.\ 6 (Discussion) and Appendix~\ref{sec:appendix_computational_cost}.
    
\item {\bf Code of ethics}
    \item[] Question: Does the research conducted in the paper conform, in every respect, with the NeurIPS Code of Ethics \url{https://neurips.cc/public/EthicsGuidelines}?
    \item[] Answer: \answerYes{}
    \item[] Justification: Our research conforms with the NeurIPS Code of Ethics in every respect.

\item {\bf Broader impacts}
    \item[] Question: Does the paper discuss both potential positive societal impacts and negative societal impacts of the work performed?
    \item[] Answer: \answerYes{}
    \item[] Justification: Our method supports IP protection for LLMs.
    
\item {\bf Safeguards}
    \item[] Question: Does the paper describe safeguards that have been put in place for responsible release of data or models that have a high risk for misuse (e.g., pre-trained language models, image generators, or scraped datasets)?
    \item[] Answer: \answerNA{}
    \item[] Justification: This paper poses no such risks.

\item {\bf Licenses for existing assets}
    \item[] Question: Are the creators or original owners of assets (e.g., code, data, models), used in the paper, properly credited and are the license and terms of use explicitly mentioned and properly respected?
    \item[] Answer: \answerYes{}
    \item[] Justification: The assets used in the paper are properly credited and respected.

\item {\bf New assets}
    \item[] Question: Are new assets introduced in the paper well documented and is the documentation provided alongside the assets?
    \item[] Answer: \answerYes{}
    \item[] Justification: New assets introduced in the paper were well documented and the documentation is provided alongside the assets.

\item {\bf Crowdsourcing and research with human subjects}
    \item[] Question: For crowdsourcing experiments and research with human subjects, does the paper include the full text of instructions given to participants and screenshots, if applicable, as well as details about compensation (if any)? 
    \item[] Answer: \answerNA{}
    \item[] Justification: No crowdsourcing or human subjects involved.

\item {\bf Institutional review board (IRB) approvals or equivalent for research with human subjects}
    \item[] Question: Does the paper describe potential risks incurred by study participants, whether such risks were disclosed to the subjects, and whether Institutional Review Board (IRB) approvals (or an equivalent approval/review based on the requirements of your country or institution) were obtained?
    \item[] Answer: \answerNA{}
    \item[] Justification: No human subjects research.

\item {\bf Declaration of LLM usage}
    \item[] Question: Does the paper describe the usage of LLMs if it is an important, original, or non-standard component of the core methods in this research? Note that if the LLM is used only for writing, editing, or formatting purposes and does \emph{not} impact the core methodology, scientific rigor, or originality of the research, declaration is not required.
    \item[] Answer: \answerNA{}
    \item[] Justification: LLMs are research subjects, not method components.

\end{enumerate}

\end{document}